# Formation of asteroid pairs by rotational fission

P. Pravec[1], D. Vokrouhlický[2], D. Polishook[3], D. J. Scheeres[4], A. W. Harris[5], A. Galád[6,1], O. Vaduvescu[7,8], F. Pozo[7], A. Barr[7], P. Longa[7], F. Vachier[9], F. Colas[9], D. P. Pray[10], J. Pollock[11], D. Reichart[12], K. Ivarsen[12], J. Haislip[12], A. LaCluyze[12], P. Kušnirák[1], T. Henych[1], F. Marchis[13,14], B. Macomber[13,14], S.A. Jacobson[15], Yu. N. Krugly[16], A.V. Sergeev[16] & A. Leroy[17]

[1]*Astronomical Institute AS CR, Fričova 1, CZ-25165 Ondřejov, Czech Republic.* [2]*Institute of Astronomy, Charles University, V Holešovičkách 2, CZ-18000 Prague, Czech Republic.* [3]*Wise Observatory and Department of Geophysics and Planetary Sciences, Tel-Aviv University, Israel.* [4]*Department of Aerospace Engineering Sciences, University of Colorado, Boulder, CO, USA.* [5]*Space Science Institute, 4603 Orange Knoll Ave., La Canada, CA 91011, USA.* [6]*Modra Observatory, Comenius University, Bratislava SK-84248, Slovakia.* [7]*Instituto de Astronomia, Universidad Catolica del Norte, Avenida Angamos 0610, Antofagasta, Chile.* [8]*Isaac Newton Group of Telescopes, E-38700 Santa Cruz de la Palma, Canary Islands, Spain.* [9]*IMCCE-CNRS-Observatoire de Paris, 77 avenue Denfert Rochereau, 75014, Paris, France.* [10]*Carbuncle Hill Observatory, W. Brookfield, MA, USA.* [11]*Physics and Astronomy Dept., Appalachian State University, Boone, NC, USA.* [12]*Physics and Astronomy Department, University of North Carolina, Chapel Hill, NC, USA.* [13]*University of California at Berkeley, Berkeley CA, USA.* [14]*SETI Institute, Mountain View CA, USA.* [15]*Department of Astrophysical and Planetary Sciences, University of Colorado, Boulder, CO, USA.* [16]*Institute of Astronomy of Kharkiv National University, Sumska Str. 35, Kharkiv 61022, Ukraine.* [17]*Observatoire Midi Pyrénées and Association T60, Pic du Midi, France.*

**Asteroid pairs sharing similar heliocentric orbits were found recently**[1–3]**. Backward integrations of their orbits indicated that they separated gently with low relative velocities, but did not provide additional insight into their formation** **mechanism. A previously hypothesized rotational fission process[4] may explain their formation – critical predictions are that the mass ratios are less than about 0.2 and, as the mass ratio approaches this upper limit, the spin period of the larger body becomes long. Here we report photometric observations of a sample of asteroid pairs revealing that primaries of pairs with mass ratios much less than 0.2 rotate rapidly, near their critical fission frequency. As the mass ratio approaches 0.2, the primary period grows long. This occurs as the total energy of the system approaches zero requiring the asteroid pair to extract an increasing fraction of energy from the primary's spin in order to escape. We do not find asteroid pairs with mass ratios larger than 0.2. Rotationally fissioned systems beyond this limit have insufficient energy to disrupt. We conclude that asteroid pairs are formed by the rotational fission of a parent asteroid into a proto-binary system which subsequently disrupts under its own internal system dynamics soon after formation.**

Analyses of the orbits of asteroid pairs reveal some common properties. Asteroid pairs are ubiquitous, found throughout the main belt and Hungarias[5]. Pair members are separated with low relative velocities on the order of metres per second in the space of proper elements ($d_{prop}$), and with the best studied asteroid pair (6070)-(54827) having a speed after escape of only 0.17 m/s (ref. 3). They are young with most pairs separated less than 1 Myr ago (Supplementary Information). The existence of a population of bound binary systems at a similar range of sizes suggests that related processes may account for both binary and pair formation. Previous investigations have indicated that binaries form from parent bodies spinning at a critical rate by some sort of fission or mass shedding process[4,6,7] and that binaries formed by fission will initially have chaotic orbit and spin evolution[8]. The free energy of a binary system formed by fission, defined as the total energy (kinetic and potential) minus the self-potentials of each component[9], is found here to play a fundamental role in the evolution of a binary. Systems with mass ratios less than ~0.2 have a positive free energy and can escape under internal dynamics, while systems with greater mass ratios have a negative free energy and cannot. As the system mass ratio approaches this limit, kinetic energy for escape is drawn from the primary rotation leaving it rotating at a slower rate. This predicts that primaries of pairs with small mass ratios rotate nearer to their critical fission period, as the mass ratio approaches 0.2 the primary period grows long, and beyond this limit disruption of the binary is not possible. (See Supplementary Information.)

We studied the relative sizes, spin rates, and shapes between pairs and binaries via a photometric observational program. Our sample consists of 35 asteroid pairs, listed in Table 1. Thirty two of them were taken from ref. 2, the pair (6070)-(54827) was from ref. 3, and the pairs (48652)-(139156) and (229401)-2005 $UY_{97}$ were identified with backward orbit integrations of the pair's components. The only selection criterion, other than the pair identification procedures used in the above works, was that the pairs occurred in favorable conditions for photometric observations with available telescopes (brightness, position in the sky). Our sample covers a range of sizes 1.9-7.0 km for primaries and 1.2-4.5 km for secondaries, with the median of 3.2 and 1.9 km, respectively, as estimated from the asteroids' absolute magnitudes[6], assuming geometric albedos according to their orbital group membership[12].

We integrated orbits of the asteroid pairs backward to 1 Myr before the present using techniques developed in refs 1–3 and achieved convergence for 31 of the 35 asteroid pairs in our sample. That strengthened their identification as real, genetically related pairs, and also provided estimates of ages of the pairs (*T*). The four pairs for which convergence was not achieved may be older than 1 Myr (see Supplementary Information).

We estimate the size ratio ($X$) and mass ratio ($q$) between the components of an asteroid pair from the difference between their absolute magnitudes ($\Delta H \equiv H_2 - H_1$) using the relation between an asteroid's absolute magnitude ($H$), effective diameter ($D$), and geometric albedo[6]. Assuming that the components have the same albedo and bulk density, the size ratio is $X \equiv D_2/D_1 = 10^{-\Delta H/5}$ and the mass ratio $q \equiv M_2/M_1 = X^3$, where $M_1$, $M_2$ is the mass of the primary and secondary (see Discussion 2 in Supplementary Information). A dominating uncertainty source in the size and mass ratio estimate are uncertainties of the estimated absolute magnitudes of the components. For all asteroids in this work, we have used absolute magnitude estimates obtained as a by-product of asteroid astrometric observations published in the AstDyS catalog (also used in ref. 2). We estimated a mean uncertainty of the catalog absolute magnitudes for our studied paired asteroids to be $\delta H \sim 0.3$ mag. This propagates to a relative uncertainty of the estimated size ratio of 20%.

Figure 1 presents the observed primary period $P_1$ vs $\Delta H$ data for the 32 primaries. The primary spin rates show a correlation with the mass ratio $q$ between members of the asteroid pairs. Specifically, primaries of pairs with small secondaries ($\Delta H > 2.0$, $q < 0.06$) rotate rapidly and their spin periods have a narrow distribution from 2.53 to 4.42 hr with a median of 3.41 hr, near the critical fission rotational period. As the mass ratio approaches the predicted approximate cutoff limit of 0.2, the primary period grows long. The correlation coefficient between $\omega_1^2 = (2\pi/P_1)^2$ and $q$ is -0.73 and the Student's $t$ statistic is -5.94 for the degrees of freedom of 30 of the sample that shows that the correlation is significant at a level higher than 99.9% (see Supplementary Information).

It is notable that the distribution of primary rotations of pairs with small secondaries is similar to rotations of primaries of orbiting binaries with similar sizes and size ratios. From a database of binary system parameters[6], main belt and Hungaria binaries with primary diameters $D_1$ = 2–10 km and size ratios $D_2/D_1 < 0.4$ ($q < 0.06$) have primary rotation periods from 2.21 to 4.41 hr, with a median of 2.92 hr.

The observed primary spin rate distribution is consistent with pair formation via the mutual escape of a transient proto-binary system formed from the rotational fission of a critically spinning parent body. In Fig. 1 we show theoretical curves that incorporate constraints on the expected spin period of a primary assuming a system that is initially fissioned and later undergoes escape. For these computations, systems are given an initial angular momentum consistent with a critical rotation rate as observed in orbiting binary systems[6]. Then, assuming conservation of energy and angular momentum we evaluate the spin period of the larger body if the binary orbit undergoes escape. For simplicity we assume planar systems. Shape ratios of the primary and secondary are incorporated into the energy and angular momentum budget[6] and in conjunction with a range of initial angular momentum values lead to the envelopes shown in the figure. (The mathematical formulation of the model is given in Supplementary Information.)

The observed data for paired asteroids show consistency with the theoretical curves from the simple model of the post-fission system of two components starting in close proximity, as explained above. This starting condition and subsequent process is consistent with the rotational fission model[4,8,13] which models the parent asteroid as a contact binary-like asteroid with components resting on each other. A mechanism to spin the asteroid up to its critical rotation frequency is provided by the Yarkovsky–O'Keefe–Radzievskii–Paddack (YORP) effect[14]. When the angular momentum of the system is increased sufficiently the components can enter orbit about each other, what we term rotational fission. The fission spin rate is a function of the mass ratio between the components, for large mass ratios or highly ellipsoidal shapes the spin period can be significantly longer than the surface disruption limit of a rapidly spinning sphere[4]. The free energy of the proto-binary system is also a strong function of the mass ratio between the components and is relatively independent of the mutual shapes of the pairs that enter fission, with the theory predicting that systems with $q$ less than about 0.20 will have a positive free energy and systems with greater mass ratios a negative free energy. Taking shapes into account, there is some variability about this mass ratio value with the dividing mass ratio ranging up to 0.28 for more distended shapes[8].

Further evidence indicates that asteroid fission as described here may not be the only process at work forming multi-component asteroid systems. There is a disparity between the lightcurve amplitudes of the primary components in asteroid pairs and binary systems (Fig. 2). This cannot be explained with the mechanism of rotational fission only, as all asteroid systems that undergo rotational fission are initially unstable, regardless of the degree of elongation[8]. Though a more elongated primary may be more efficient at ejecting a secondary, numerical simulations show that ejection can occur even for systems with moderate elongation (Supplementary Information). It suggests that either some process occurs during the formation process of a stable binary asteroid to form a nearly symmetric shape of the primary or that a different formation mechanism is at work[7]. Further observational and theoretical studies must be carried out to discover a cause for this pattern.

The fission mechanism that describes our observations is independent of mineralogical properties and only depends on mechanical/gravitational interactions between two mass components. Given this and the ubiquity of asteroid pairs as well as binary asteroids, and the occurrence of binaries among the major taxonomic types, S and C, we can speculate that the formation of asteroid binaries is driven by mechanical and not mineralogical properties.

1. Vokrouhlický, D. & Nesvorný, D. Pairs of asteroids probably of a common origin. *Astron. J.* **136,** 280–290 (2008).

2. Pravec, P. & Vokrouhlický, D. Significance analysis of asteroid pairs. *Icarus* **204,** 580–588 (2009).

3. Vokrouhlický, D. & Nesvorný, D. The common roots of asteroids (6070) Rheinland and (54827) 2001 NQ8. *Astron. J.* **137,** 111–117 (2009).

4. Scheeres, D. J. Rotational fission of contact binary asteroids. *Icarus* **189,** 370–385 (2007).

5. Milani, A., Knežević, Z., Novaković, B. & Cellino, A. Dynamics of the Hungaria asteroids. *Icarus* **207**, 769-794 (2010).

6. Pravec, P. & Harris, A. W. Binary asteroid population. 1. Angular momentum content. *Icarus* **190,** 250–259 (2007).

7. Walsh, K. J., Richardson, D. C. & Michel, P. Rotational breakup as the origin of small binary asteroids. *Nature* **454,** 188–191 (2008).

8. Scheeres, D. J. Stability of the planar full 2-body problem. *Celest. Mech. Dyn. Astr.* **104,** 103–128 (2009).

9. Scheeres, D. J. Stability in the Full Two-Body Problem. *Celest. Mech. Dyn. Astr.* **83,** 155–169 (2002).

10. Warner, B. D. Asteroid lightcurve analysis at the Palmer Divide Observatory: 2008 May – September. *Minor Planet Bull.* **36,** 7–13 (2009).

11. Galád, A., Kornoš, L. & Világi, J. An ensemble of lightcurves from Modra. *Minor Planet Bull.* **37,** 9–15 (2009).

12. Warner, B. D., Harris, A. W. & Pravec, P. The asteroid lightcurve database. *Icarus* **202,** 134–146 (2009).

13. Scheeres, D. J. Minimum energy asteroid reconfigurations and catastrophic disruptions. *Planetary and Space Science* **57,** 154–164 (2009).

14. Bottke, W. F. Jr, Vokrouhlický, D., Rubincam, D. P. & Nesvorný, D. The Yarkovsky and Yorp Effects: Implications for asteroid dynamics. *Annu. Rev. Earth Planet. Sci.* **34,** 157–191 (2006).

**Supplementary Information** is linked to the online version of the paper at

http://www.nature.com/nature/journal/v466/n7310/extref/nature09315-s1.pdf.

**Acknowledgements** Research at Ondřejov was supported by the Grant Agency of the Czech Republic. D.V. was supported by the Czech Ministry of Education. D.P. was supported by an *Ilan Ramon* grant from the Israeli Ministry of Science, and is grateful for the guidance of N. Brosch and D. Prialnik. D.J.S. and S.A.J. acknowledge support by NASA's PG&G and OPR research programs. A.W.H. was supported by NASA and NSF. Work at Modra Observatory was supported by the Slovak Grant Agency for Science. The observations at Cerro Tololo were performed with the support of CTIO and Joselino Vasquez, using telescopes operated by SMARTS Consortium. Work at Pic du Midi Observatory was supported by CNRS – Programme de Planétologie. Operations at Carbuncle Hill Observatory were supported by the Planetary Society's Gene Shoemaker NEO Grant. Support for PROMPT has been provided by the NSF. F.M. and B.M. were supported by the NSF. We thank to O. Bautista, T. Moulinier and P. Eclancher for assistance with observations with the T60 on Pic du Midi.

**Author Contributions** All named authors made significant contributions to this work, often to more than just one task, thus this work is a result of the team and individual contributions cannot always be isolated. In following we only specify the most significant contributions by individual authors. P.P. led the project and worked on most of its parts, except SI sections 1 and 4. D.V. performed the backward orbit integrations and contributed to interpretations of the results. D.P. ran the observations and data reductions and contributed to interpretations. D.J.S. developed the fission theory and worked out its implications for the observation data A.W.H. contributed to interpretations and implications of the data. A.G., O.V., F.P., A.B., P.L. F.V., F.C., D.P.P., J.P., D.R., K.I., J.H., A.L., O.K., T.H., F.M., B.M., Yu.N.K., A.V.S. and A.L. carried out the observations, data reductions and analyses. S.A.J. ran simulations of the satellite ejection process in a proto-binary after fission.

**Author Information** Correspondence and requests for materials should be addressed to P.P. (ppravec@asu.cas.cz).

**Table 1: Parameters of asteroid pairs**

| Asteroid pair | $d_{prop}$ (m s$^{-1}$) | $T$ (kyr) | $\Delta H$ | $P_1$ (h) | $\delta P_1$ (h) | $U_1$ | $A_1$ (mag) | $P_2$ (h) | $\delta P_2$ (h) | $U_2$ | $A_2$ (mag) |
|---|---|---|---|---|---|---|---|---|---|---|---|
| 1979-13732 | 9.02 | >1,000 | 0.7 | | | | | 8.2987 | 0.0004 | 3 | 0.28 |
| 2110-44612 | 3.36 | >1,000 | 2.2 | 3.34474 | 0.00002 | 3 | 0.38 | 4.9070 | 0.0002 | 3 | 0.44 |
| 4765-2001XO105 | 3.49 | >90 | 3.8 | 3.6260 | 0.0002 | 3 | 0.56 | | | | |
| 5026-2005WW113 | 13.99 | 17 ± 2 | 4.1 | 4.4243 | 0.0003 | 3 | 0.49 | | | | |
| 6070-54827 | 8.09 | 17.2 ± 0.3 | 1.5 | 4.2733 | 0.0004 | 3 | 0.42 | 5.8764 | 0.0005 | 3 | 0.25 |
| 7343-154634 | 19.91v | >800 | 2.9 | 3.7547 | 0.0004 | 3 | 0.20 | | | | |
| 9068-2002OP28 | 34.15v | ~32 | 4.1 | 3.406 | 0.004 | 3 | 0.20 | | | | |
| 10484-44645 | 0.43 | >130 | 0.9 | 5.508 | 0.002 | 3 | 0.21 | | | | |
| 11842-228747 | 0.71 | >150 | 2.6 | 3.68578 | 0.00009 | 3 | 0.13 | | | | |
| 15107-2006AL54 | 2.07 | >300 | 2.6 | 2.530 | 0.002 | 3 | 0.14 | | | | |
| 17198-229056 | 0.93 | >100 | 2.6 | 3.2430 | 0.0002 | 3 | 0.13 | | | | |
| 19289-2006YY40 | 6.31v | 640 ± 50 | 2.3 | 2.85 | 0.01 | 3 | 0.16 | | | | |
| 21436-2003YK39 | 5.88 | $70^{+70}_{-35}$ | 3.3 | 2.87 | 0.03 | 3 | 0.08 | | | | |
| 23998-205383 | 3.25 | >300 | 1.2 | 13.526 | 0.004 | 3 | 1.0 | 5.554 | 0.004 | 3 | 0.30 |
| 38707-32957 | 1.01 | >1,000 | 1.1 | 6.1509 | 0.0004 | 3 | 0.36 | | | | |
| 40366-78024 | 27.22v | >350 | 1.2 | | | | | >17 | | 2 | >0.12 |
| 48652-139156 | 1.38 | >650 | 1.1 | 13.829 | 0.004 | 3 | 0.63 | | | | |
| 51609-1999TE221 | 1.49 | >300 | 1.5 | 6.767 | 0.002 | 3 | 0.42 | | | | |
| 52773-2001HU24 | 2.07 | >250 | 2.1 | 3.7083 | 0.0003 | 3 | 0.35 | | | | |
| 52852-2003SC7 | 1.51 | >300 | 2.0 | 5.432 | 0.002 | 2 | 0.19 | | | | |
| 54041-220143 | 0.56 | >125 | 1.8 | 18.86 | 5 | 2 | 0.23 | 3.502 | 0.004 | 3 | 0.10 |
| 56232-115978 | 40.31v | >60 | 1.1 | 5.6 | 1 | 2 | 0.47 | 2.9 | 0.4 | 2 | 0.11 |
| 60744-218099 | 8.18 | 350 ± 50 | 1.1 | | | | | 5.03 | 0.04 | 2 | 0.27 |
| 63440-2004TV14 | 0.21 | $50^{+55}_{-20}$ | 2.3 | 3.2969 | 0.0002 | 3 | 0.17 | | | | |
| 69142-127502 | 4.81 | >400 | 1.0 | 7.389 | 0.002 | 3 | 0.55 | | | | |

| | | | | | | | |
|---|---|---|---|---|---|---|---|
| 76111-2005JY103 | 0.50 | >120 | 1.8 | 5.3 | 0.2 | 2 | 0.13 |
| 84203-2000SS4 | 3.46 | >100 | 1.0 | 17.73 | 1 | 2 | 0.62 |
| 88259-1999VA117 | 0.61 | $60^{+50}_{-15}$ | 2.2 | 4.166 | 0.002 | 3 | 0.12 |
| 88604-60546 | 1.61 | >1,000 | 1.3 | 7.178 | 0.001 | 3 | 0.55 |
| 92336-143662 | 2.82 | >300 | 1.1 | 28.212 | 0.002 | 3 | 0.44 |
| 101065-2002PY103 | 1.34 | >300 | 1.8 | 4.977 | 0.002 | 3 | 0.42 |
| 101703-142694 | 13.82v | 700 ± 100 | 1.9 | 3.899 | 0.001 | 3 | 0.29 |
| 139537-210904 | 6.61 | >400 | 1.5 | 45 | 15 | 2 | 0.1 |
| 226268-2003UW156 | 11.23v | >350 | 1.3 | 31.2 | 1.4 | 2 | 0.36 |
| 229401-2005UY97 | 11.57v | $17^{+50}_{-15}$ | 1.0 | 28 | 11 | 2 | 0.8 |

Data on spin rates of paired asteroids were taken with our photometric observations from the following observatories: Carbuncle Hill (W. Brookfield, MA, USA), Cerro Tololo, Dark Sky (University of North Carolina, USA), La Silla, Lick, Maidanak, Modra, Observatoire de Haute-Provence, Pic du Midi, and Wise (Tel-Aviv University, Israel). We also included published data for the paired asteroid (9068) 1993 OD (refs 10,11). In total we estimated rotation periods (P) and amplitudes (A) for 32 primaries and 8 secondaries of the 35 asteroid pairs in our sample. (Details given in Supplementary Information.) The asteroid designations in the 1$^{st}$ column are given in a compact form, i.e., the asteroid numbers without parentheses and the designations without space between the year and the additional numbers plain and not as subscript, to facilitate readability. The ''v'' flags in the 2nd column indicate a long-term instability of the heliocentric orbits that affected (increased) the estimated $d_{prop}$ value (ref. 2). The error bars of the $T$ estimates represent a 90% probability interval estimated as described in Supplementary Information. In the 7$^{th}$ and the 11$^{th}$ column, we give quality code ratings (U) for the estimated periods as described in ref. 12.

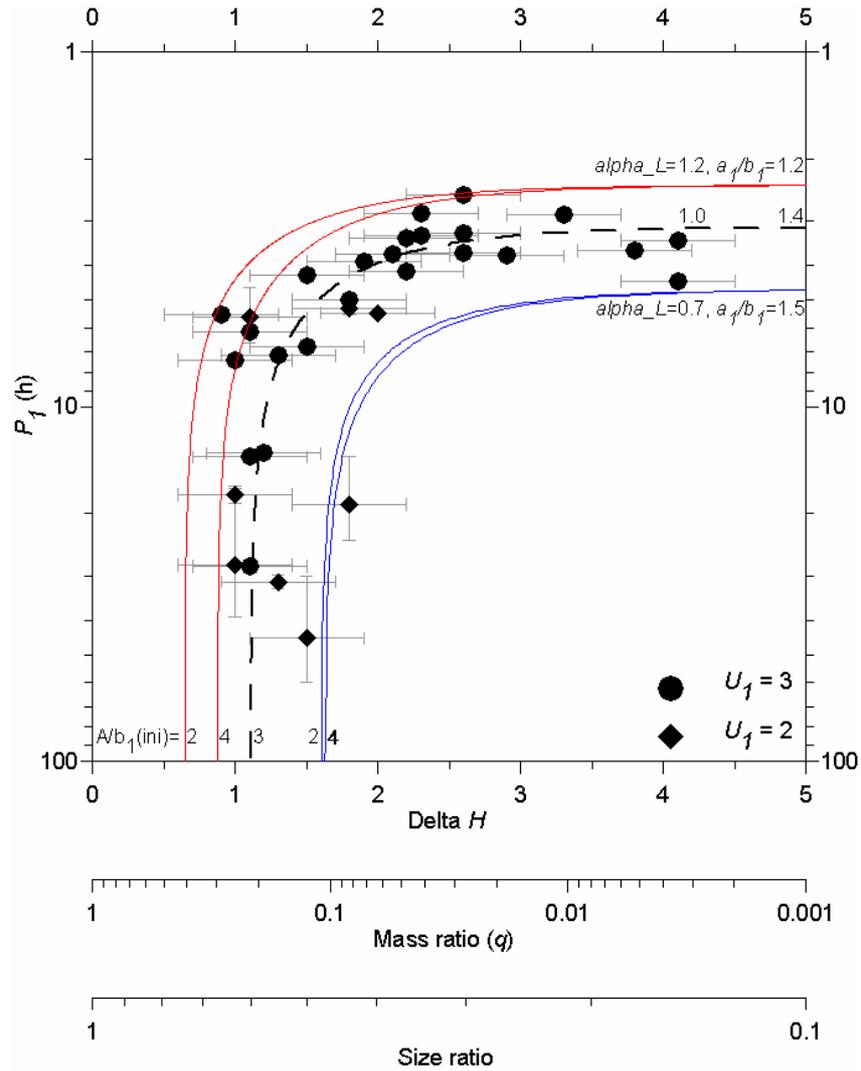

Figure 1. Primary rotation periods $P_1$ vs mass ratios $q$ of asteroid pairs. The mass ratio values were estimated from the differences between the absolute magnitudes of the pair components $\Delta H$. Error bars represent estimated standard errors of the values. The curves were generated with the model of pair separation from the post-fission transient proto-binary. The dashed curve is for the set of parameters best representing the pairs properties and the red and blue curves represent upper and lower limit cases. The upper curves are for the system's scaled angular momentum $\alpha_L = 1.2$, primary's axial ratio $a_1/b_1 = 1.2$, and initial orbit's normalized semi-major axis $A_{ini}/b_1 = 2$ and 4. The lower curves are for $\alpha_L = 0.7$, $a_1/b_1 = 1.5$, and $A_{ini}/b_1 = 2$ and 4. The choice of $a_1/b_1 = 1.2$ for the upper limit cases is because the four primaries closest to the upper limit curve have low amplitudes $A_1 = 0.1$–0.2 mag. Similarly, the choice of $a_1/b_1 = 1.5$ for the lower limit cases is because the point closest to the lower limit curve has the amplitude $A_1 = 0.49$, suggesting the equatorial elongation ~ 1.5.

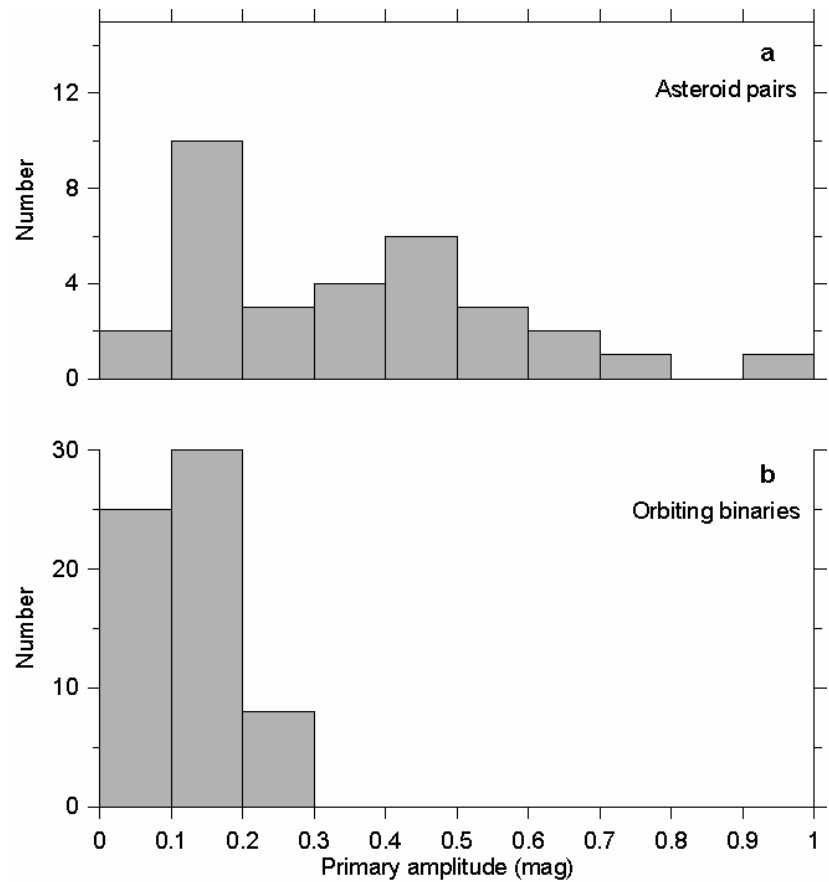

Figure 2. A disparity between the lightcurve amplitudes of the primary components in asteroid pairs and binary systems. The amplitudes of primaries of asteroid pairs (a) are distributed relatively randomly, and achieve high values in general while primaries of binary asteroids (b) have more subdued amplitudes. Either asteroids with shapes closer to rotational symmetry are more prone to form stable binaries, or some process occurs during the formation process of a binary asteroid to create such a shape.

# Supplementary Information

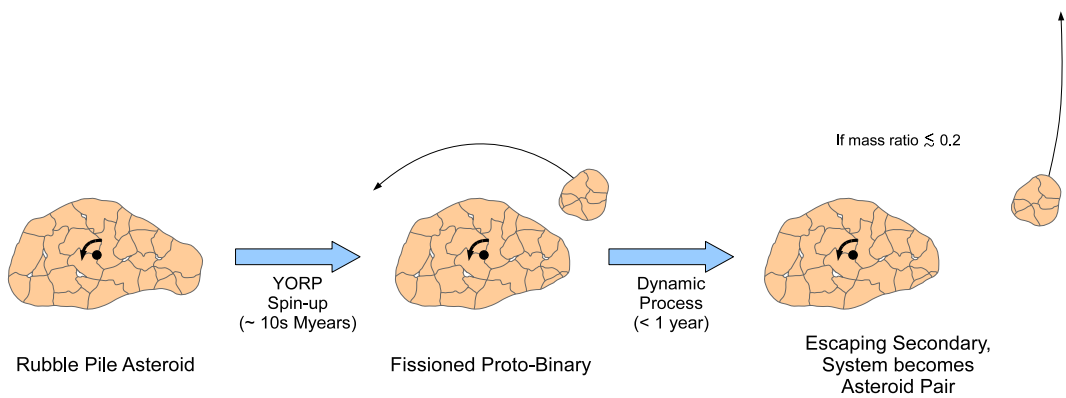

Supplementary Figure 1: A parent asteroid consisting of small component pieces that can be pulled apart without tensile resistance is spun up to the critical fission frequency by the Yarkovsky-O'Keefe-Radzievskii-Paddack (YORP) effect, forms a proto-binary system which subsequently disrupts under its own internal dynamics and becomes an asteroid pair. This is a conceptual sketch based on important model assumptions/limitations.

# 1 Pair ages estimation

## 1.1 Initial data and integrator used

The orbit integrations were nominally run back to the time 500 ky before present. For some pairs though, the backward integrations showed a possibility for convergence of the orbits not far beyond the nominal integration time, so we extended it to 1 My for them.

As described in refs 1, 3, it is insufficient to perform backward integration of the nominal solutions for the paired-asteroids as given by the orbit determination from observations. There are at least two reasons: (i) observation uncertainties that directly project onto the uncertainty of the orbit determination, and (ii) uncertainty in the force model that may not corrupt the orbit determination procedure but may affect the orbit evolution on a long term (thousands of years and more). The first topic is accounted for by considering statistically equivalent past evolution of the integrated orbits that all reside inside the uncertainty hyper-ellipsoid at the current epoch. This is typically a rather confined zone in the orbital elements space[*] but it may quickly stretch to the past. For instance, just taking the typical uncertainty factor $\delta a/a \sim 10^{-7} - 10^{-5}$ in the determination of the semimajor axis, the simple Keplerian shear would cause lost of determinism in computing the mean longitude in orbit in $\sim \left(\frac{a}{3\delta a}\right) P_{\text{orb}} \sim 0.1 - 10$ My ($P_{\text{orb}}$ is the orbital period). Things are, however, worse because the lost of determinism is actually not dominated by the initial orbital uncertainty but more by the underlying chaoticity of the dynamical problem of gravitational perturbations due to the planets. The relevant timescale is typically expressed in terms of the Lyapunov time that ranges from $\sim 10$ ky to $\sim 1$ My for the orbits we are interested in, depending on whether they are close to some prominent orbital resonances or not. To account for these effects, we have to consider a statistically equivalent multitude of the past orbital evolutions of a given orbit randomly sampling the uncertainty ellipsoid at the current epoch (taken into account as the initial data). In practice, we typically take $50-70$ such initial data and call them "geometric clones".

---

[*] In our work we consider the best-fit orbital elements and their uncertainties as determined by the `OrbFit` software and free-available through the `AstDyS` website at http://hamilton.dm.unipi.it/astdys/

The problem of the propagation model uncertainty (ii) above is dominated by our lack of information about the thermal (Yarkovsky) forces acting differentially on both components in the pair[14,15]. These forces make primarily the semimajor axis of the orbit secularly drift with $(da/dt)$ as large as $\sim 10^{-4}\,\mathrm{AU/My} \sim 5 \times 10^{-8}$ km/s for kilometer sized objects in the inner main belt (the $(da/dt)$ may be positive or negative within this range of amplitude). The effect may be though smaller for larger objects and/or favorably oriented spin axis. Because we do not know anything about the strength of the Yarkovsky effect on the asteroids in the pairs at this moment we have to consider different past orbital histories with different values of the Yarkovsky drift in the semimajor axis. Since the Yarkovsky forces are likely to be dominated by the diurnal variant[14,15], for which $(da/dt) \propto \cos\gamma$ ($\gamma$ is the obliquity of the spin axis), we should consider a uniform distribution of $(da/dt)$ value within the limits set by $\cos\gamma = \pm 1$ values. Different variants of the past histories with different strength of the Yarkovsky forces, hereafter called "Yarkovsky clones", quickly diverge from each other. The effect is again fastest in the longitude in orbit and the total lost of predictability occurs on a timescale $\sim \left(\frac{v_{\mathrm{orb}}}{3\pi\,(da/dt)}\right)^{1/2} P_{\mathrm{orb}} \sim 20$ ky only for $v_{\mathrm{orb}} \sim 20$ km/s and the maximum $(da/dt)$ estimated above. Note that this value is (i) very short, for most pairs shorter than the Lyapunov timescale, and (ii) is not affected by improvements in the orbit determination by acquiring new astrometry observations, but only can improve by observations that constrain the strength of the Yarkovsky effect (such as the size and/or pole determination). For that reason the uncertainty due to dynamical model incompleteness appears to be the prime reason for our inability to accurately reconstruct the past fate of the asteroid-pairs orbits.

In practice, we typically used $40-70$ Yarkovsky clones *for each of the geometric clones*, uniformly sampling the interval of $(da/dt)$ values between the minimum and maximum values estimated for the asteroid's size. This provided between 2000 and 5000 clones altogether for each of the components in a given pair.

The past orbit of each of these clones was propagated using the SWIFT_MVS integrator (e.g., http://www.boulder.swri.edu/~hal/swift.html and ref. 16). All planetary perturbations were included and planets' positions, together with physical parameters such as masses, taken from the JPL DE405 ephemeris file. The Yarkovsky acceleration was modeled as an along-track acceleration with an amplitude $\frac{1}{2}n\frac{na}{v_{\rm orb}}\left(\frac{da}{dt}\right)$, where $n$ is the mean motion and $v_{\rm orb}$ orbital velocity[1,3]. Integration timestep was 5 days and we stored the state vectors of all integrated clones every 10 years (only in some cases of intense search for the age limit of the pair we had a denser output every 1 year).

## 1.2  How close do we want the components to converge?

Performing the integrations described above we end-up, for each of the pairs we are interested in, with a typically 10-30 GB output file that keeps track of the state vectors for each of the clones for the two asteroids. In the analysis phase we want to judge whether the possible past orbits of the asteroids converged to each other and how well. For the latter we need some quantitative criterion that would sift through our output data and suggest success or failure.

Obviously, most of the pairs have at the starting epoch quite different values of the longitude in orbit for the two asteroids. So the two clouds of clones (for each of the component in the pair) initially start well separated in space, typically of the order of astronomical unit or more. However, as the Keplerian shear, chaoticity effects and accumulated differential motion in longitude of different Yarkovsky clones start to spread the clones along the whole Keplerian ellipses some of the clones approach closer. Eventually, the two clouds of clones may overlap (or near-overlap) in some parts of space bringing some of the clones very close each other. We are primarily interested to know the minimum distance in space to which some clones of the two asteroids in a pair can be brought. But, how close is close in quantitative terms?

For instance, given the fact that asteroids in pairs have frequently their initial semimajor axes close to few times $10^{-4}$ AU, it might not be surprising to get the closest clones at a distance of $\sim 50 - 100$ thousand kilometers which would be just the opposition distance if all other orbital elements are equal. Obviously a factor of few may be expected because of slight differences in other orbital elements. It has been suggested in refs 1, 3 that the ultimate-goal criterion to be met for the past convergence of the paired-asteroids orbits is to bring them closer in space than the estimated Hill radius of the parent asteroid:† $R_{\text{Hill}} \sim aD\frac{1}{2}\left(\frac{4\pi}{9}\frac{G\rho}{\mu}\right)^{-1/3}$ where $a$ is the semimajor axis of the pair-components orbits, $D$ the estimated size of the parent body (roughly obtained to have a volume equal to sum of volumes of the two components), $G$ is the gravitational constant, $\rho$ is the bulk density and $\mu$ is the gravitational parameter of the Sun. Assuming $a$ in AU and $D$ in kilometers, we have $R_{\text{Hill}} \sim 90\,aD$ km. For a multikilometer parent object in the inner zone of the asteroid belt we have $R_{\text{Hill}}$ of the order of several hundreds to a thousand of kilometers.

However, bringing the two clones at the $R_{\text{Hill}}$ distance is just part of the condition we want to meet. The case of an incredible fluke might be recognized by checking the relative velocity of the two clones when their distance is smaller than $R_{\text{Hill}}$. Having in mind the model of initially gentle separation of the two asteroids in the pair, the relative velocity must be smaller (in fact a fraction) of the escape velocity from the parent object: $v_{\text{esc}} \sim D\frac{1}{2}\left(\frac{8\pi}{3}G\rho\right)^{-1/2}$. Again, plugging in characteristic parameters we obtain $v_{\text{esc}} \sim 0.6\,D$ m/s, where the size $D$ of the parent object is in kilometers. For $D$ in the kilometer size range the expected relative velocities are of the order of m/s. Assuming for simplicity again two nearby orbits with a semimajor axis separation of $\sim 10^{-3}$ AU we may expect their relative velocity at opposition $\sim v_{\text{orb}}\frac{\delta a}{2a}$, about 10 m/s. So a few tens of m/s relative velocities between two nearby orbits at opposition, hence close encounter, is not that demanding condition. In reality, though, we shall see below that really good solutions we shall reach will be characterized by relative velocities of decimeters per second or less, a really small fraction of the estimated escape velocities.

---

†This is where actually our model should fail to represent the true orbital history of the clones because we consider them massless particles.

## 1.3 Examples

Hereafter we give some examples of successful past-convergence simulations for orbits of asteroids in pairs and therefore constrains on their age.

The couple (21436) Chaoyichi and 2003 YK$_{39}$ is a very tight pair residing in the inner part of the main asteroid belt. Luckily the orbits happen to fall aside the nearby mean motion resonance M7/12 with Mars and their past orbital reconstruction is not largely troubled by this source (the Lyapunov timescale for Chaoyichi's orbit is $\sim 22$ ky; http://hamilton.dm.unipi.it/astdys/). The orbits are reasonably well constrained such that the semimajor axis values have relative uncertainties of $\sim 2.5 \times 10^{-8}$ and $\sim 2.5 \times 10^{-7}$, respectively. The larger component in the pair, (21436) Chaoyichi, is estimated to have roughly $\sim 2$ km size (for $0.2$ albedo) and the smaller component, 2003 YK$_{39}$, is a sub-kilometer object with a size $\sim 0.7$ km (for $0.2$ albedo). The composite parent object thus might have a size of $\sim 2.1$ km and we roughly estimate the Hill radius of its gravitational influence to be $R_{\text{Hill}} \sim 600$ km and the escape velocity $v_{\text{esc}} \sim 1.2$ m/s.

Supplementary Figure 2 summarizes the principal information from our propagation of 2500 clones for each of the two asteroids. At each of the output instants, separated by $10$ y, we randomly selected 5 million trial identifications between the clones of the primary and secondary components and determined their minimum distance in space and the relative velocity of these closest pair of clones. These data are shown in the middle and bottom parts of Suppl. Fig. 2. We note the minimum distances reach about $\sim 50$ km, well inside the Hill sphere of influence of either the parent object or the primary. Additionally, these very close clones move with a typical relative velocity of few centimeters per second only, with minima reaching below a centimeter per second. At the top panel of Suppl. Fig. 2 we show a distribution of successful pair matches for which a relative distance was smaller than $R_{\rm Hill}$ and their relative velocity was smaller than $v_{\rm esc}$. We note this distribution is quasi-Maxwellian with a median value of 70 ky. Constructing a cumulative distribution of these successful matches we may estimate the characteristic range of ages for this pair by neglecting the first and the last 5% cases in the distribution. With that we obtain the best estimated age of this pair to be $70^{+70}_{-35}$ ky. The distribution is skewed toward the younger ages, as also seen in Suppl. Fig. 2. This is a real effect in the simulation that has to do with dilution of the clone clouds for the two components as the time increases to the past.

We now turn to another example, namely a pair (88259) 2001 HJ$_7$ and 1999 VA$_{117}$ which resides in the Hungaria zone. As in the previous case, the strongest near-by mean motion resonance M7/10 with Mars does not perturb these asteroid orbits and we can reliably reconstruct their past evolution (the Lyapunov timescale for 2001 HJ$_7$ is 1.3 Myr; ref. 5 and `http://hamilton.dm.unipi.it/astdys/`). The orbits are well constrained with relative uncertainties of the semimajor axis value of $\sim 2 \times 10^{-8}$ and $\sim 2.5 \times 10^{-8}$, respectively. The larger component in the pair, (88259) 2001 HJ$_7$, is estimated to have roughly $\sim 2.4$ km size (for $0.3$ albedo appropriate for the Hungaria population) and the smaller component, 1999 VA$_{117}$, has a size $\sim 1$ km (for $0.3$ albedo). The composite parent object thus might have a size of $\sim 2.5$ km and we roughly estimate the Hill radius of its gravitational influence to be $R_{\rm Hill} \sim 600$ km and the escape velocity $v_{\rm esc} \sim 1$ m/s.

The past orbital evolution of the two asteroids, and their clones, has been analysed using the same approach as above and the result is shown in Suppl. Fig. 3. The situation is very similar in the previous and this pair such that the minimum distances reached between the trial-pairs of different clones we near to $\sim 50$ km with relative velocities of a few centimeters per second only. This is less then 10% of the estimated escape velocity from the parent body and this shows that the smaller component, 1999 VA$_{117}$, obtained just barely enough relative energy to escape the gravitational well of the primary component separating from it very gently. The analysis of the distribution of successful clone identifications at different times, top panel in Suppl. Fig. 3, indicates an age of $60^{+50}_{-15}$ ky for this pair. This age corresponds well to the result found in ref. 5, where they obtained more distant close approaches of these two asteroids because their analysis did not include the Yarkovsky forces.

Our last example is that of a pair (229401) 2005 $SU_{152}$ and 2005 $UY_{97}$ that resides in the middle zone of the asteroid belt. Proper elements of the primary‡ indicate its location in the asteroid Iannini family, that itself is very young[17]. Our analysis below indicates, that this pair should be much younger than the family itself (with an estimated age between 2-5 My; ref. 17). On a longer term, the past evolution of orbits in this pair is affected by the nearby J11/4 mean motion resonance with Jupiter, but luckily this does not seen to greatly affect shorter-term propagation. The orbit of the smaller component, 2005 $UY_{97}$ has only been recovered at a second opposition in September 2009 and thus is not very accurate. The same is actually true for the primary. The relative uncertainties of the semimajor axis values are $\sim 3 \times 10^{-7}$ and $\sim 9 \times 10^{-7}$, respectively, in this case. Henceforth, the possibilities of predicting/constraining the past orbital evolution for this pair will improve after more astrometry observations are taken during the next few years. The larger component in the pair, (229401) 2005 $SU_{152}$, is estimated to have roughly $\sim 1.7$ km size (for $0.15$ albedo) and the smaller component, 2005 $UY_{97}$, has probably a size of $\sim 1.1$ km (for $0.15$ albedo). The composite parent object thus might have a size of $\sim 1.9$ km and we roughly estimate the Hill radius of its gravitational influence to be $R_{\text{Hill}} \sim 500$ km and the escape velocity $v_{\text{esc}} \sim 1$ m/s.

We propagated about 5000 geometric and Yarkovsky clones for both asteroids in this pair over the past 100 ky time interval (the youth of this pair is suggested by very close correspondence of the osculating orbital elements in this case). Using the same approach as in the previous two cases we give our results in Suppl. Fig. 4 (because of larger number of clones we now used 10 million of trial identifications of the clone pairs at each output time, sampled by 1 y step). The minimum distances we could reach between the trial pairs of clones were as small as $\sim 10$ km with a relative velocity of millimeters per second (note this is already comparable to the physical size of the two asteroids). Obviously the success here goes along with the suggested young age of $17^{+25}_{-10}$ ky for this pair.

‡See http://hamilton.dm.unipi.it/astdys/.

Unfortunately, in most of the other cases orbital uncertainties prevented us to reach such a satisfactory result as above. Very often we are able to prove possible convergence of a subset of clones to distances comparable to the Hill sphere of the parent object and comfortably small relative speed (typically smaller than meter per second), but we set only the lower limit for the pair's age. The upper limits, such as seen of Suppl. Figs. 2 to 4, are in these other cases pushed well beyond the $500$ ky limit of our integration. Future orbital constrains, more astrometry and physical observations able to constrain the Yarkovsky forces, and using more orbital clones for pairs with ages larger than $\sim 100$ ky could lead to improvements in the age determination for these pairs.

## 1.4  Implications from the backward integrations

The fact that for nearly all of the pairs in our sample we found the convergence of a subset of their clones to the level of (or close to, for some older ones) the estimated Hill radius of the parent object at small relative velocity strengthens the case that they are real, genetically related pairs. While in refs 1, 2 we made efforts to characterize the statistical significances of the pairs and demonstrate they are real, some issues in the method remained debated[5]. The ability to converge to the level that strongly distincts real pairs from coincidental associations therefore increases credibility of most of the pairs in our sample. We note that we did not reach a convergence during the past $1$ Myr (the maximum backward integration time in this work) for four pairs: (1979) Sakharov and (13732) Woodall, (2110) Moore-Sitterly and (44612) 1999 RP$_{27}$, (32957) 1996 HX$_{20}$ and (38707) 2000 QK$_{89}$, and (60546) 2000 EE$_{85}$ and (88604) 2001 QH$_{293}$. We consider two possible reasons for not reaching convergence for a pair: (i) it is a random coincidental couple of asteroids from the background population, or (ii) its age is greater than the integration interval of 1 Myr. At this moment we cannot resolve between these two alternatives with the backward integrations for the four pairs. Their low $d_{\text{prop}}$ values in the range 1 to 9 m/s as well as mostly low probabilities of chance coincidence of asteroids from the background population suggest that they may be real pairs older than 1 Myr, but it needs to be confirmed with future studies. In any case, we note that the data on primary spin rates and $\Delta H$ for all the four suspect old pairs fit with the model of pair separation presented in our paper.

An important implication from the pair age estimation effort is estimating or constraining a magnitude of change of the primary spin rate by the YORP effect. The premise used when comparing the data for the pairs with outcomes of the model in this work is that the primary spin rate did not change significantly since separation of the pair, so the current observed $P_1$ value can be taken as representing a final rotation rate that the primary reached at the end of the pair separation. If a magnitude of change of the spin rate due to the YORP effect was not much less than the spin rate itself, we would have to take that additional effect into account in the comparison of the data with the model.

As an example, we present an estimation of the magnitude of the YORP effect on the primary of the pair (229401)-2005 UY$_{97}$. The primary (229401) 2005 SU$_{152}$ rotates slowly, with $P_1 \sim 28$ hr, i.e., $\omega_1 \sim 6.2 \times 10^{-5}$ s$^{-1}$. Rescaling the YORP spin evolution rate estimated in ref. 18 to the size and heliocentric distance of (229401), we estimate the secular rate of change of the angular velocity due to YORP of $(\mathrm{d}\omega/\mathrm{d}t) \simeq 5 \times 10^{-5}$ s$^{-1}$ My$^{-1}$. During the pair's age of $\sim 17$ kyr, the YORP effect might have changed the angular frequency of (229401) by no more than $\sim 10^{-6}$ s$^{-1}$, i.e., by less than 2%. This change of the spin rate due to YORP is negligible for the purpose of the comparison of the data with the model (and actually in this particular case where there is present also a substantial uncertainty in the $P_1$ estimate itself, the possible change due to YORP is also much less than the observational uncertainty $\delta P_1$). We conclude that YORP could not affect the rotation of the primary (229401) significantly and the observed $P_1$ value is a very good proxy for a final rotation rate that the primary reached at the end of the pair separation. Similar results of the relative (in)significance of the YORP effect were obtained also for the other pairs in our sample.

## 2  Photometric observations of paired asteroids

We carried out photometric observations using our standard asteroid lightcurve photometry techniques and obtained estimates for rotation periods and amplitudes for 32 primaries (i.e., larger members) and 8 secondaries (smaller members) of the 35 asteroid pairs in our sample. In Supplementary Table 1, there are listed participating observatories and instruments used. Supplementary Table 2 gives observational circumstances of the observations. We give references and descriptions of observational procedures on individual observatories in following.

*CarbH0.35, 0.50* - Observational and reduction procedure at Carbuncle Hill Observatory is described in ref. 19.

*CTIO0.9* - General information about the system is available at `http://www.ctio.noao.edu/telescopes/36/0-9m.html`. The telescope was operated in service mode and we used the full chip setting with the FOV $13' \times 13'$, the readout noise $\sim 5$ ADU = $3e^-$ and V filter. Integration times were mostly 120 s. *MaxImDL* was used to process and reduce the observations.

*CTIO1.0* - General information about the system is available at `http://www.astronomy.ohio-state.edu/Y4KCam/detector.html`. We used the $2 \times 2$ binning mode. Integration times were mostly 120 s. *MaxImDL* was used to process and reduce the observations.

*Danish1.54* - Technical information on the telescope is available at `http://www.eso.org/lasilla/telescopes/d1p5/`. Most of the observations were done in the Cousins R filter. Integration times were 60 to 180 s, depending on apparent sky motion of targeted asteroid so that to get an image streak not longer than 2 pixels (i.e., comparable to a typical seeing at the site). The data were processed and reduced with *MaxImDL* and our photometric reduction software package *Aphot32*, both using the aperture photometry tech-nique. A part of the observations was calibrated in the Johnson-Cousins system using Landolt standard stars[20].

*DarkSky and PROMPT* - Appalachian State University's Dark Sky Observatory is located in the mountains of northwestern North Carolina at an elevation of 1000 meters. The 0.8-m telescope is equipped a Photometrics CCD camera with a $1024 \times 1024$ 25-micron pixel Tek chip. The field of view is $8' \times 8'$ with 0.50 arcsec/pixel.

The University of North Carolina at Chapel Hill's PROMPT observatory (Panchromatic Robotic Optical Monitoring and Polarimetry Telescopes) is on Cerro Tololo. PROMPT consists of six 0.41-m outfitted with Alta U47+ cameras by Apogee, which make use of E2V CCDs. The field of view is $10' \times 10'$ with 0.59 arcsec/pixel.

All raw image frames were processed (master dark, master flat, bad pixel correction) using the software package MIRA. Aperture photometry was then performed on the asteroid and three comparison stars. A master image frame was created to identify any faint stars in the path of the asteroid. Data from images with background contamination stars in the asteroid's path were then eliminated.

*Lick* - Observations were collected using the Lick Observatory 1m-Nickel telescope and its Direct Imaging Camera at the f/17 Cassegrain focus in R band. The detector is a thinned, Loral, $2048 \times 2048$ CCD with 15-micron pixels, corresponding to 0.184 arcsec/pixel, so a FOV of $6.3 \times 6.3$ arcmin. The observations were remotely conducted from a control room located at the Department of Astronomy of the University of California at Berkeley. The relative photometry measurements were made using an automatic software developed in Python 2.5. It detects and reduces the asteroid and three selected nearby bright comparison stars on each processed frame (after dark subtraction, badpixel removal, and flat-field correction). The flux is estimated with an aperture photometry technique using a Gaussian fit function. A reducer checks the frames for a possible contamination of the images of the asteroid and the comparison stars by a remnant bad pixel, cosmic rays, or a background star and such affected data points are discarded.

*Maidanak* - Observations were carried out at Maidanak Astronomical Observatory (Uzbekistan) with 1.5-m telescope AZT-22 (Cassegrain f/7.7), equipped with back-illuminated Fairchild 486 CCD camera (4096 × 4096 CCD, 15 × 15 $\mu m$ pixels, 0.27 arcsec/pixel, FOV 18.4 × 18.4 arcmin) and Bessell UBVRI standard filters. The observations were carried out in the R band and were reduced in the standard way with master-bias subtracting and median flat-field dividing. The aperture photometry of the asteroid and comparison stars in the images was done with the ASTPHOT package developed at DLR[21]. The effective radius of aperture was equal to $1 - 1.5\times$ the seeing that included more than 90% of the flux of a star or the asteroid. The relative photometry of the asteroid was done with typical errors in a range of $0.01 - 0.02$ mag using an ensemble of comparision stars.

*Modra* - Observational system, data analysis and reduction process are described in ref. 22.

*PdM0.6* - Details about the telescope are available at `http://astrosurf.com/t60`. Reduction was performed using the Prism V7 software.

*T120-OHP* - We used the 1.2-m Newton f/6 telescope equipped with a CCD TeK $1024 \times 1024$ in the R band, with the field of view of 12 arcmin and 0.7 arcsec/pix. The standard reduction (flat-field correction, dark substraction, badpixel and cosmic removal) was performed. Relative photometry was performed with a custom software written in IDL, using a fiting of the Gauss function. Images affected by close encounters with background stars were discarded by the user.

*Wise0.46+1.0* - Observations were performed using the two telescopes of the Wise Observatory (Tel-Aviv University) in the Israeli desert (MPC code 097): A 1-m Ritchey-Chrétien telescope and a 0.46-m Centurion telescope (see ref. 23 for a description of the telescope and its performance). The 1-m telescope is equipped with a cryogenically-cooled Princeton Instruments CCD. At the f/7 focus of the telescope this CCD covers a field of view of $13' \times 13'$ with $1340 \times 1300$ pixels (0.58 arcsec per pixel, unbinned). The 0.46-m telescope was used with an SBIG STL-6303E CCD at the f/2.8 prime focus. This CCD covers a field of view of $75' \times 50'$ with $3072 \times 2048$ pixels, with each pixel subtending 1.47 arcsec, unbinned. R and V filters were used on the 1-m telescope while observations with the 0.46-m telescope were unfiltered. Integration times were 120-300 s, all with auto-guider. The reduction, measurements, calibration and analysis methods of the photometric data are fully described in refs 24, 25.

We analysed the observations using the standard Fourier series method[26,27,28].

Data for most of the asteroids given in Supplementary Table 2 and presented in plots with composite lightcurves (Suppl. Figs. 6 to 44) are self-explanatory, but comments on four of them are given below. In a few cases where lightcurve amplitude changes were observed, we give an amplitude at the lowest observed solar phase in Table 1 as it was least affected by the amplitude-phase effect[29].

(6070) Rheinland was observed over a 4-month long interval during July-November 2009. The observations showed variations of synodic period and amplitude of the asteroid that will be useful data for future spin axis and shape modeling. In Suppl. Fig. 10, we plot the data and fit with a constant synodic period.

(78024) 2002 $JO_{70}$: A lower limit on its period of 17 hr was estimated from the data of 2009 Sept. 26 that showed a continuous increase of brightness during the 4.4-hr long observing interval. The data from the previous night of 2009 Sept. 25 showed no variation greater than 0.02 mag and they are consistent with being taken around an extremum of long-period lightcurve. See Suppl. Fig. 32.

(101703) 1999 $CA_{150}$: A few first measurements taken on 2009 Sept. 20 showed an attenuation that resembles mutual events observed in orbiting binaries[30]. The possibility that this primary is actually a bound orbiting binary needs to be confirmed with more observations in future. See Suppl. Fig. 38.

(139537) 2001 QE$_{25}$: A continuous brightness decrease by 0.10 mag observed during the 7.7-hr long observing interval on 2009 March 19 gives a lower limit on the asteroid's period of 30 hr. In Supplementary Table 1, we give a period of $45 \pm 15$ hr, as periods longer by a factor of 2 of the lower limit are less likely. The shorter session of 2009 March 18 showed no brightness variation greater than 0.04 mag during the 5.4-hr long interval, consistent with being taken around an extremum of long-period lightcurve. See Suppl. Fig. 40.

# 3 Correlation between primary spin rate and pair mass ratio

The primary spin rate is correlated with the mass ratio between components of the asteroid pair (Fig. 1). Here we give a test of the statistical significance of the correlation between the two variables.

The model of a proto-binary separation given in Sect. 5 predicts that there is a linear correlation between $\omega_1^2 \equiv (2\pi/P_1)^2$ and $q$ (Eq. 15). We computed the correlation coefficient for the $\omega_1^2$ and $q$ data of the sample of 32 asteroid pairs, $r = -0.7349$. This sample correlation coefficient is a point estimate of the population parameter $r_{\rm pop}$, the correlation coefficient in the population that was sampled. To investigate whether the observed correlation is statistically significant, i.e., whether the sample correlation coefficient is different from 0 at a high confidence level, we test the null hypothesis $r_{\rm pop} = 0$ using Student's $t$ statistics,

$$t = \frac{r}{s_r}, \text{ where } s_r = \sqrt{\frac{1-r^2}{n-2}}, \tag{1}$$

and the degrees of freedom $n - 2 = 30$. We get $t = -5.9351$. Since $t_{0.001} = 3.646$ for the degrees of freedom of 30, we get that the correlation between the primary spin rate and the pair mass ratio is significant at a level higher than 99.9%.

## 4 Fission mechanics

Rotating bodies can be characterized by their total energy and rotational angular momentum. When the body is a single entity, the rotational angular momentum vector is simply computed as:

$$\boldsymbol{L} = \boldsymbol{I} \cdot \boldsymbol{\omega} \quad (2)$$

where $\boldsymbol{I}$ is the rotational inertia dyad and $\boldsymbol{\omega}$ is the angular velocity of the body. The total energy of a rotating body is also driven by its rotation rate, but is also a function of its mass distribution through its self-potential ($\mathcal{U}$):

$$E = \frac{1}{2}\boldsymbol{\omega} \cdot \boldsymbol{I} \cdot \boldsymbol{\omega} + \mathcal{U} \quad (3)$$

The rotational inertia tensor and the self-potential are defined through the mass distribution of the body:

$$\boldsymbol{I} = \int_{\beta} \left[ (\boldsymbol{\rho} \cdot \boldsymbol{\rho}) \boldsymbol{U} - \boldsymbol{\rho}\boldsymbol{\rho} \right] dm \quad (4)$$

$$\mathcal{U} = -\frac{G}{2} \int_{\beta} \int_{\beta} \frac{dm\,dm'}{|\boldsymbol{\rho} - \boldsymbol{\rho}'|} \quad (5)$$

where $\beta$ represents the mass distribution, $\boldsymbol{U}$ is the identity dyad, $\boldsymbol{\rho}, \boldsymbol{\rho}'$ is the location in the body of a mass element $dm, dm'$, and $G$ is the gravitational constant.

Due to the well-documented YORP effect the angular velocity of asteroids of size $< 10$km in diameter can be changed in time spans short relative to their lifetime. As a rigid or rubble-pile body undergoes changes in its rotation rate the mass distribution parameters can remain constant over a relatively wide range of rates, unlike a fluidic body which will change its shape incrementally with changes in total angular momentum. Despite this, if large enough changes in the total angular momentum of the object occur, even collections of rigid components can undergo shifts into configurations that have a lower total energy, with excess energy being dissipated thermally or through seismic waves[4,13,31]. This occurs as the angular momentum of a collection of rigid components resting on each other changes, and represents different configurations of the mass that may have a lower energy at that given angular momentum.

If the total system angular momentum increases sufficiently, the minimum energy configuration for the body will eventually involve components of the body entering orbit about each other[4,13]. Once this occurs a different regime of physics takes over and the system can evolve rapidly and dynamically with the components in orbit about each other. A detailed analysis of the fission process shows that such systems will invariably enter a highly unstable dynamical state, and the orbit and rotations of each component will vary chaotically as they interchange energy and angular momentum between each other[8]. The transition from a collection of rigid components resting on each other to one where two of the collections are in orbit liberates energy that can drive the system dynamically.

The total angular momentum and energy are, in general, conserved across fission but becomes decomposed into multiple components:

$$\bm{I}\cdot\bm{\omega} = \bm{I}_1\cdot\bm{\omega}_1 + \bm{I}_2\cdot\bm{\omega}_2 + \frac{M_1 M_2}{M_1+M_2}\bm{r}\times\bm{v} \qquad (6)$$

$$\frac{1}{2}\bm{\omega}\cdot\bm{I}\cdot\bm{\omega} + \mathcal{U} = \frac{1}{2}\bm{\omega}_1\cdot\bm{I}_1\cdot\bm{\omega}_1 + \frac{1}{2}\bm{\omega}_2\cdot\bm{I}_2\cdot\bm{\omega}_2 + \frac{1}{2}\frac{M_1 M_2}{M_1+M_2}\bm{v}\cdot\bm{v} + \mathcal{U}_{11} + \mathcal{U}_{22} + \mathcal{U}_{12} \qquad (7)$$

where $M_1$ and $M_2$ are the masses of the two components, $\bm{r}$ and $\bm{v}$ are the relative position and velocity vector between these two components, $\mathcal{U}_{ii}$ is the self-potential of the new components and $\mathcal{U}_{12}$ is the mutual potential between the components.

$$\mathcal{U}_{12} = -G\int_{\beta_1}\int_{\beta_2}\frac{dm_1 dm_2}{|\bm{\rho}_1-\bm{\rho}_2|} \qquad (8)$$

The mutual potential represents a conduit for energy being transferred from rotational to translational energy and vice-versa and can be surprisingly effective.

The fundamental characteristic of the system after it undergoes fission is its free energy, defined as the total energy minus the self-potentials of each component. If the free energy is positive, the two components can escape from each other and become an "asteroid pair". If the free energy is negative, then such a mutual escape is impossible, barring release of energy from changes in the self-potentials of the components or an exogenous source of angular momentum and energy. A positive free energy does not necessarily dictate that the system disrupt. Constancy of the free energy requires that the self-potentials of the system after fission will not change, i.e., that the mass distribution is fixed after fission. This is a reasonable assumption as the components are under lower stresses when in orbit than they were in when in close proximity to each other. However, even if changes in the self potentials occur, the free energy will still provide a crucial parameter for this system.

The free energy can be represented as:

$$E_{\text{Free}} = \frac{1}{2}\boldsymbol{\omega}_1 \cdot \boldsymbol{I}_1 \cdot \boldsymbol{\omega}_1 + \frac{1}{2}\boldsymbol{\omega}_2 \cdot \boldsymbol{I}_2 \cdot \boldsymbol{\omega}_2 + \frac{1}{2}\frac{M_1 M_2}{M_1 + M_2}\boldsymbol{v} \cdot \boldsymbol{v} + \mathcal{U}_{12} \qquad (9)$$

If the system undergoes disruption, we note that the mutual potential will decrease to $0$ and the translational kinetic energy will approach hyperbolic escape speeds. Thus, if disruption occurs we find:

$$E_{\text{Free}} = \frac{1}{2}\boldsymbol{\omega}_1 \cdot \boldsymbol{I}_1 \cdot \boldsymbol{\omega}_1 + \frac{1}{2}\boldsymbol{\omega}_2 \cdot \boldsymbol{I}_2 \cdot \boldsymbol{\omega}_2 + \frac{1}{2}\frac{M_1 M_2}{M_1 + M_2}v_\infty^2 \qquad (10)$$

We note that all of these terms are positive by definition, and thus if the free energy is negative, a mutual disruption cannot occur. Further, escape speeds can be quite small, indicating that

$$E_{\text{Free}} > \frac{1}{2}\boldsymbol{\omega}_1 \cdot \boldsymbol{I}_1 \cdot \boldsymbol{\omega}_1 + \frac{1}{2}\boldsymbol{\omega}_2 \cdot \boldsymbol{I}_2 \cdot \boldsymbol{\omega}_2 \qquad (11)$$

Thus if the free energy is positive but small, escape is possible but requires that the rotational kinetic energy of the components must be reduced, indicating that energy has been drawn from these modes to enable escape. As the majority of the rotational kinetic energy will reside in the larger component, a consequence is that the larger component will naturally be rotating at a slower rate, with the resulting rotation rate approaching zero with the free energy. The rotation rate of the smaller component can still remain relatively high, as its much lower mass allows it to "hide" relative to the more massive body.

Being equipped with the results of the theory of fission mechanics, we constructed a simple model of the post-fission system interpreting the observed spin rates of asteroid pairs primaries. The model, its assumptions and mathematical formulation are given in the next section.

# 5  Model of a proto-binary separation

To quantify our asteroid pairs we model the post-fission system as a binary of two components starting in close proximity and with the total angular momentum in the range of critical values as observed in close binary systems[6]. Assuming the mass distribution in the components of the system is fixed in this post-fission evolution phase, the free energy of the system is constant. Energy is transferred between the rotational and orbital energy by a conduit of the mutual potential between the components. The model is following:

- The initial state is a close binary.

- The end state is a barely escaping satellite (parabolic orbit).

- Both the free energy and the total angular momentum of the system are conserved.

- The total angular momentum is close to critical ($\alpha_L \sim 1$), as we observe in small binary systems[6].

- The spin vectors of the components are coplanar with their mutual orbit, i.e., rotation and orbit poles are aligned. The rotations are prograde and around the principal axes of the bodies.

- We assume a constant secondary period, neglecting possible changes in the secondary's rotational angular momentum due to its small size, and we take it to be $P_2 = 6$ hr (approximately a mean of the observed secondary rotation periods).

- Bulk density of both components is $\rho = 2$ g/cm$^3$.

The first five assumptions are fundamental, whereas the last two ones are less critical as outcomes of the model are less sensitive to variations of the two parameters within observed or plausible ranges.

The mathematical formulation of our model follows. The free energy of the system is

$$E_{\text{Free}} = \frac{1}{2}I_1\omega_1^2 + \frac{1}{2}I_2\omega_2^2 - G\frac{M_1 M_2}{2A}, \tag{12}$$

where $I_i, \omega_i, M_i$ are the moment of inertia around the principal axis, the angular velocity and the mass of the $i$-th body (1 for primary, 2 for secondary), respectively, and $A$ is the system's orbit semimajor axis.

Since the free energy and $\omega_2$ are constant, we get

$$\frac{1}{2}I_1\omega_{1\text{ini}}^2 - G\frac{M_1 M_2}{2A_{\text{ini}}} = \frac{1}{2}I_1\omega_{1\text{final}}^2, \tag{13}$$

where the subscripts "ini" and "final" denote initial and end state values of the parameters. Note that $1/A_{\text{final}} = 0$ for the end state of a barely escaping satellite (parabolic orbit).

In Eq. 13, we substitute $M_2 \equiv qM_1$, the moment of inertia of the primary

$$I_1 = \frac{M_1}{5}(a_1^2 + b_1^2), \tag{14}$$

and $M_1 = V_1\rho$, where $V_1$ is the volume of the primary. We assume that $V_1$ is equal to the volume of the dynamically equivalent equal mass ellipsoid (DEEME) of the primary, i.e., $V_1 = a_1 b_1 c_1 \pi 4/3$. The parameters $a_1, b_1, c_1$ are semiaxes of the DEEME of the primary. After the substitutions, we get

$$\omega_{1\text{final}}^2 = \omega_{1\text{ini}}^2 - \frac{\frac{20}{3}\pi q G \frac{a_1}{b_1}\frac{c_1}{b_1}\rho}{\left[1 + \left(\frac{a_1}{b_1}\right)^2\right]\frac{A_{\text{ini}}}{b_1}}. \tag{15}$$

The initial angular velocity of the primary $\omega_{1\text{ini}}$ is computed from the normalized total angular momentum of the binary system. It was defined in ref. 6:

$$\alpha_L \equiv \frac{L_1 + L_2 + L_{\text{orb}}}{L_{\text{eqsph}}}, \tag{16}$$

where $L_1$ and $L_2$ are rotational angular momentum of the primary and the secondary, respectively, $L_{orb}$ is the orbital angular momentum, and $L_{\text{eqsph}}$ is the angular momentum of the equivalent sphere spinning at the critical spin rate. The quantities in the numerator in Eq. 16 are given with following formulas:

$$L_1 = \frac{M}{5(1+q)}(a_1^2 + b_1^2)\omega_1, \qquad (17)$$

$$L_2 = \frac{qM}{5(1+q)}(a_2^2 + b_2^2)\omega_2, \qquad (18)$$

$$L_{\text{orb}} = \frac{qM}{(1+q)^2}\sqrt{GMA(1-e^2)}, \qquad (19)$$

where $M \equiv M_1 + M_2$ is the total mass of the system, and $a_i, b_i, c_i$ are semiaxes of the DEEME of the $i$th body. The quantity in the denominator in Eq. 16 is the angular momentum of the equivalent sphere spinning at the critical spin rate. It is

$$L_{\text{eqsph}} = \frac{2}{5}M\left(\frac{3}{4\pi}V_1\right)^{2/3}(1+q)^{2/3}\omega_{c\text{sph}}, \qquad (20)$$

where $\omega_{c\text{sph}}$ is the critical spin rate for the sphere with the angle of friction of 90° and it is

$$\omega_{c\text{sph}} = \sqrt{\frac{4}{3}\pi\rho G}. \qquad (21)$$

We have evaluated this disruption scenario for a number of values of scaled angular momentum ($\alpha_L$), normalized initial semi-major axis ($A_{\text{ini}}/b_1$), and ratios of the primary's axes ($a_1/b_1$ and $b_1/c_1$), where $a_1, b_1, c_1$ are the long, intermediate, and short axis of the dynamically equivalent equal mass ellipsoid of the primary. For $b_1/c_1$, we assumed a value of 1.2, corresponding to a low primary polar flattening (cf. with values $b_1/c_1$ between 1.1 and 1.2 estimated for primaries of NEA binaries 1999 KW$_4$ and 2002 CE$_{26}$ (refs 32, 33).Values of the primary's equatorial ratio $a_1/b_1$ were varied between 1.1 and 2.0, i.e., in the range of primary elongations suggested by the observed primary amplitudes. We assumed a value of 1.3 for the equatorial axis ratio of the secondary ($a_2/b_2$), corresponding to the observed low to moderate secondary amplitudes.

The normalized total angular momentum $\alpha_L$ was varied over an interval 0.7 to 1.3, which is the range observed for orbiting binaries with $D_1 < 10$ km (ref. 6). The initial relative semi-major axis $A_{\mathrm{ini}}/b_1$ values were taken in the range from 2 to 4. Values in the upper half of the range, 3 to 4, have been observed in known very close asteroid binaries, while values between 2 and 3 are extremely close initial orbits near or below the Roche's limit for strengthless satellites.

For each set of parameters we generated a corresponding rotation period of the primary, assuming a separated pair, as a function of mass ratio $q$ between the pair components. Over the varied ranges of parameters, this model shows the same general character as the asteroid pairs data plotted in Fig. 1. The dashed line plotted there is for $\alpha_L = 1.0, a_1/b_1 = 1.4$ and $A_{\mathrm{ini}}/b_1 = 3$. This set of parameters can be considered as the best representation of pair parameters. In particular, the total angular momentum content of 1.0 is about the mean of the distribution of $\alpha_L$ values in small binaries, and the axial ratio of 1.4 is about a mean of equatorial elongations of pair primaries suggested by their observed amplitudes. We have also plotted four additional curves that represent limit cases.

## *Discussion 1*
## Predictions and implications of the fission theory

The results of the fission theory[4,8,13] predict the observational data presented in Fig. 1. The primary rotation periods become long with increasing mass ratio, consistent with the free energy of a binary system formed by rotational fission approaching zero and the non-existence of separated pairs with mass ratios appreciably larger than the predicted cutoff. The limit curves of our model for $\alpha_L = 1.2$ go up to $q = 0.4$, i.e., a bit behind the upper limit $q \sim 0.2$ estimated in ref. 8. It is because the value of $\alpha_L = 1.2$ is actually a super-critical amount of angular momentum for a body with the low elongation of the upper limit case and such system can form from an original elongated body that is re-shaped during satellite's formation with a resulted low-elongation primary (see ref. 6, Fig. 1). The limit of $q \sim 0.2$ is valid if the components behave like rigid bodies during a fission of the original body. An alternative explanation is that the few systems closest to the upper limit may have a higher bulk density than the assumed value of 2 g/cm$^3$, with the total angular momentum not exceeding the critical value for a low-elongation body.

It is important to note that the theory predicts that fissioned binary asteroids are always initially unstable, independent of the free energy and mass ratio, and thus undergo chaotic variations immediately after fission[8]. The time scale for these systems to transition from their initial fission proto-binary state to an orbitally disrupted, asteroid pair state (for those systems with positive free energy) is not immediate but occurs after a characteristic time span of tens to hundreds of days, with a median time estimated to be 0.6 yr (see Suppl. Fig. 5).

It is important to note the existence of binary asteroids with mass ratios larger than this threshold[6]. Thus, such objects exist but, if formed from the rotational fission process described here, these binary asteroids should not be able to disrupt by themselves. The existence of this population across the mass ratio threshold, and the fact that asteroid pairs are found consistent with the disruption threshold[§], shows a remarkable consistency with the rigid body rotational fission model for the creation of binary asteroids and asteroid pairs.

However, this model may not cover all aspects of asteroid rotational fission. The effect of adding angular momentum via YORP to a continuum model of ellipsoidal asteroids modeled as a cohesionless soil was studied and it was found that, theoretically, these bodies should reshape themselves as the angular momentum increases[34]. While the transition of such a body to actual fission has not been studied in the work, they indicate that such fission could occur. In their model the transition to a binary asteroid may be distinct from the rigid body fission model discussed above, although no clear predictions on the initial configuration of such a system have been made. The current results do provide a target state for such a fission process, however.

An alternate formation process for binary asteroids was proposed in ref. 7. This model was motivated by simulations of asteroid fission using a numerical model of a parent body, consisting of cohesionless, hard spheres resting on each other, with its spin rate increasing (simulating YORP) until material left the surface. It is significant to note that the numerical runs were only allowed for evolution over less than 10 orbits following the fission of material before the spin rate of the primary was incremented again. From their computations they found that the secondary satellite was formed in orbit following inter-particle collisions. However, such a formation mechanism should dissipate excess energy in the system, and hence would not lead to a system that can spontaneously undergo escape. Additionally, in this model the satellites form at a distance where orbital ejection does not occur. Finally, we note that this model of stable binary formation does not predict the observed correlation between the primary period and mass ratio of asteroid pairs and thus is likely not the source of these asteroid pairs.

---

[§]While most asteroid pairs identified as probably genetically related pairs in ref. 2 have $\Delta H > 1$, implying $q$ about or less than 0.2, a smaller fraction of pairs have low estimated $\Delta H$ values. We suspect that the absolute magnitude estimates of one or both members of the asteroid pairs with the anomalous low $\Delta H$ values may be in error. Accurate absolute magnitude estimates for the asteroids are needed to check the theoretical prediction.

# *Discussion 2*
# Assumptions and uncertainties in the size and mass ratios estimates

The assumption that the components of an asteroid pair have the same albedo and bulk density is plausible for the bodies formed by a fission or breakup of the original parent asteroid of a presumably homogeneous composition. As a result of details of the pair formation mechanism, the two components might differ in a size distribution of pieces of the regolith and its surface fraction coverage, and cosmic ray exposure age of the surface material, but we have little knowledge on these possible differences and how large albedo difference they could cause. Photometric observations of orbiting binaries are consistent with the primary and secondary albedos being same to within 20% (refs 30, 35). If it is valid also for components of asteroid pairs, then the uncertainty of the albedo ratio less than 20% propagates to an uncertainty of the ratio of effective diameters between pair components of less than 10%. In this work we neglect this possible uncertainty and assume that the components have the same albedo.

It was found that bulk densities of the secondary and primary of 1999 KW$_4$ differ by a factor of 1.43, but with a relative uncertainty of about 30%, so it was only a marginally significant difference[32]. Another, implicit assumption in the conversion between the size and mass ratios is that the effective diameter $D$ estimated from the absolute magnitude is proportional to a mean diameter $D_{\mathrm{mean}}$ that is related to the volume of the asteroid $V = \pi D_{\mathrm{mean}}^3/6$. Deviations from this assumption can cause some systematic errors in the mass ratio estimates, because the effective diameter is actually related to the cross-section rather than volume of an asteroid, and the cross-section-to-volume ratio is a function of the asteroid's shape and viewing aspect, but we neglect this additional uncertainty source in this study.

15. Bottke, W. F., Jr., Vokrouhlický, D., Rubincam, D. P. & Brož, M. in *Asteroids III* (eds Bottke Jr., W. F., Cellino, A., Paolicchi, P. & Binzel, R. P.) 395–408 (Univ. Arizona Press, 2002).

16. Levison, H. F. & Duncan, M. J. The long-term dynamical behavior of short-period comets. *Icarus* **108**, 18–36 (1994).

17. Nesvorný, D., Bottke, W. F., Levison, H. F. & Dones, L. Recent Origin of the Solar System Dust Bands. *Astrophys. J.* **591**, 486–497 (2003).

18. Čapek, D. & Vokrouhlický, D. The YORP effect with finite thermal conductivity. *Icarus* **172**, 526–536 (2004).

19. Warner, B. D. & Pray, D. P. Analysis of the Lightcurve of (6179) Brett. *Minor Planet Bull.* **36**, 166–168 (2009).

20. Landolt, A. U. UBVRI photometric standard stars in the magnitude range $11.5 < V < 16.0$ around the celestial equator. *Astron. J.* **104**, 340–371 (1992).

21. Mottola, S. *et al.* The Near-Earth objects follow-up program: first results. *Icarus* **117**, 62–70 (1995).

22. Galád, A., Pravec, P., Gajdoš, Š., Kornoš, L. & Világi, J. Seven asteroids studied from Modra observatory in the course of binary asteroid photometric campaign. *Earth Moon Planets* **101**, 17–25 (2007).

23. Brosch *et al.* The Centurion 18 telescope of the Wise Observatory. *Astrophys. Space Sci.* **314**, 163–176 (2008).

24. Polishook, D. & Brosch, N. Photometry of Aten asteroids–More than a handful of binaries. *Icarus* **194**, 111–124 (2008).

25. Polishook, D. & Brosch, N. Photometry and spin rate distribution of small main belt asteroids. *Icarus* **199**, 319–332 (2009).

26. Harris, A. W. *et al.* Photoelectric observations of asteroids 3, 24, 60, 261, and 863. *Icarus* **77**, 171–186 (1989).

27. Pravec, P., Šarounová, L. & Wolf, M. Lightcurves of 7 near-Earth asteroids. *Icarus* **124**, 471–482 (1996).

28. Pravec, P. *et al.* Fast rotating asteroids 1999 TY$_2$, 1999 SF$_{10}$, and 1998 WB$_2$. *Icarus* **147**, 477–486 (2000).

29. Zappalà, V., Cellino, A., Barucci, A. M., Fulchignoni, M. & Lupishko, D. F. An analysis of the amplitude-phase relationship among asteroids. *Astron. Astrophys.* **231**, 548–560 (1990).

30. Pravec, P. *et al.* Photometric survey of binary near-Earth asteroids. *Icarus* **181**, 63–93 (2006).

31. Guibout, V. & Scheeres, D. J. Stability of Surface Motion on a Rotating Ellipsoid. *Celest. Mech. Dyn. Astr.* **87**, 263–290 (2003).

32. Ostro, S. J. *et al.* Radar imaging of binary near-Earth asteroid (66391) 1999 KW$_4$. *Science* **314**, 1276–1280 (2006).

33. Shepard, M. K. *et al.* Radar and infrared observations of binary near-Earth Asteroid 2002 CE$_{26}$. *Icarus* **184**, 198–210 (2006).

34. Holsapple, K. A. On YORP-induced spin deformations of asteroids. *Icarus* **205**, 430–442 (2010).

35. Scheirich, P. & Pravec, P. Modeling of lightcurves of binary asteroids. *Icarus* **200**, 531–547 (2009).

**Acknowledgements** Research at Ondřejov was supported by the Grant Agency of the Czech Republic, Grants 205/09/1107 and 205/08/H005. D.V. was supported by the Research Program MSM0021620860 of the Czech Ministry of Education. D.P. was supported by an *Ilan Ramon* doctoral scholarship from the Israeli Ministry of Science, and he is grateful for guidance and support provided by Dr. N. Brosch and Prof. D. Prialnik. D.J.S. acknowledges support by NASA's PG&G program grant NNX 08AL51G. A.W.H. was supported by NASA grant NNX 09AB48G and by National Science Foundation grant AST-0907650. Work at Modra Obser-vatory was supported by the Slovak Grant Agency for Science VEGA, Grant 2/0016/09. The observations on Cerro Tololo were performed with the support of CTIO and Joselino Vasquez, using telescopes operated by SMARTS Consortium. Work at Pic du Midi Observatory has been supported by CNRS – Programme de Planétologie. Operations at Carbuncle Hill Observatory were supported by a Gene Shoemaker NEO Grant from the Planetary Society. Support for PROMPT has been provided by the National Science Foundation under awards CAREER-0449001, AAG-0707634, and MRI-0836187. F.M. and B.M. were supported by the National Science Foundation under award number AAG-0807468. We thank to O. Bautista, T. Moulinier and P. Eclancher for assistance with observations with the T60 on Pic du Midi.

Supplementary Table 1: Observatories, Instruments, and Observers

| Obs/Tel Code | Observatory | Telescope | Diameter (m) | Observers |
|---|---|---|---|---|
| CarbH0.35, 0.50 | Carbuncle Hill | | 0.35, 0.50 | Pray |
| CTIO0.9 | CTIO | | 0.90 | Longa, Pozo |
| CTIO1.0 | CTIO | | 1.0 | Pozo, Barr |
| DarkSky | Dark Sky | DSO-32 | 0.81 | Pollock |
| Danish1.54 | La Silla | Danish | 1.54 | Galád, Pravec, Henych |
| Lick | Lick | Nickel | 1.0 | Marchis, Macomber |
| Maidanak | Maidanak | AZT-22 | 1.50 | Krugly, Sergeev |
| Modra | Modra | | 0.60 | Galád |
| OHP | OHP | OHP-120 | 1.20 | Vachier |
| PdM0.6 | Pic du Midi | | 0.60 | Leroy, Bautista, Moulinier, Eclancher |
| PdM1.0 | Pic du Midi | | 1.0 | Colas, Maquet |
| PROMPT | PROMPT | Prompt5 | 0.41 | Pollock |
| Wise0.46, 1.0 | Wise | | 0.46, 1.0 | Polishook |

Supplementary Table 2: Observational circumstances of paired asteroids. Given are the mid-time of observing interval rounded to the nearest tenth of a day, telescope code from Table 1, filter (C for clear or undefined), duration of the observing interval, the helio– and geocentric distances and solar phase of the asteroid at the mid-time.

| Dates | Obs/Tel | Filter | Int(hr) | r(AU) | Δ(AU) | phase(°) |
|---|---|---|---|---|---|---|
| **(2110) Moore-Sitterly** | | | | | | |
| 2008-11-26.3 | PROMPT | R | 1.2 | 2.224 | 1.281 | 9.8 |
| 2008-11-28.3 | PROMPT | R | 3.6 | 2.228 | 1.276 | 8.7 |
| 2008-12-06.3 | PROMPT | R | 3.2 | 2.245 | 1.268 | 4.3 |
| 2008-12-07.2 | PROMPT | R | 2.1 | 2.246 | 1.268 | 3.8 |
| 2008-12-07.3 | PROMPT | R | 2.5 | 2.247 | 1.268 | 3.8 |
| 2008-12-08.2 | PROMPT | R | 2.3 | 2.248 | 1.268 | 3.3 |
| 2008-12-09.2 | PROMPT | R | 0.6 | 2.250 | 1.269 | 2.7 |
| 2008-12-09.3 | PROMPT | R | 3.2 | 2.251 | 1.269 | 2.7 |
| 2008-12-10.2 | PROMPT | R | 2.4 | 2.252 | 1.270 | 2.2 |
| 2008-12-10.3 | PROMPT | R | 2.1 | 2.253 | 1.270 | 2.1 |
| 2008-12-20.0 | Wise0.46 | C | 2.8 | 2.272 | 1.294 | 3.6 |
| 2008-12-23.8 | Modra | C | 3.9 | 2.279 | 1.310 | 5.6 |
| 2008-12-24.0 | Modra | C | 3.0 | 2.280 | 1.311 | 5.7 |
| 2008-12-25.8 | Modra | C | 3.2 | 2.283 | 1.320 | 6.7 |
| 2008-12-27.9 | Modra | C | 3.6 | 2.287 | 1.332 | 7.7 |
| 2009-01-02.9 | Wise0.46 | C | 7.0 | 2.299 | 1.372 | 10.7 |
| 2009-01-03.9 | Wise0.46 | C | 7.3 | 2.300 | 1.380 | 11.1 |
| **(4765) Wasserburg** | | | | | | |
| 2008-08-23.2 | CTIO0.9 | V | 3.2 | 2.008 | 1.229 | 23.7 |
| 2008-08-25.1 | CTIO0.9 | V | 4.3 | 2.009 | 1.244 | 24.1 |
| 2008-08-26.1 | CTIO0.9 | V | 4.1 | 2.010 | 1.252 | 24.3 |
| **(5026) Martes** | | | | | | |
| 2008-02-13.2 | PdM1.0 | C | 4.4 | 2.874 | 1.996 | 10.8 |
| 2008-02-14.2 | PdM1.0 | C | 3.1 | 2.873 | 1.987 | 10.5 |
| 2009-07-15.9 | Wise0.46 | C | 1.4 | 1.802 | 0.959 | 24.9 |
| 2009-07-17.0 | Wise0.46 | C | 1.7 | 1.802 | 0.952 | 24.6 |
| 2009-07-18.0 | Wise0.46 | C | 4.3 | 1.802 | 0.945 | 24.2 |
| 2009-08-13.9 | Wise0.46 | C | 3.0 | 1.804 | 0.819 | 11.2 |
| 2009-08-14.9 | Wise0.46 | C | 1.1 | 1.805 | 0.817 | 10.6 |
| 2009-08-24.0 | PdM0.6 | C | 3.9 | 1.809 | 0.805 | 5.3 |
| 2009-09-17.9 | OHP | R | 3.1 | 1.831 | 0.858 | 11.5 |

**Supplementary Table 2 – continued from previous page**

| Dates | Obs/Tel | Filter | Int(hr) | r(AU) | $\Delta$(AU) | phase(°) |
|---|---|---|---|---|---|---|
| **(6070) Rheinland** | | | | | | |
| 2009-07-22.0 | Wise0.46 | C | 2.5 | 1.981 | 1.267 | 26.4 |
| 2009-07-24.0 | Wise0.46 | C | 1.8 | 1.978 | 1.247 | 26.0 |
| 2009-07-25.0 | Wise0.46 | C | 3.8 | 1.976 | 1.237 | 25.7 |
| 2009-07-26.0 | Wise0.46 | C | 2.0 | 1.975 | 1.227 | 25.5 |
| 2009-08-16.3 | CarbH0.35 | C | 2.9 | 1.943 | 1.043 | 18.8 |
| 2009-08-17.9 | Wise0.46 | C | 1.6 | 1.941 | 1.032 | 18.1 |
| 2009-08-18.3 | CarbH0.35 | C | 3.8 | 1.941 | 1.029 | 17.9 |
| 2009-09-20.9 | Wise0.46 | C | 6.6 | 1.905 | 0.904 | 3.7 |
| 2009-10-20.8 | Wise0.46 | C | 6.4 | 1.888 | 0.994 | 18.4 |
| 2009-10-23.8 | Wise0.46 | C | 4.8 | 1.887 | 1.012 | 19.7 |
| 2009-11-20.8 | Wise0.46 | C | 3.8 | 1.887 | 1.237 | 28.2 |
| 2009-11-22.7 | Wise0.46 | C | 1.3 | 1.888 | 1.255 | 28.5 |
| **(7343) Ockeghem** | | | | | | |
| 2009-07-18.9 | Wise0.46 | C | 3.4 | 2.255 | 1.244 | 3.7 |
| 2009-07-23.9 | Wise0.46 | C | 4.3 | 2.247 | 1.235 | 3.0 |
| 2009-07-24.9 | Wise0.46 | C | 0.6 | 2.246 | 1.234 | 3.1 |
| 2009-07-25.9 | Wise0.46 | C | 2.7 | 2.244 | 1.233 | 3.3 |
| 2009-07-28.9 | Wise0.46 | C | 1.3 | 2.239 | 1.232 | 4.4 |
| **(10484) Hecht** | | | | | | |
| 2008-11-29.1 | Wise1.0 | R | 2.3 | 2.142 | 1.401 | 21.7 |
| 2008-11-30.0 | Wise1.0 | R | 5.9 | 2.142 | 1.393 | 21.4 |
| 2008-12-01.0 | Wise1.0 | R | 6.1 | 2.143 | 1.384 | 21.1 |
| 2008-12-02.0 | Wise1.0 | R | 5.3 | 2.143 | 1.375 | 20.8 |
| **(11842) Kap'bos** | | | | | | |
| 2009-08-20.0 | OHP | R | 6.6 | 2.242 | 1.398 | 18.2 |
| 2009-08-21.0 | OHP | R | 6.5 | 2.241 | 1.389 | 17.8 |
| 2009-08-28.0 | Wise0.46 | C | 2.8 | 2.233 | 1.332 | 15.2 |
| 2009-09-02.1 | Modra | C | 1.1 | 2.228 | 1.296 | 13.1 |
| 2009-09-17.8 | Wise0.46 | C | 1.7 | 2.211 | 1.219 | 5.7 |
| 2009-09-22.8 | Wise0.46 | C | 3.5 | 2.205 | 1.207 | 3.7 |
| 2009-09-23.0 | OHP | R | 8.2 | 2.205 | 1.207 | 3.7 |
| 2009-09-26.1 | CarbH0.35 | C | 3.5 | 2.202 | 1.203 | 3.1 |
| **(13732) Woodall** | | | | | | |
| 2009-08-27.1 | Wise1.0 | C | 1.4 | 2.258 | 1.387 | 16.6 |
| 2009-08-28.0 | Wise1.0 | C | 5.2 | 2.257 | 1.380 | 16.3 |
| 2009-08-30.9 | Wise0.46 | C | 3.6 | 2.254 | 1.357 | 15.2 |
| 2009-09-18.2 | CarbH0.50 | C | 2.9 | 2.236 | 1.251 | 6.8 |

| Dates | Obs/Tel | Filter | Int(hr) | r(AU) | $\Delta$(AU) | phase(°) |
|---|---|---|---|---|---|---|
| 2009-09-19.2 | CarbH0.50 | C | 4.8 | 2.235 | 1.248 | 6.3 |
| 2009-09-19.9 | Wise0.46 | C | 8.7 | 2.235 | 1.245 | 5.9 |
| 2009-09-20.2 | CarbH0.50 | C | 6.5 | 2.235 | 1.244 | 5.7 |
| 2009-09-21.2 | CarbH0.50 | C | 5.5 | 2.234 | 1.241 | 5.2 |
| 2009-09-24.0 | OHP | C | 7.5 | 2.231 | 1.234 | 3.8 |
| **(15107) Toepperwein** | | | | | | |
| 2008-09-23.4 | Lick | R | 8.3 | 2.230 | 1.282 | 11.1 |
| 2008-09-24.4 | Lick | R | 7.8 | 2.232 | 1.279 | 10.7 |
| 2008-11-02.9 | Wise1.0 | V | 2.8 | 2.311 | 1.381 | 11.1 |
| 2008-11-04.0 | Wise1.0 | V | 1.7 | 2.314 | 1.390 | 11.5 |
| **(17198) Gorjup** | | | | | | |
| 2008-07-27.9 | Wise1.0 | R | 3.0 | 2.083 | 1.078 | 5.6 |
| 2008-07-28.9 | Wise1.0 | R | 4.3 | 2.083 | 1.081 | 6.1 |
| 2008-08-01.9 | Wise1.0 | R | 3.5 | 2.086 | 1.094 | 8.1 |
| 2008-08-02.9 | Wise1.0 | R | 3.0 | 2.087 | 1.098 | 8.6 |
| 2008-08-22.1 | CTIO0.9 | V | 4.2 | 2.102 | 1.212 | 17.3 |
| 2008-08-24.1 | CTIO0.9 | V | 4.5 | 2.103 | 1.228 | 18.1 |
| **(19289) 1996 HY$_{12}$** | | | | | | |
| 2009-11-12.1 | Danish1.54 | R | 5.5 | 2.080 | 1.228 | 18.2 |
| **(21436) Chaoyichi** | | | | | | |
| 2009-11-16.1 | Danish1.54 | R | 5.0 | 2.233 | 1.390 | 16.9 |
| **(23998) 1999 RP$_{29}$** | | | | | | |
| 2009-02-14.3 | CTIO1.0 | V | 4.6 | 1.839 | 0.898 | 13.4 |
| 2009-02-15.3 | CTIO1.0 | V | 2.7 | 1.839 | 0.894 | 12.9 |
| 2009-02-18.0 | Wise0.46 | C | 7.1 | 1.838 | 0.885 | 11.7 |
| 2009-02-18.9 | Wise0.46 | C | 3.8 | 1.838 | 0.882 | 11.3 |
| **(38707) 2000 QK$_{89}$** | | | | | | |
| 2009-02-16.3 | CTIO1.0 | V | 1.9 | 2.468 | 1.559 | 11.3 |
| 2009-02-17.3 | CTIO1.0 | V | 3.8 | 2.469 | 1.553 | 10.9 |
| 2009-02-25.2 | CarbH0.50 | C | 2.9 | 2.476 | 1.518 | 7.4 |
| 2009-03-02.2 | PROMPT | C | 8.1 | 2.480 | 1.504 | 5.2 |
| 2009-03-04.0 | Wise0.46 | C | 6.3 | 2.481 | 1.501 | 4.4 |
| **(44612) 1999 RP$_{27}$** | | | | | | |
| 2009-10-09.3 | Danish1.54 | R | 2.2 | 1.984 | 1.107 | 18.6 |
| 2009-10-11.3 | Danish1.54 | R | 1.7 | 1.987 | 1.098 | 17.7 |
| 2009-10-12.3 | Danish1.54 | R | 2.5 | 1.989 | 1.094 | 17.2 |
| 2009-10-23.0 | Wise1.0 | C | 4.7 | 2.010 | 1.058 | 11.4 |

**Supplementary Table 2 – continued from previous page**

| Dates | Obs/Tel | Filter | Int(hr) | r(AU) | $\Delta$(AU) | phase(°) |
|---|---|---|---|---|---|---|
| 2009-10-25.0 | Wise1.0 | C | 2.1 | 2.014 | 1.054 | 10.2 |
| **(48652) 1995 VB** | | | | | | |
| 2009-11-12.3 | Danish1.54 | R | 2.6 | 2.000 | 1.197 | 21.5 |
| 2009-11-13.1 | OHP | C | 6.3 | 2.000 | 1.190 | 21.2 |
| 2009-11-13.3 | Danish1.54 | R | 2.3 | 2.000 | 1.188 | 21.2 |
| 2009-11-14.3 | Danish1.54 | R | 1.9 | 1.999 | 1.180 | 20.8 |
| 2009-11-15.3 | Danish1.54 | R | 2.8 | 1.999 | 1.172 | 20.4 |
| 2009-11-16.0 | Maidanak | R | 1.1 | 1.998 | 1.166 | 20.2 |
| 2009-11-16.3 | PROMPT | C | 1.8 | 1.998 | 1.164 | 20.1 |
| 2009-11-16.4 | DarkSky | R | 2.2 | 1.998 | 1.163 | 20.0 |
| 2009-11-17.1 | OHP | C | 2.9 | 1.998 | 1.157 | 19.7 |
| **(51609) 2001 HZ$_{32}$** | | | | | | |
| 2009-10-26.0 | Wise1.0 | C | 5.8 | 2.152 | 1.185 | 8.3 |
| 2009-10-30.0 | Modra | C | 6.5 | 2.159 | 1.180 | 6.0 |
| 2009-10-30.9 | Modra | C | 4.7 | 2.161 | 1.180 | 5.5 |
| 2009-10-31.1 | Modra | C | 2.1 | 2.161 | 1.180 | 5.4 |
| **(52773) 1998 QU$_{12}$** | | | | | | |
| 2008-07-28.0 | Wise1.0 | V | 2.0 | 1.794 | 0.949 | 25.0 |
| 2008-07-29.0 | Wise1.0 | V | 2.2 | 1.793 | 0.942 | 24.6 |
| 2008-08-02.0 | Wise1.0 | V | 2.7 | 1.789 | 0.913 | 23.1 |
| 2008-08-03.0 | Wise1.0 | V | 3.3 | 1.788 | 0.906 | 22.7 |
| 2008-10-06.2 | CTIO0.9 | V | 2.0 | 1.770 | 0.826 | 15.6 |
| 2008-10-07.1 | CTIO0.9 | V | 3.5 | 1.770 | 0.830 | 16.1 |
| **(52852) 1998 RB$_{75}$** | | | | | | |
| 2008-09-06.3 | CTIO0.9 | V | 1.3 | 2.118 | 1.137 | 8.5 |
| 2008-09-07.1 | CTIO0.9 | V | 3.9 | 2.117 | 1.139 | 8.9 |
| 2008-09-11.1 | CTIO0.9 | V | 4.7 | 2.113 | 1.151 | 10.9 |
| **(54041) 2000 GQ$_{113}$** | | | | | | |
| 2008-10-08.1 | CTIO0.9 | V | 3.8 | 2.034 | 1.183 | 19.4 |
| 2008-11-03.8 | Wise0.46 | C | 4.7 | 2.052 | 1.438 | 26.2 |
| 2008-11-04.8 | Wise1.0 | R | 4.8 | 2.052 | 1.449 | 26.4 |
| 2008-11-16.7 | Wise0.46 | C | 2.5 | 2.062 | 1.585 | 27.7 |
| **(54827) 2001 NQ$_8$** | | | | | | |
| 2009-11-14.3 | Danish1.54 | R | 3.7 | 2.040 | 1.163 | 17.1 |
| 2009-11-16.3 | Danish1.54 | R | 3.0 | 2.044 | 1.154 | 16.1 |
| 2009-11-17.3 | Danish1.54 | R | 3.8 | 2.046 | 1.149 | 15.6 |
| 2009-11-18.8 | Maidanak | R | 5.1 | 2.049 | 1.143 | 14.9 |
| 2009-11-22.0 | Wise1.0 | C | 7.1 | 2.055 | 1.132 | 13.2 |

**Supplementary Table 2 – continued from previous page**

| Dates | Obs/Tel | Filter | Int(hr) | r(AU) | $\Delta$(AU) | phase(°) |
|---|---|---|---|---|---|---|
| 2009-11-25.9 | Maidanak | R | 3.8 | 2.063 | 1.121 | 11.1 |
| **(56232) 1999 JM$_{31}$** | | | | | | |
| 2009-06-27.9 | Wise1.0 | R | 3.6 | 1.942 | 0.934 | 5.6 |
| **(63440) 2001 MD$_{30}$** | | | | | | |
| 2009-08-17.9 | Wise1.0 | C | 3.5 | 1.988 | 1.092 | 18.5 |
| 2009-08-26.9 | Wise1.0 | C | 7.1 | 1.978 | 1.095 | 19.2 |
| 2009-09-30.1 | Danish1.54 | R | 3.5 | 1.942 | 1.223 | 26.2 |
| 2009-10-01.1 | Danish1.54 | R | 3.9 | 1.941 | 1.229 | 26.4 |
| 2009-10-02.1 | Danish1.54 | R | 3.5 | 1.940 | 1.235 | 26.6 |
| **(69142) 2003 FL$_{115}$** | | | | | | |
| 2009-05-25.9 | Wise1.0 | R | 6.6 | 1.914 | 1.052 | 21.6 |
| 2009-05-28.0 | PdM1.0 | C | 6.5 | 1.915 | 1.052 | 21.6 |
| 2009-05-29.0 | PdM1.0 | C | 6.4 | 1.915 | 1.052 | 21.6 |
| **(76111) 2000 DK$_{106}$** | | | | | | |
| 2009-03-17.0 | Wise1.0 | R | 2.6 | 2.686 | 1.701 | 3.8 |
| 2009-03-20.0 | PdM1.0 | C | 9.7 | 2.687 | 1.706 | 4.5 |
| 2009-03-20.3 | Lick | R | 7.5 | 2.688 | 1.706 | 4.6 |
| **(84203) 2002 RD$_{133}$** | | | | | | |
| 2009-01-19.8 | Wise0.46 | C | 3.8 | 2.013 | 1.132 | 16.6 |
| 2009-01-25.9 | Wise0.46 | C | 3.5 | 2.020 | 1.138 | 16.5 |
| 2009-01-26.8 | Wise0.46 | C | 4.8 | 2.021 | 1.139 | 16.6 |
| 2009-02-01.9 | Wise1.0 | R | 11.0 | 2.027 | 1.154 | 17.1 |
| 2009-02-03.0 | Wise1.0 | R | 4.8 | 2.028 | 1.157 | 17.3 |
| **(88259) 2001 HJ$_{7}$** | | | | | | |
| 2009-02-13.2 | CTIO1.0 | V | 7.8 | 1.810 | 1.008 | 24.6 |
| 2009-02-14.1 | CTIO1.0 | V | 2.9 | 1.810 | 1.002 | 24.3 |
| 2009-02-15.2 | CTIO1.0 | V | 3.9 | 1.810 | 0.996 | 24.1 |
| **(88604) 2001 QH$_{293}$** | | | | | | |
| 2009-06-18.0 | Wise1.0 | R | 3.4 | 2.949 | 1.989 | 7.8 |
| 2009-06-19.0 | Wise1.0 | R | 3.2 | 2.950 | 1.983 | 7.5 |
| 2009-06-23.0 | Wise1.0 | R | 4.9 | 2.950 | 1.965 | 6.0 |
| **(92336) 2000 GY$_{81}$** | | | | | | |
| 2009-08-13.9 | Wise1.0 | R | 2.8 | 1.943 | 1.031 | 17.9 |
| 2009-08-19.9 | Modra | C | 6.1 | 1.936 | 1.020 | 17.6 |
| 2009-08-21.0 | Modra | C | 6.8 | 1.935 | 1.018 | 17.6 |
| 2009-08-23.9 | Modra | C | 6.1 | 1.932 | 1.016 | 17.6 |
| 2009-08-24.0 | OHP | R | 6.7 | 1.932 | 1.016 | 17.6 |

**Supplementary Table 2 – continued from previous page**

| Dates | Obs/Tel | Filter | Int(hr) | r(AU) | $\Delta$(AU) | phase($°$) |
|---|---|---|---|---|---|---|
| 2009-08-25.0 | OHP | R | 5.5 | 1.931 | 1.016 | 17.7 |
| 2009-08-27.9 | Wise1.0 | R | 7.5 | 1.928 | 1.016 | 17.9 |
| 2009-09-01.9 | Modra | C | 4.8 | 1.922 | 1.020 | 18.6 |
| 2009-09-21.1 | Danish1.54 | R | 3.9 | 1.902 | 1.079 | 23.2 |
| 2009-09-22.1 | Danish1.54 | R | 5.1 | 1.901 | 1.084 | 23.5 |
| 2009-09-23.1 | Danish1.54 | R | 3.5 | 1.900 | 1.089 | 23.7 |
| 2009-09-24.1 | Danish1.54 | R | 4.8 | 1.899 | 1.095 | 24.0 |
| 2009-09-25.1 | Danish1.54 | R | 4.9 | 1.898 | 1.100 | 24.3 |
| 2009-09-26.1 | Danish1.54 | R | 4.4 | 1.897 | 1.105 | 24.6 |
| 2009-09-27.1 | Danish1.54 | R | 4.3 | 1.896 | 1.111 | 24.8 |
| **(101065) 1998 $RV_{11}$** | | | | | | |
| 2009-11-19.1 | Danish1.54 | R | 4.5 | 2.274 | 1.524 | 19.9 |
| 2009-11-20.1 | Danish1.54 | R | 3.8 | 2.276 | 1.537 | 20.2 |
| 2009-11-21.1 | Danish1.54 | R | 2.4 | 2.279 | 1.549 | 20.4 |
| **(101703) 1999 $CA_{150}$** | | | | | | |
| 2009-09-20.2 | Danish1.54 | R | 3.9 | 2.372 | 1.373 | 3.5 |
| 2009-09-21.3 | Danish1.54 | R | 4.1 | 2.370 | 1.371 | 2.9 |
| 2009-09-22.3 | Danish1.54 | R | 1.2 | 2.369 | 1.368 | 2.4 |
| 2009-09-24.3 | Danish1.54 | R | 3.4 | 2.367 | 1.364 | 1.4 |
| **(115978) 2003 $WQ_{56}$** | | | | | | |
| 2009-06-27.0 | Wise1.0 | R | 3.0 | 1.860 | 0.888 | 13.3 |
| 2009-06-28.9 | Wise1.0 | R | 1.5 | 1.857 | 0.878 | 12.3 |
| **(139537) 2001 $QE_{25}$** | | | | | | |
| 2009-03-18.1 | PdM1.0 | C | 5.4 | 2.454 | 1.482 | 6.5 |
| 2009-03-19.0 | PdM1.0 | C | 7.7 | 2.451 | 1.476 | 6.0 |
| **(205383) 2001 $BV_{47}$** | | | | | | |
| 2009-02-16.2 | CTIO1.0 | V | 5.8 | 1.850 | 0.867 | 4.6 |
| 2009-02-16.9 | Wise1.0 | R | 6.1 | 1.849 | 0.868 | 5.1 |
| 2009-02-17.1 | CTIO1.0 | V | 4.1 | 1.849 | 0.868 | 5.2 |
| 2009-02-17.9 | Wise1.0 | R | 9.6 | 1.849 | 0.869 | 5.8 |
| **(218099) 2002 $MH_3$** | | | | | | |
| 2009-09-21.9 | Wise1.0 | C | 4.9 | 1.776 | 0.795 | 10.4 |
| 2009-10-22.8 | Wise1.0 | C | 5.7 | 1.812 | 0.965 | 22.8 |
| 2009-10-23.8 | Wise1.0 | C | 4.8 | 1.813 | 0.973 | 23.1 |
| **(220143) 2002 $TO_{134}$** | | | | | | |
| 2009-09-22.2 | Danish1.54 | R | 5.6 | 2.043 | 1.048 | 5.4 |
| 2009-09-23.2 | Danish1.54 | R | 5.7 | 2.043 | 1.050 | 5.6 |

**Supplementary Table 2 – continued from previous page**

| Dates | Obs/Tel | Filter | Int(hr) | r(AU) | $\Delta$(AU) | phase($°$) |
|---|---|---|---|---|---|---|
| **(226268) 2003 AN$_{55}$** | | | | | | |
| 2009-11-18.3 | Danish1.54 | R | 3.8 | 1.948 | 1.023 | 14.3 |
| 2009-11-19.3 | Danish1.54 | R | 4.0 | 1.949 | 1.020 | 13.7 |
| 2009-11-20.3 | Danish1.54 | R | 3.6 | 1.951 | 1.017 | 13.2 |
| 2009-12-17.1 | PROMPT | C | 2.8 | 2.000 | 1.021 | 4.0 |
| 2009-12-17.2 | DarkSky | C | 5.9 | 2.001 | 1.022 | 4.0 |
| 2009-12-18.9 | Wise1.0 | C | 9.1 | 2.004 | 1.028 | 5.0 |
| 2009-12-19.2 | PROMPT | C | 4.7 | 2.004 | 1.029 | 5.2 |
| 2009-12-19.8 | Wise1.0 | C | 6.4 | 2.006 | 1.031 | 5.5 |
| 2009-12-20.2 | PROMPT | C | 6.3 | 2.006 | 1.033 | 5.7 |
| **(229401) 2005 SU$_{152}$** | | | | | | |
| 2009-11-17.1 | Danish1.54 | R | 4.3 | 1.820 | 1.343 | 32.2 |
| 2009-11-18.1 | Danish1.54 | R | 4.0 | 1.821 | 1.352 | 32.2 |
| 2009-11-19.1 | Danish1.54 | R | 3.9 | 1.822 | 1.362 | 32.3 |
| 2009-11-20.1 | Danish1.54 | R | 3.7 | 1.823 | 1.372 | 32.3 |
| 2009-11-21.0 | Danish1.54 | R | 1.2 | 1.824 | 1.381 | 32.3 |

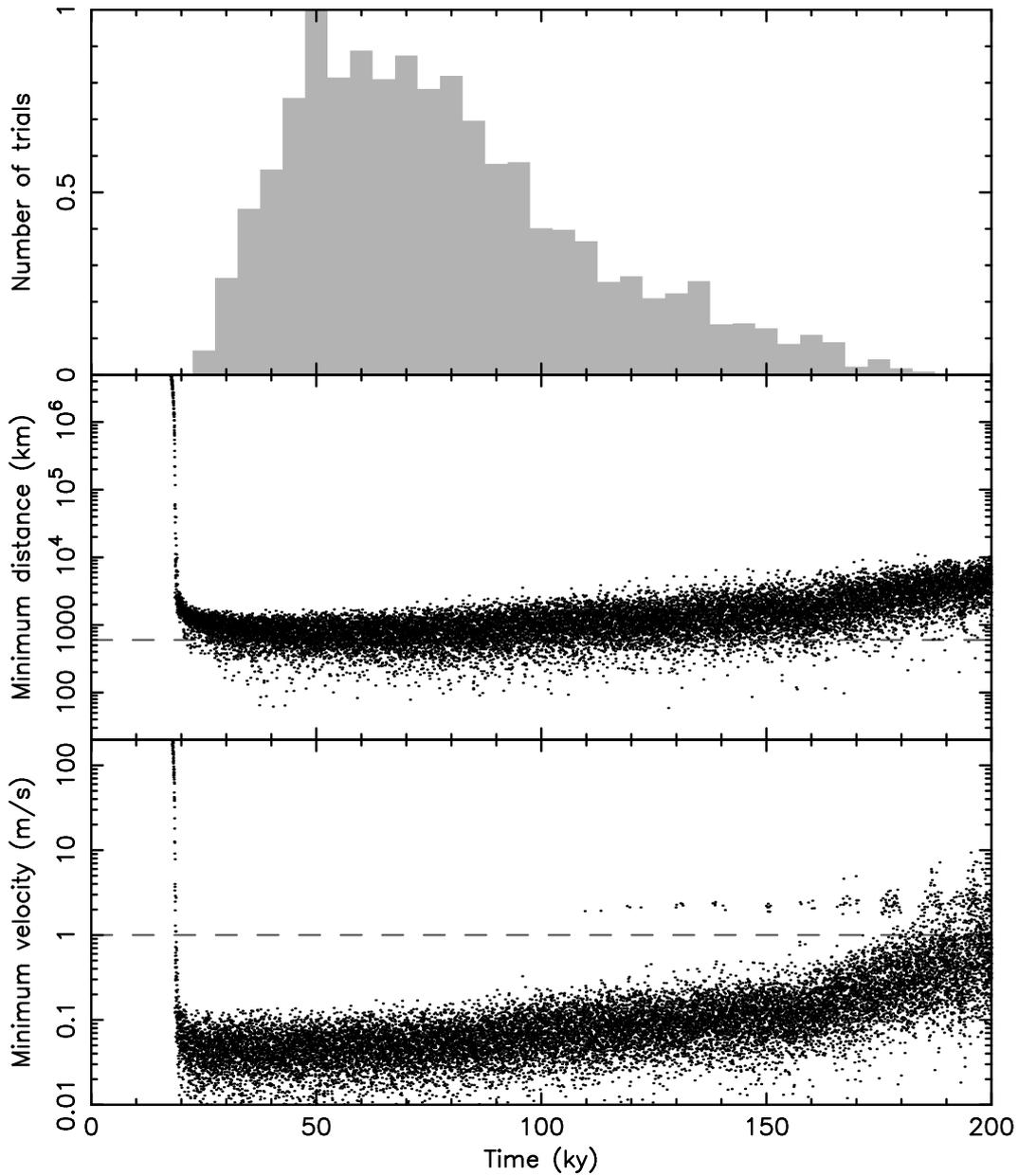

Supplementary Figure 2: Results of the backward integration of 2500 geometric and Yarkovsky clones for each of the components in the pair (21436) Chaoyichi and 2003 YK$_{39}$. At each of the 10 y-separated outputs of the propagation we show the minimum distance of the clones (middle panel) and the relative orbital velocity of the minimum-distant clones (bottom panel). The top panel shows distribution of success rate at each of the output steps such that distance of the trial pair of clones was smaller than the Hill radius of the estimated parent object for the pair (600 km in this case, dashed line in the middle panel) *and* in the same time their relative velocity was smaller than the escape velocity from the parent body (1.2 m/s in this case, dashed line in the bottom panel). The histogram collects data in 5 ky bins and uses a normalization of the maximum to unity.

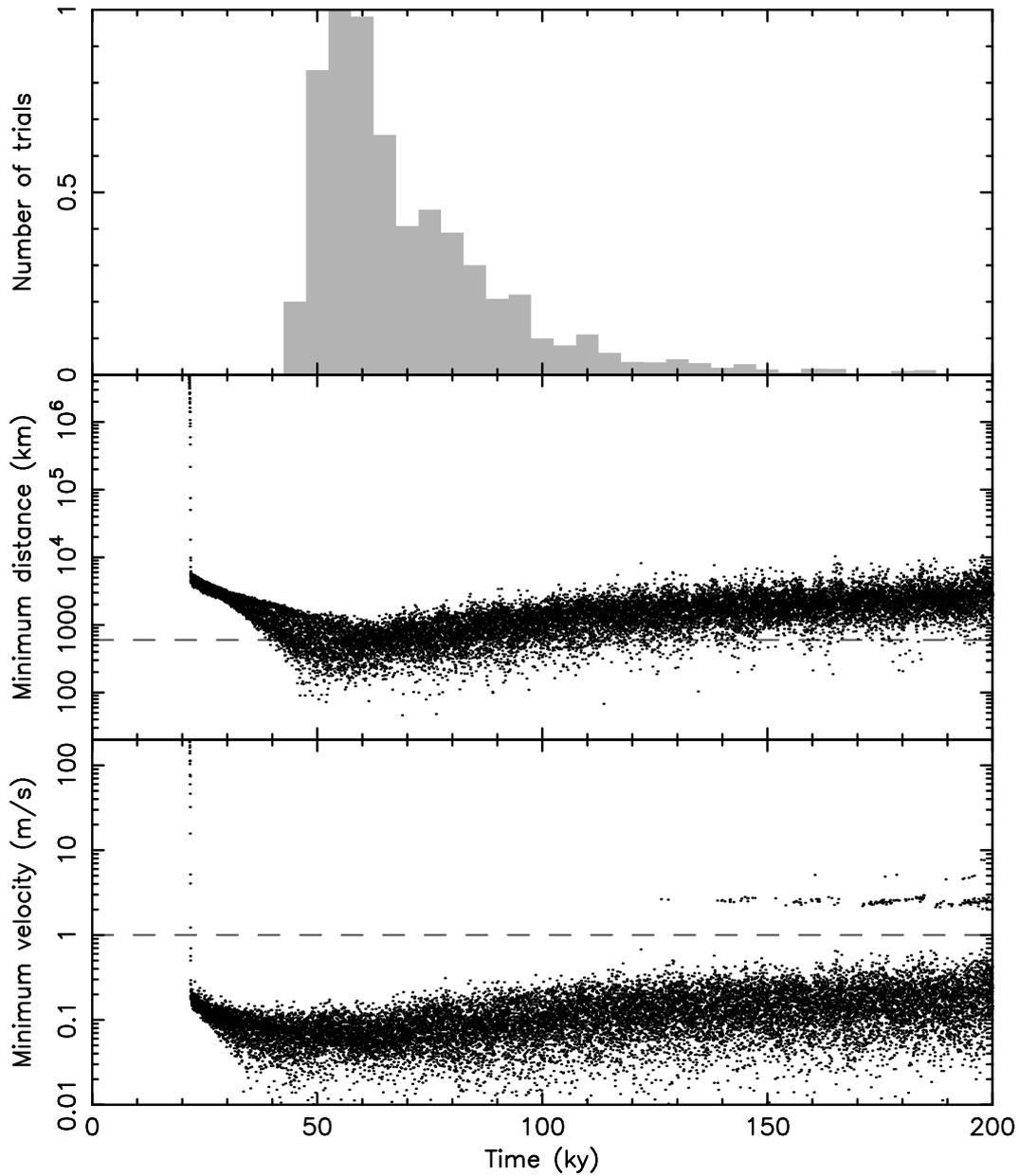

Supplementary Figure 3: Results of the backward integration of 2500 geometric and Yarkovsky clones for each of the components in the pair (88259) 2001 HJ$_7$ and 1999 VA$_{117}$. At each of the 10 y-separated outputs of the propagation we show the minimum distance of the clones (middle panel) and the relative orbital velocity of the minimum-distant clones (bottom panel). The top panel shows distribution of success rate at each of the output steps such that distance of the trial pair of clones was smaller than the Hill radius of the estimated parent object for the pair (600 km in this case, dashed line in the middle panel) *and* in the same time their relative velocity was smaller than the escape velocity from the parent body (1 m/s in this case, dashed line in the bottom panel). The histogram collects data in 5 ky bins and uses a normalization of the maximum to unity.

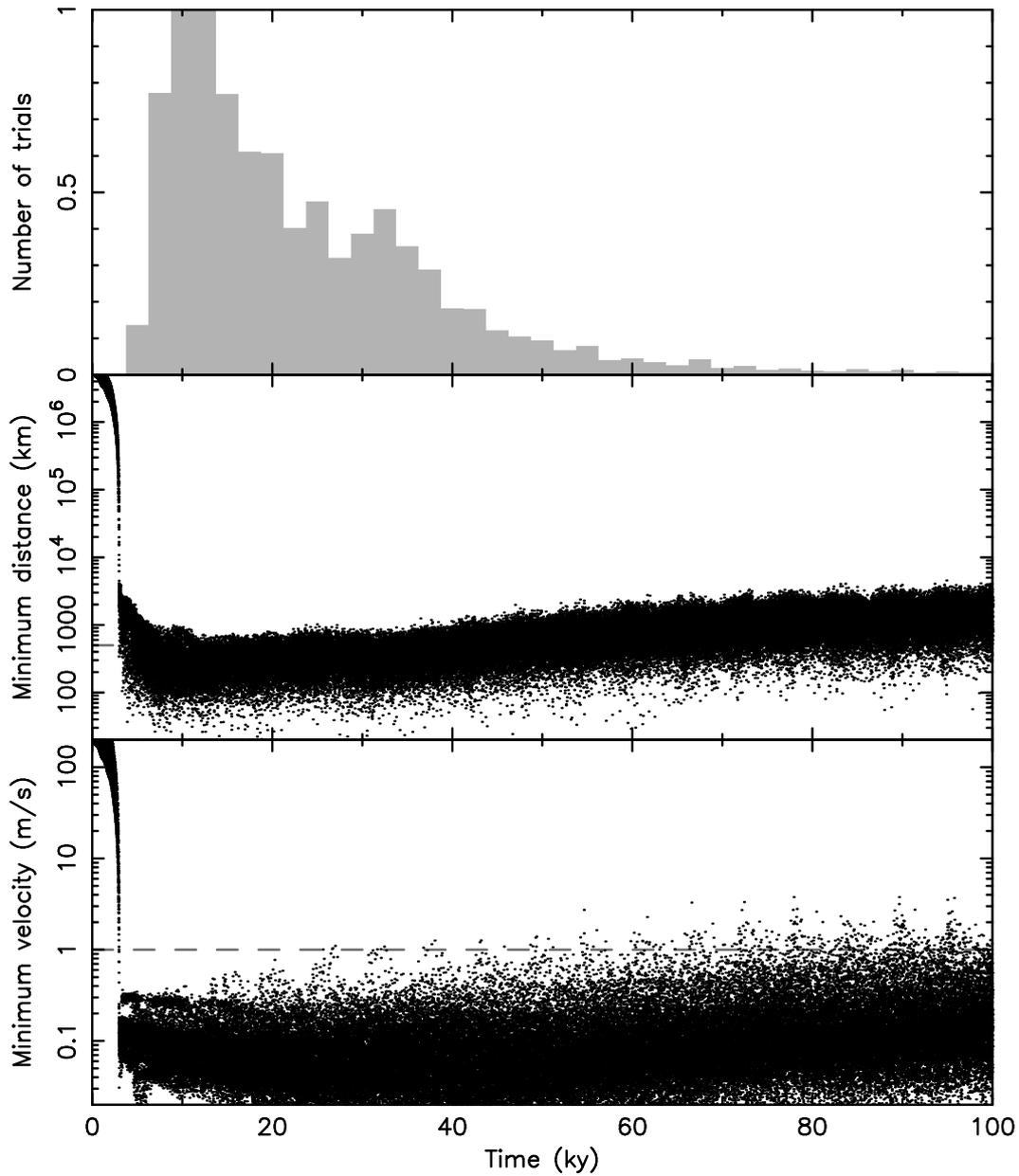

Supplementary Figure 4: Results of the backward integration of 5000 geometric and Yarkovsky clones for each of the components in the pair (229401) 2005 $SU_{152}$ and 2005 $UY_{97}$. At each of the 1 y-separated outputs of the propagation we show the minimum distance of the clones (middle panel) and the relative orbital velocity of the minimum-distant clones (bottom panel). The top panel shows distribution of success rate at each of the output steps such that distance of the trial pair of clones was smaller than the Hill radius of the estimated parent object for the pair (500 km in this case, dashed line in the middle panel) *and* in the same time their relative velocity was smaller than the escape velocity from the parent body (1 m/s in this case, dashed line in the bottom panel). The histogram collects data in 2.5 ky bins and uses a normalization of the maximum to unity.

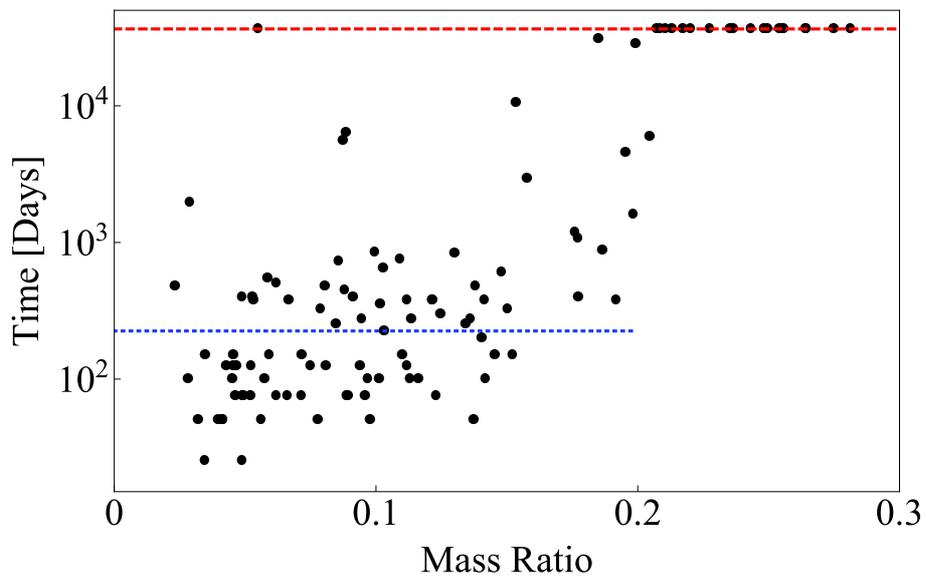

Supplementary Figure 5: Timescale for disruption of rotationally fissioned asteroids. 121 systems with mass ratios varying up to 0.3 and tri-axial ellipsoid shapes were evolved after a rotational fission event. Each point represents either the time of disruption or the length of the simulation (100 years marked by a red line). The blue line indicates the median escape time (0.6 years) for secondaries in systems with a mass ratio below 0.2.

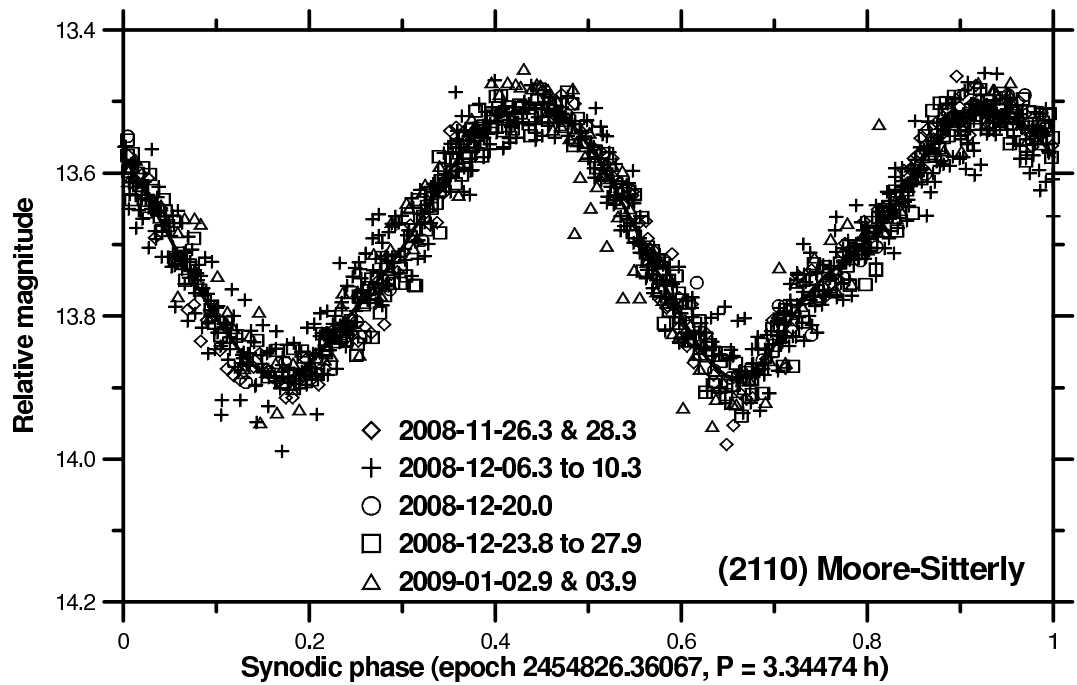

Supplementary Figure 6: Composite lightcurve of (2110) Moore-Sitterly.

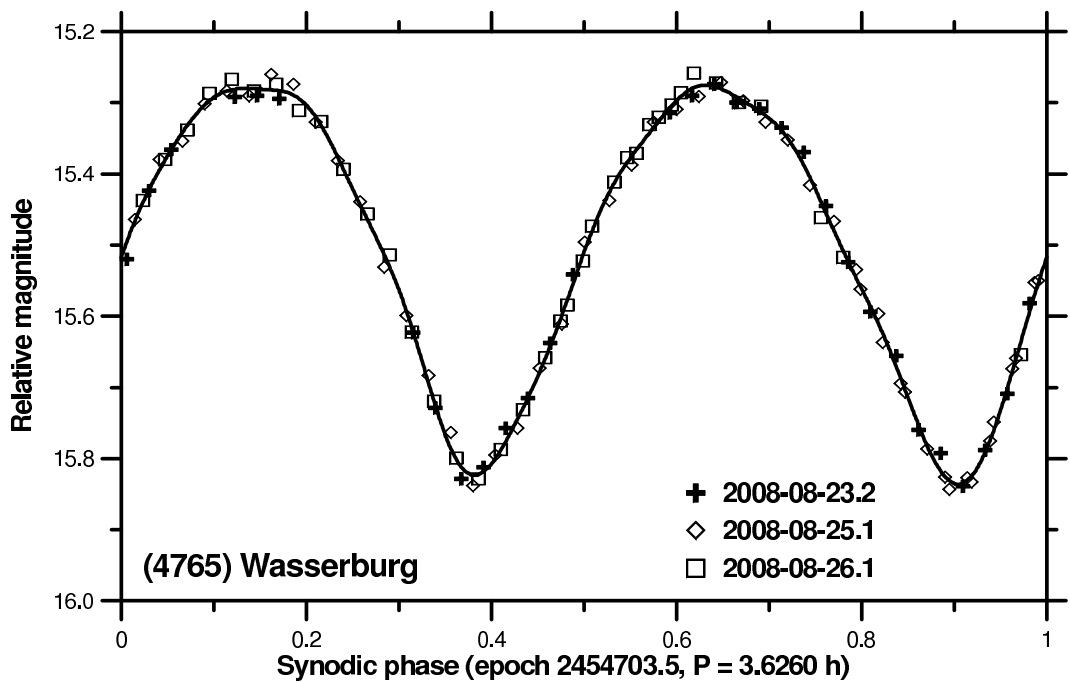

Supplementary Figure 7: Composite lightcurve of (4765) Wasserburg.

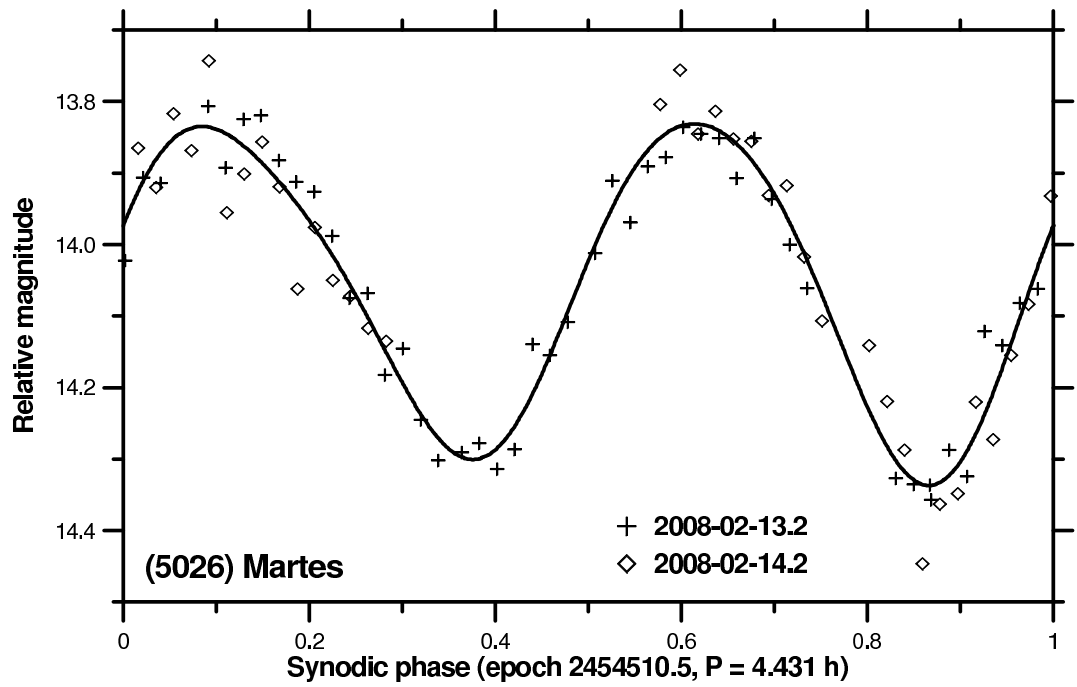

Supplementary Figure 8: Composite lightcurve of (5026) Martes in 2008.

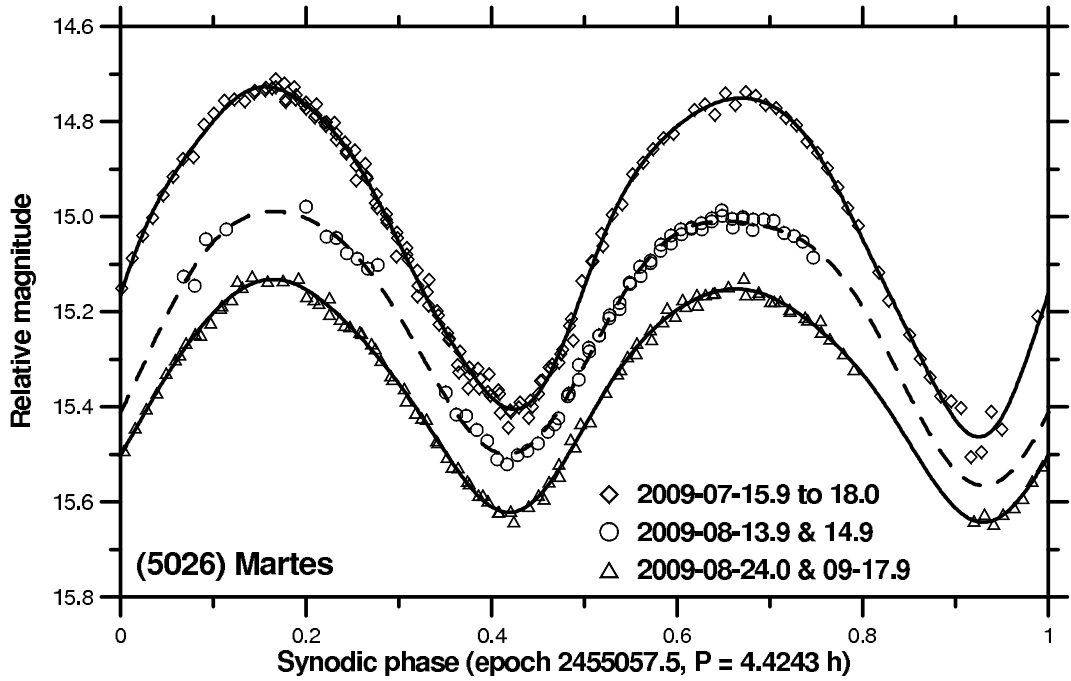

Supplementary Figure 9: Composite lightcurve of (5026) Martes during July-September 2009. The amplitude was changing during the observing campaign, and we present separate fits for three different observational intervals. They are plotted with arbitrary vertical offsets for clarity.

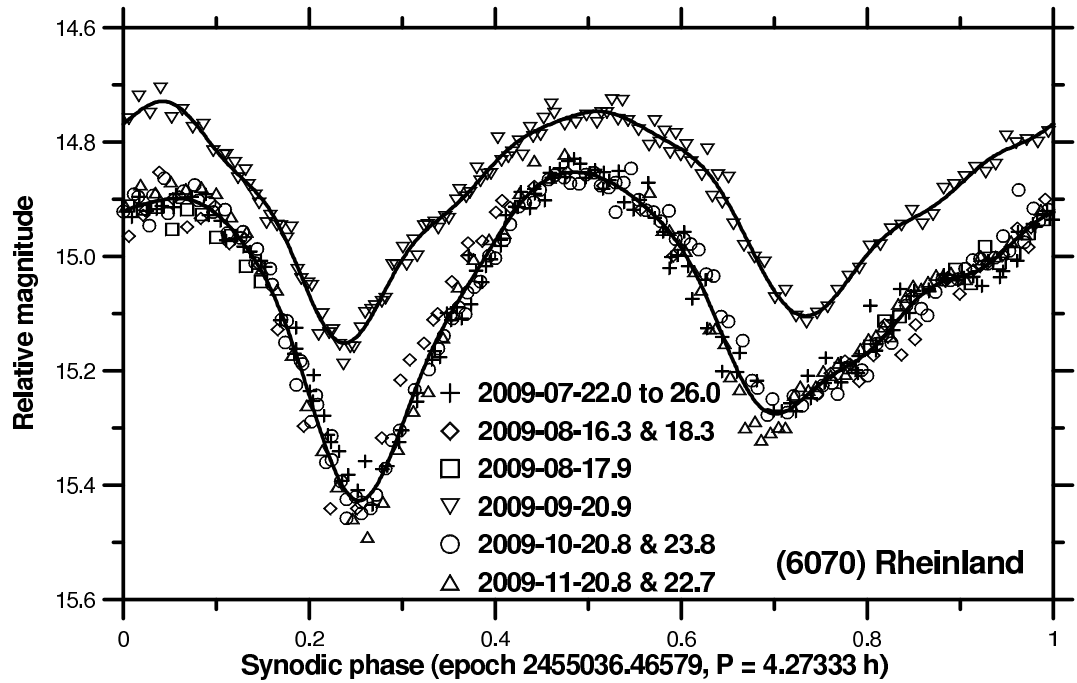

Supplementary Figure 10: Composite lightcurve of (6070) Rheinland. The amplitude was lower on 2009 Sept. 20 than on the other observing nights, and we present the data with a separate fit and with an arbitrary vertical offset for clarity. See also comments in the text.

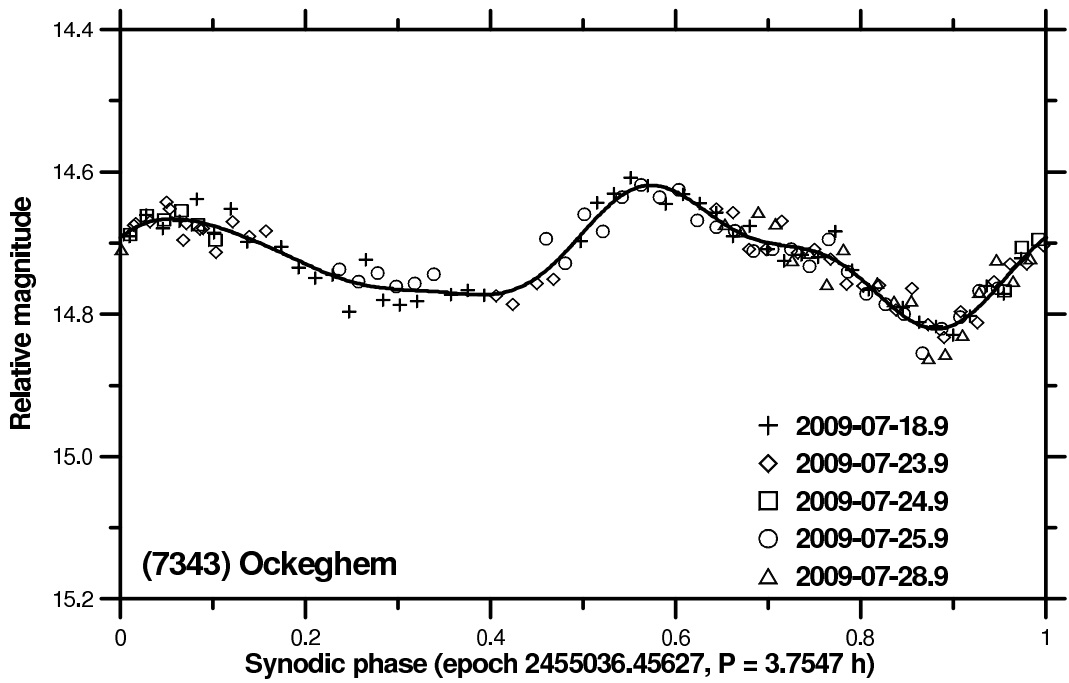

Supplementary Figure 11: Composite lightcurve of (7343) Ockeghem.

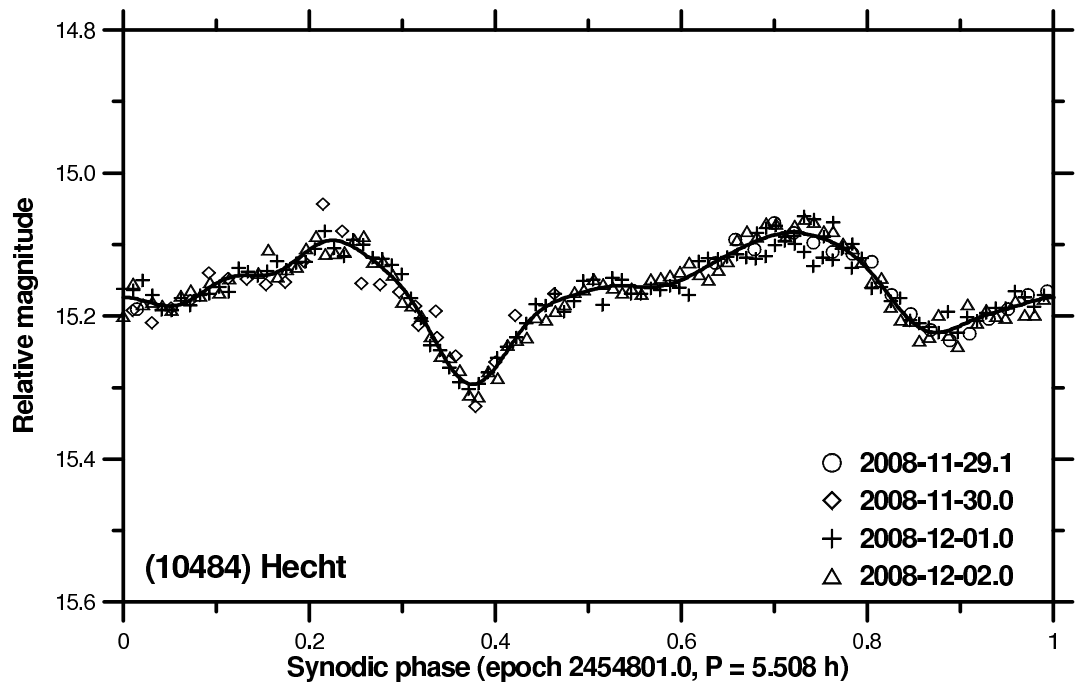

Supplementary Figure 12: Composite lightcurve of (10484) Hecht.

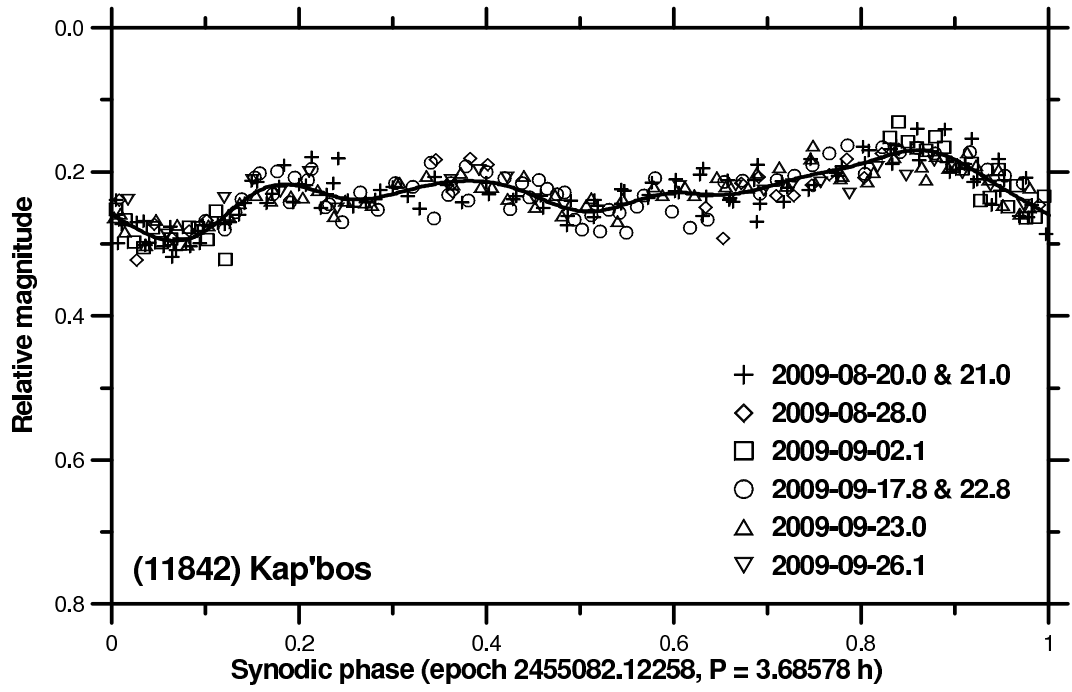

Supplementary Figure 13: Composite lightcurve of (11842) Kap'bos.

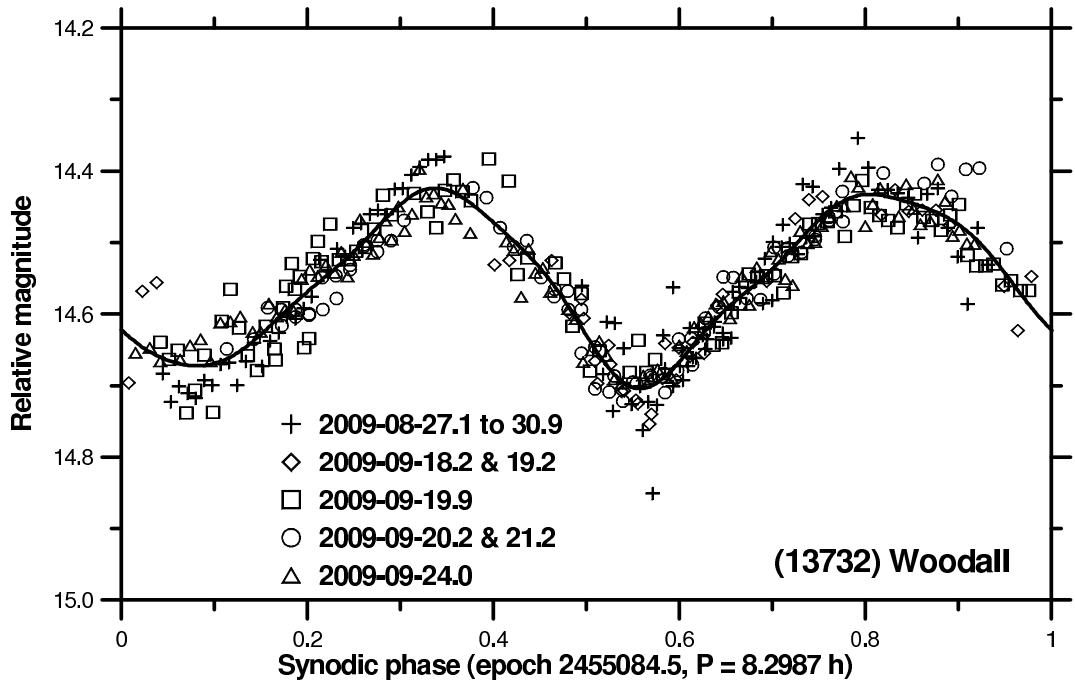

Supplementary Figure 14: Composite lightcurve of (13732) Woodall.

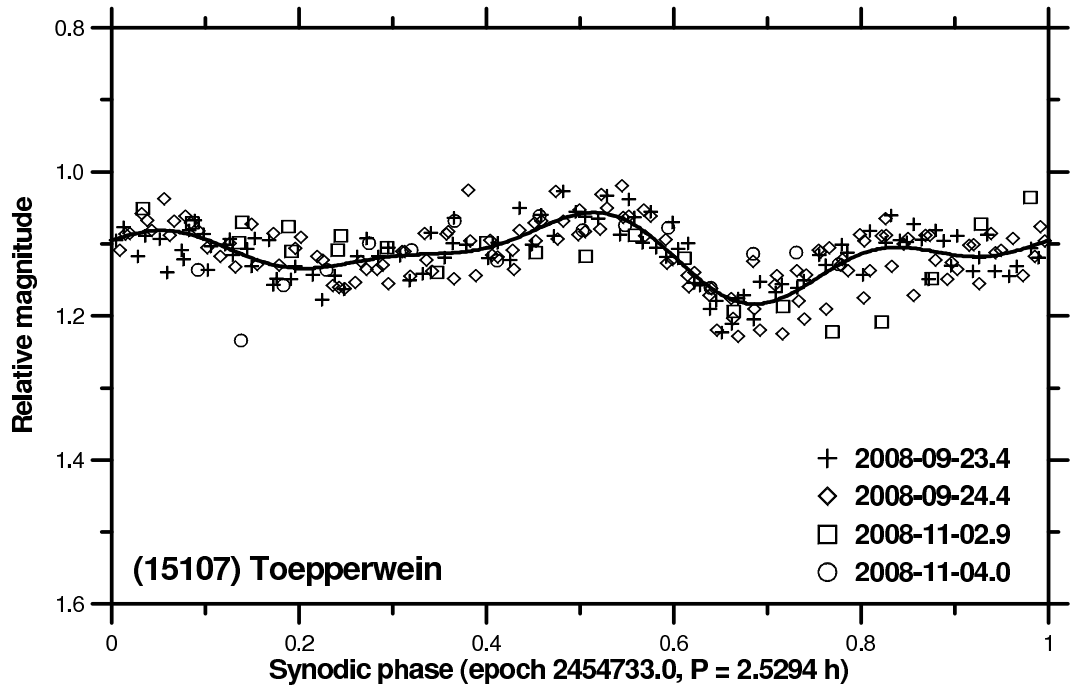

Supplementary Figure 15: Composite lightcurve of (15107) Toepperwein.

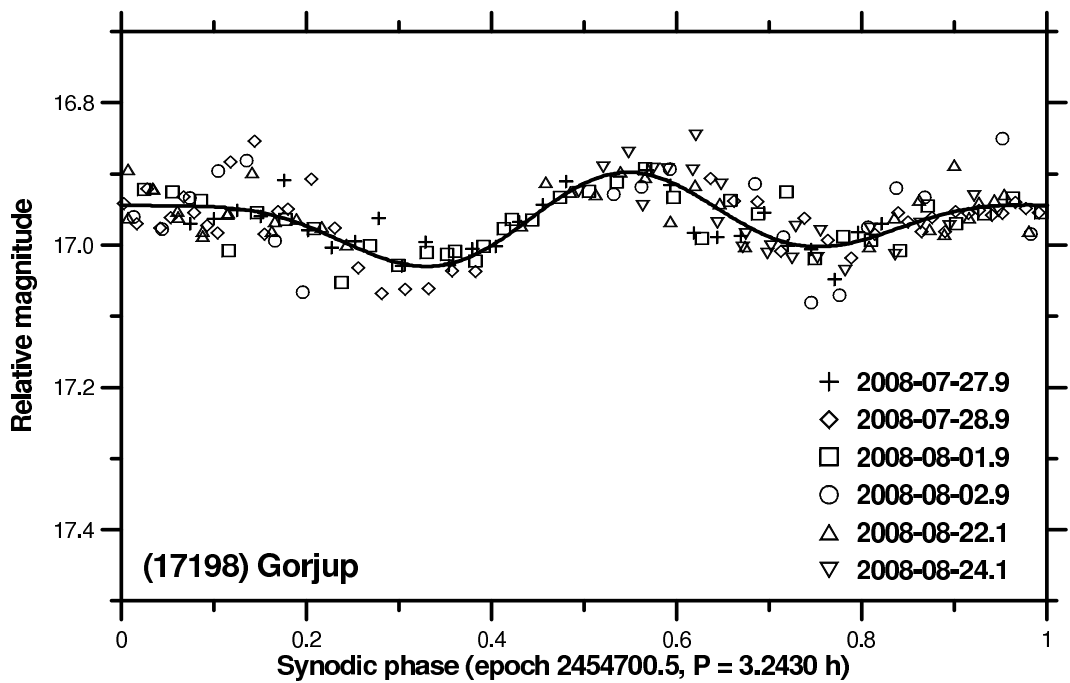

Supplementary Figure 16: Composite lightcurve of (17198) Gorjup.

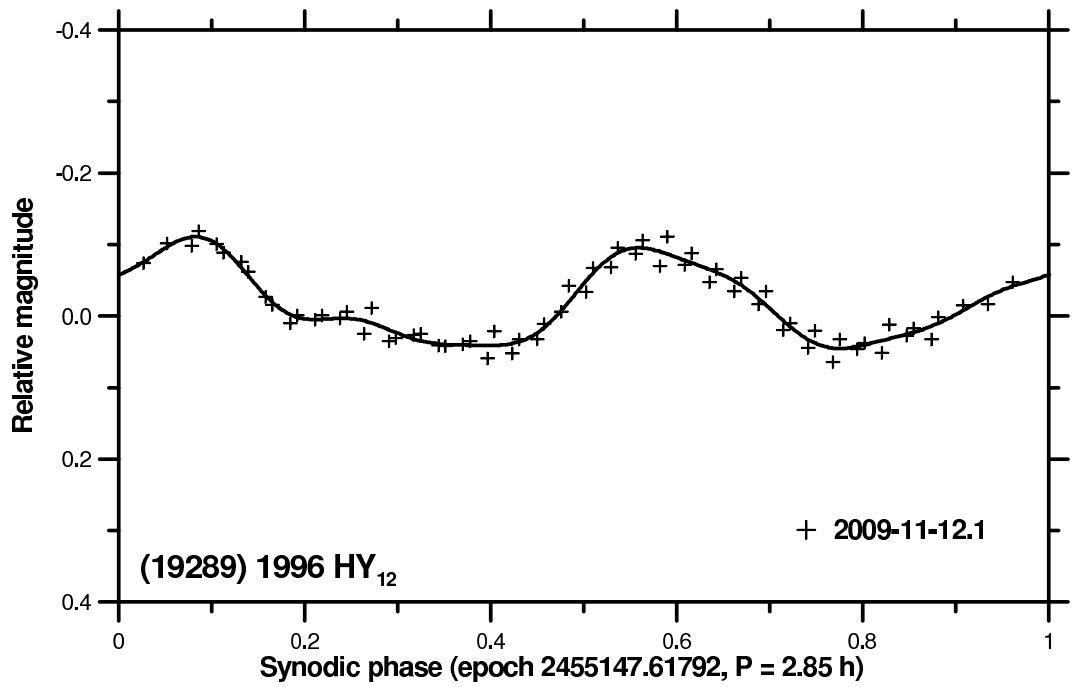

Supplementary Figure 17: Lightcurve of (19289) 1996 HY$_{12}$.

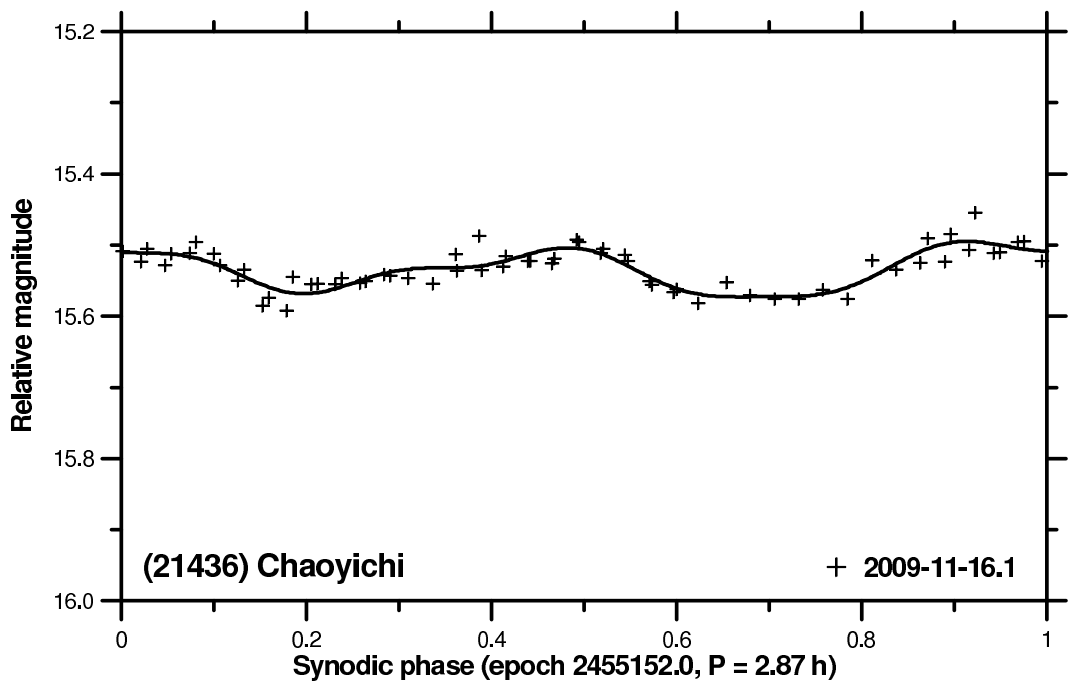

Supplementary Figure 18: Lightcurve of (21436) Chaoyichi.

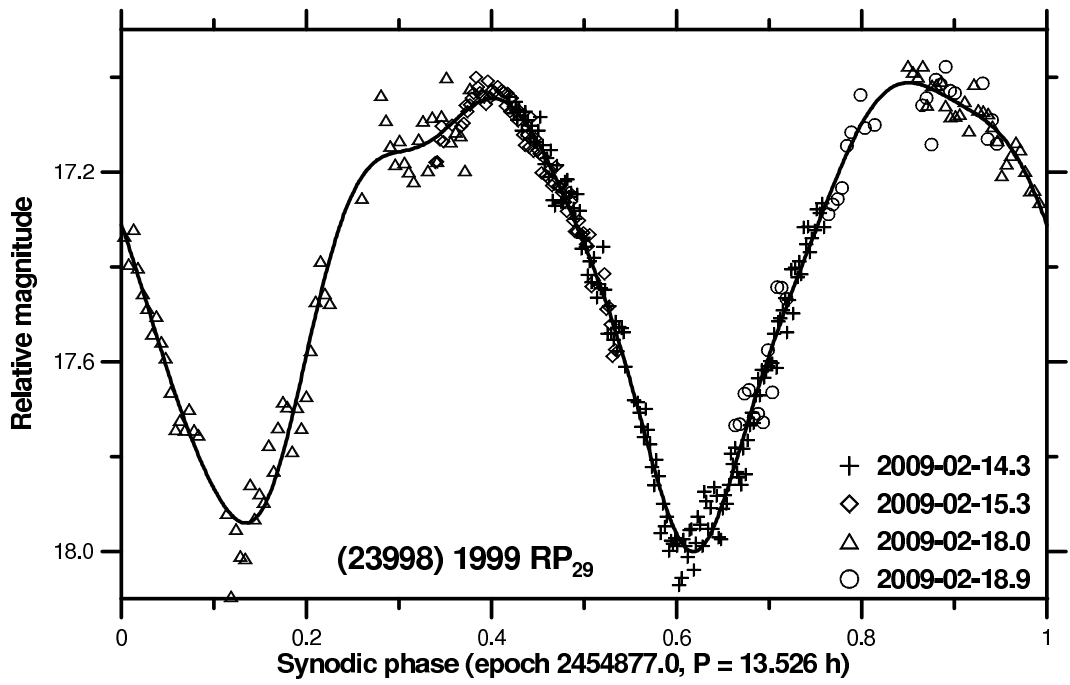

Supplementary Figure 19: Composite lightcurve of (23998) 1999 RP$_{29}$.

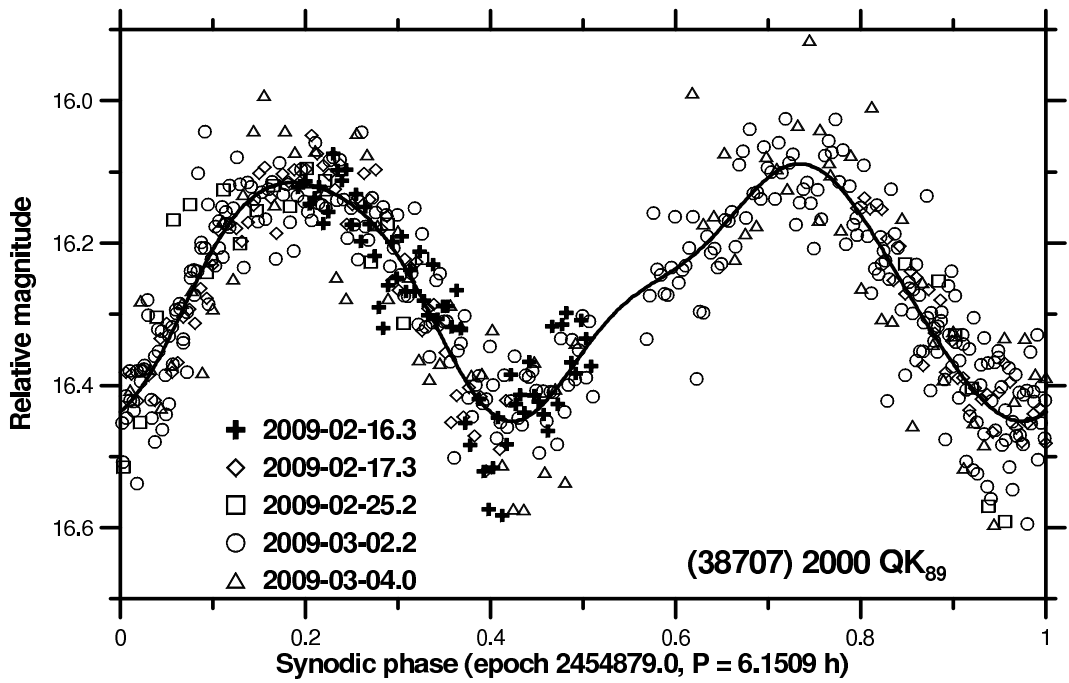

Supplementary Figure 20: Composite lightcurve of (38707) 2000 QK$_{89}$.

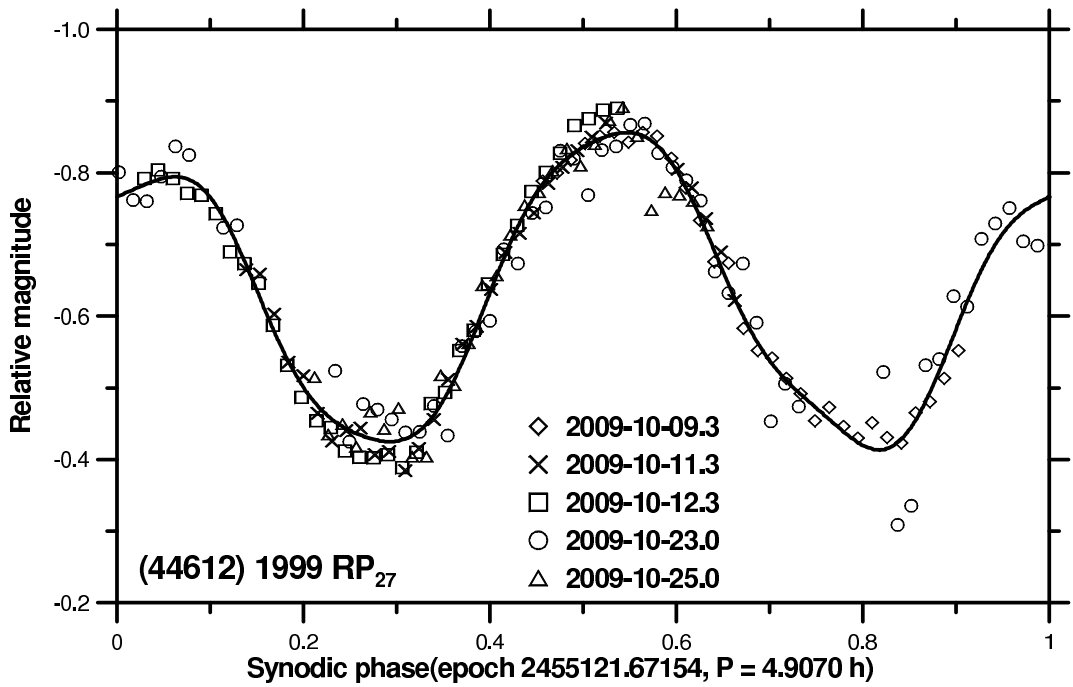

Supplementary Figure 21: Composite lightcurve of (44612) 1999 RP$_{27}$.

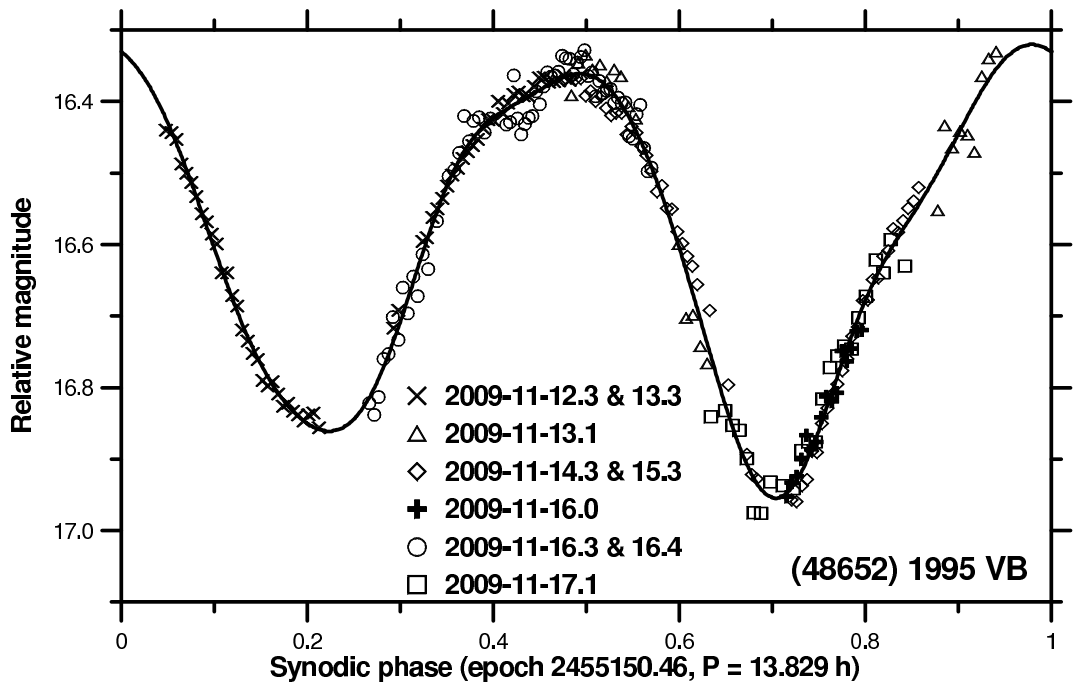

Supplementary Figure 22: Composite lightcurve of (48652) 1995 VB.

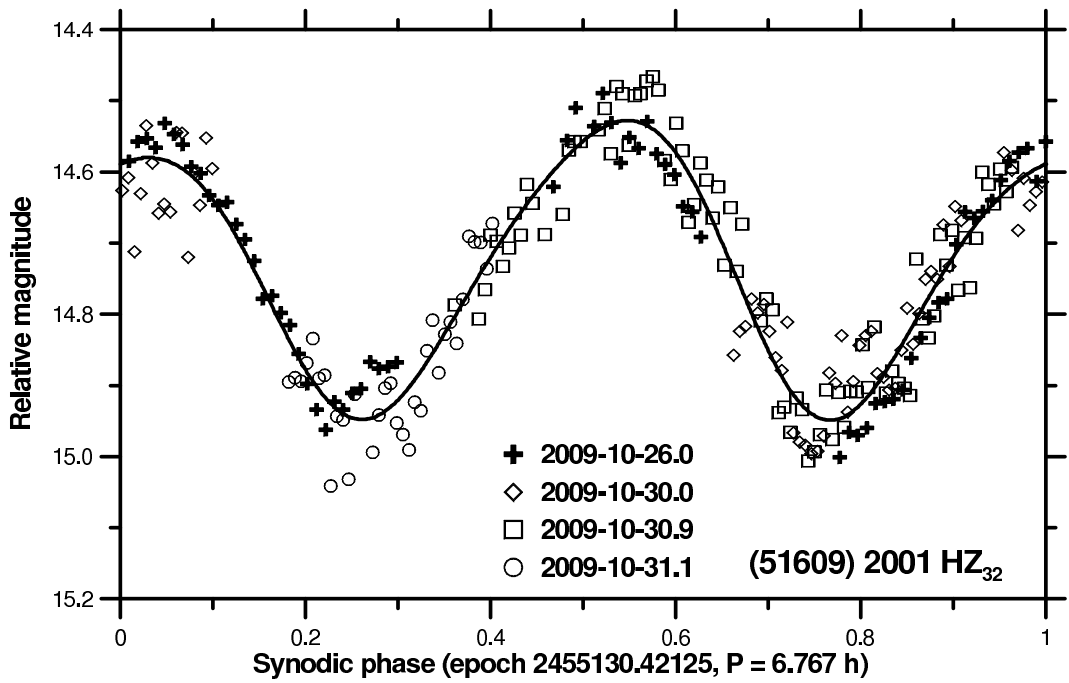

Supplementary Figure 23: Composite lightcurve of (51609) 2001 HZ$_{32}$.

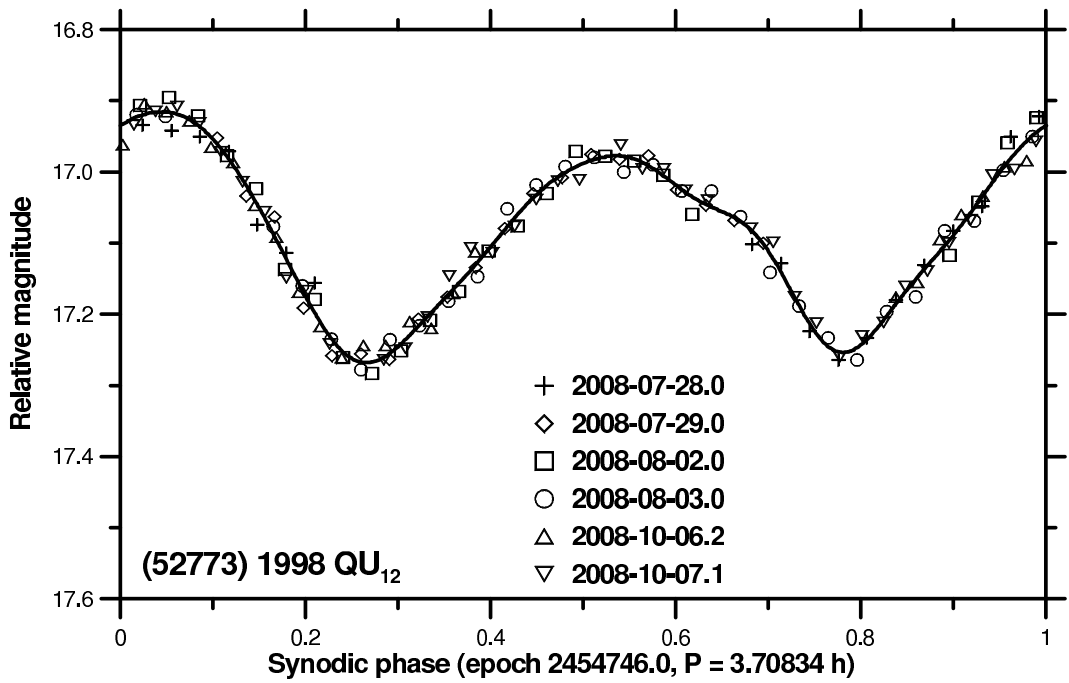

Supplementary Figure 24: Composite lightcurve of (52773) 1998 QU$_{12}$.

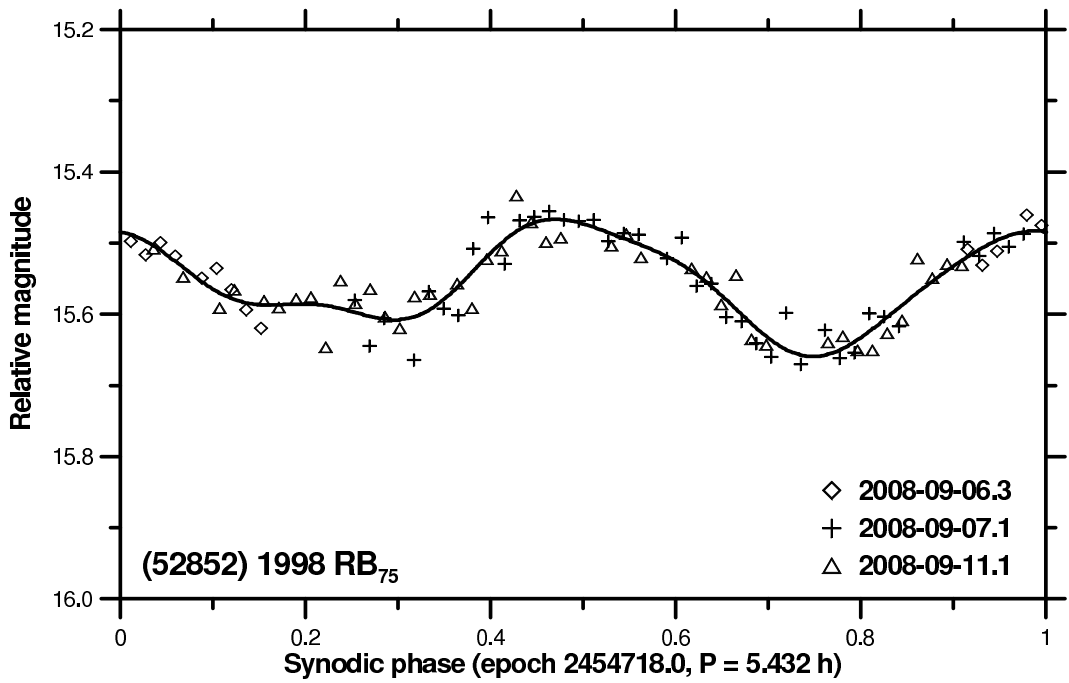

Supplementary Figure 25: Composite lightcurve of (52852) 1998 RB$_{75}$.

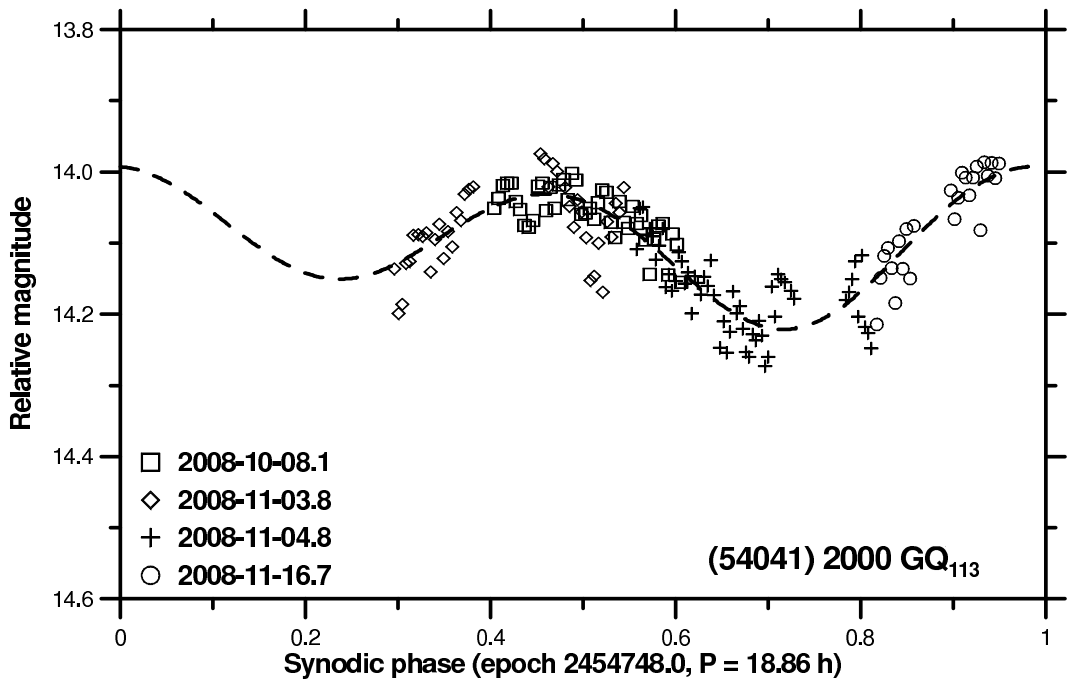

Supplementary Figure 26: Composite lightcurve of (54041) 2000 GQ$_{113}$.

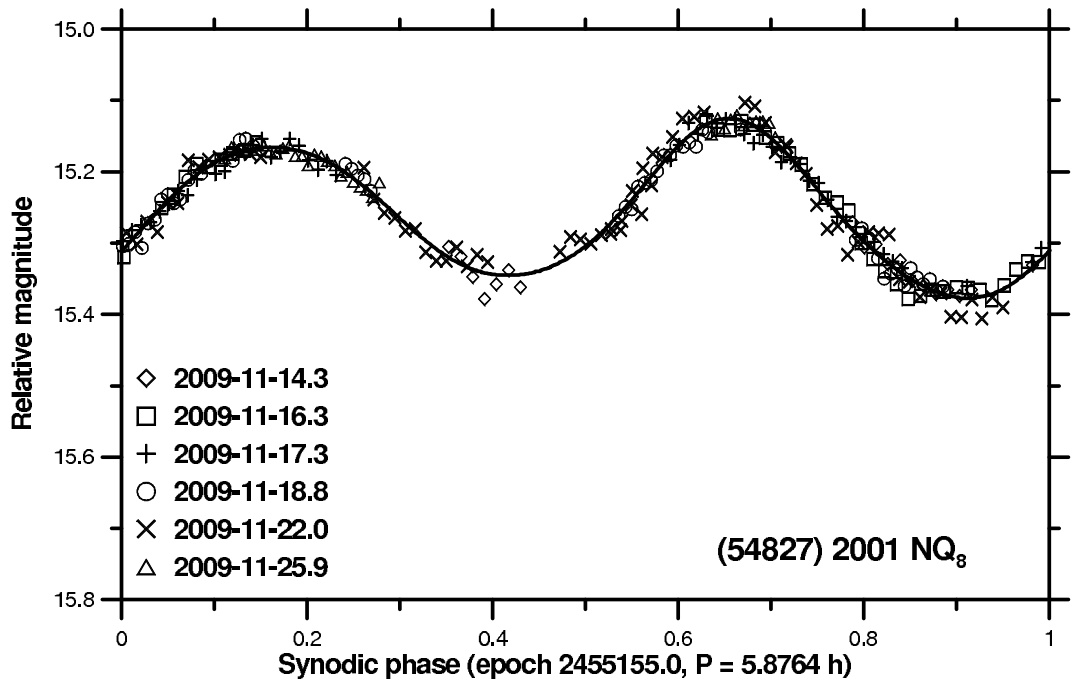

Supplementary Figure 27: Composite lightcurve of (54827) 2001 NQ$_8$.

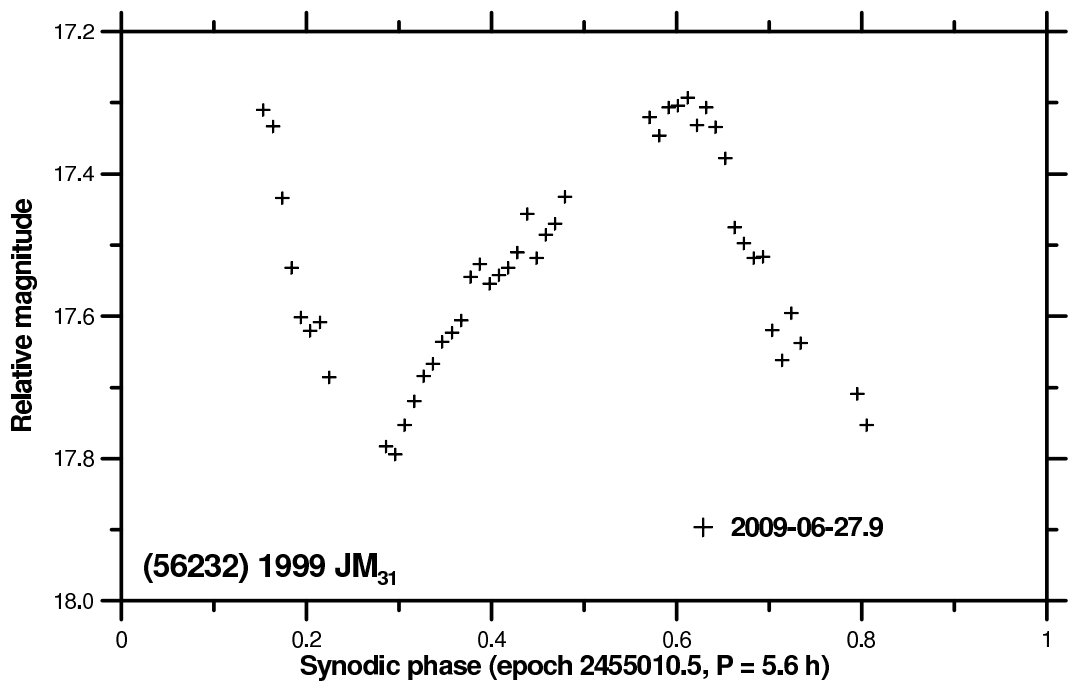

Supplementary Figure 28: Lightcurve of (56232) 1999 JM$_{31}$.

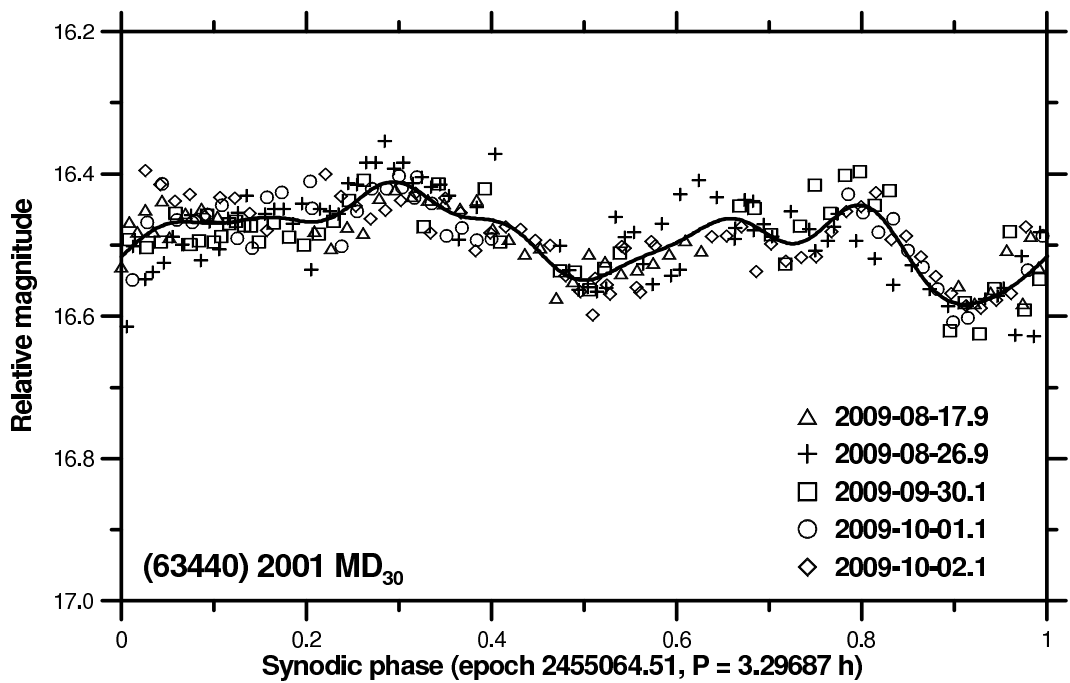

Supplementary Figure 29: Composite lightcurve of (63440) 2001 MD$_{30}$.

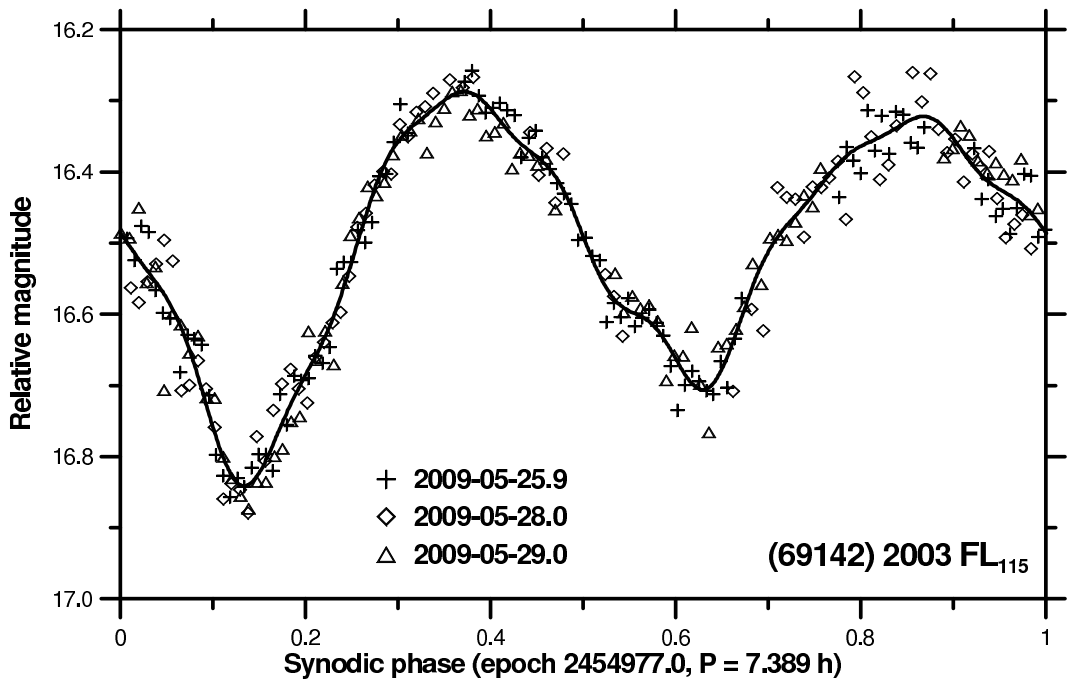

Supplementary Figure 30: Composite lightcurve of (69142) 2003 FL$_{115}$.

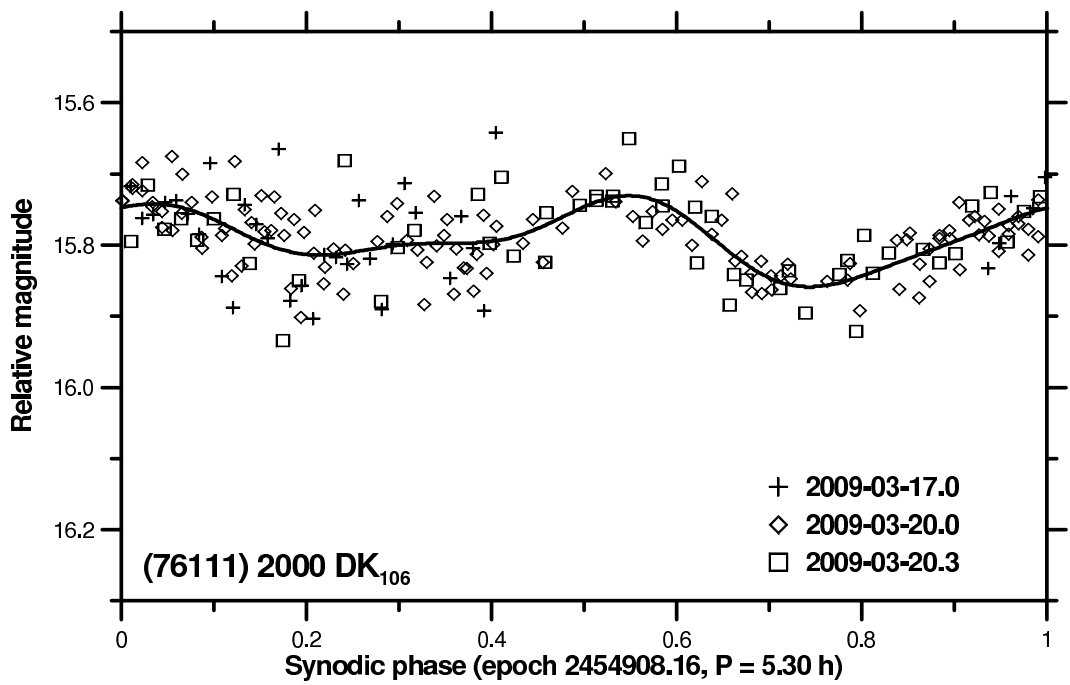

Supplementary Figure 31: Composite lightcurve of (76111) 2000 DK$_{106}$.

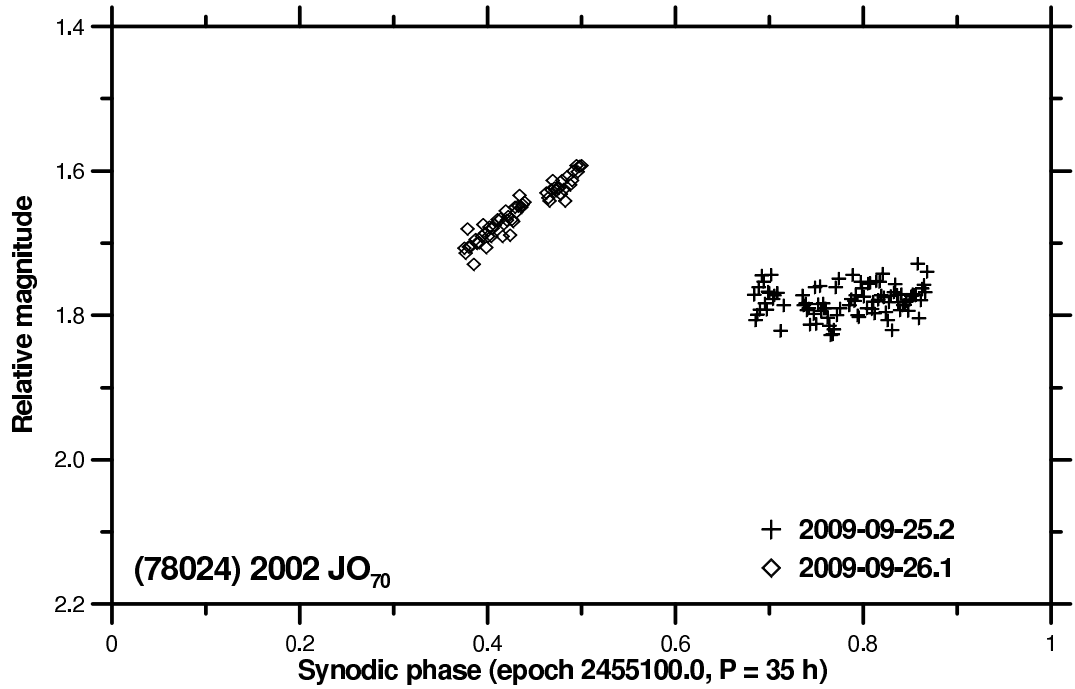

Supplementary Figure 32: Lightcurve of (78024) 2002 JO$_{70}$. Only a lower limit on the asteroid's period was estimated, see text. The period and magnitude scale offset used in this plot of the two relative nightly sessions are arbitrary.

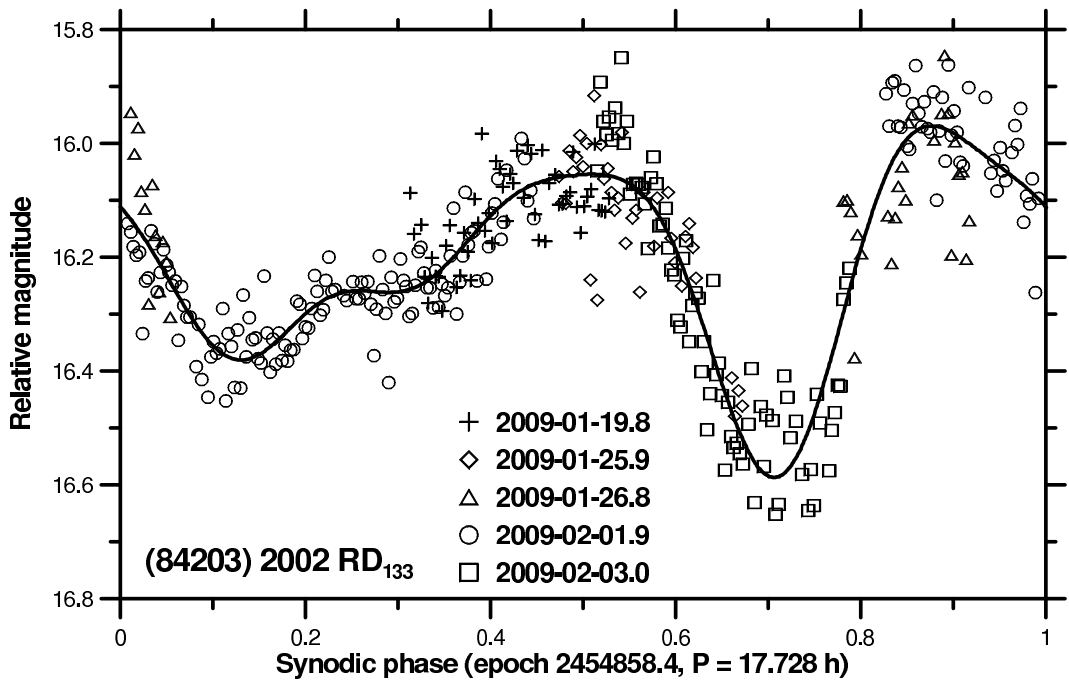

Supplementary Figure 33: Composite lightcurve of (84203) 2002 RD$_{133}$.

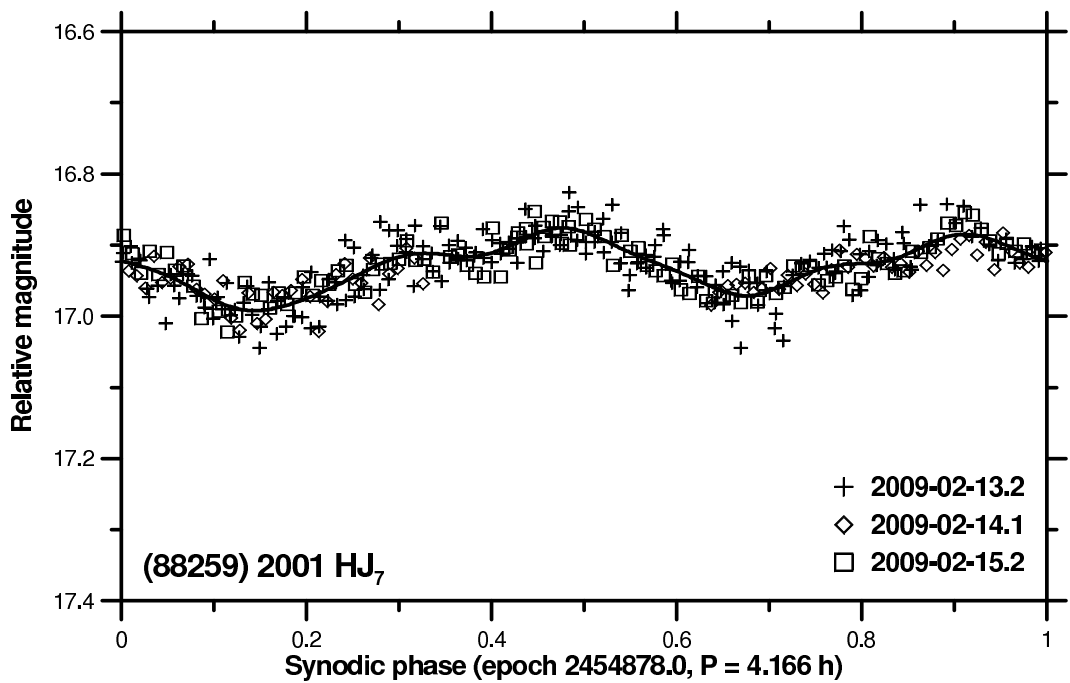

Supplementary Figure 34: Composite lightcurve of (88259) 2001 HJ$_7$.

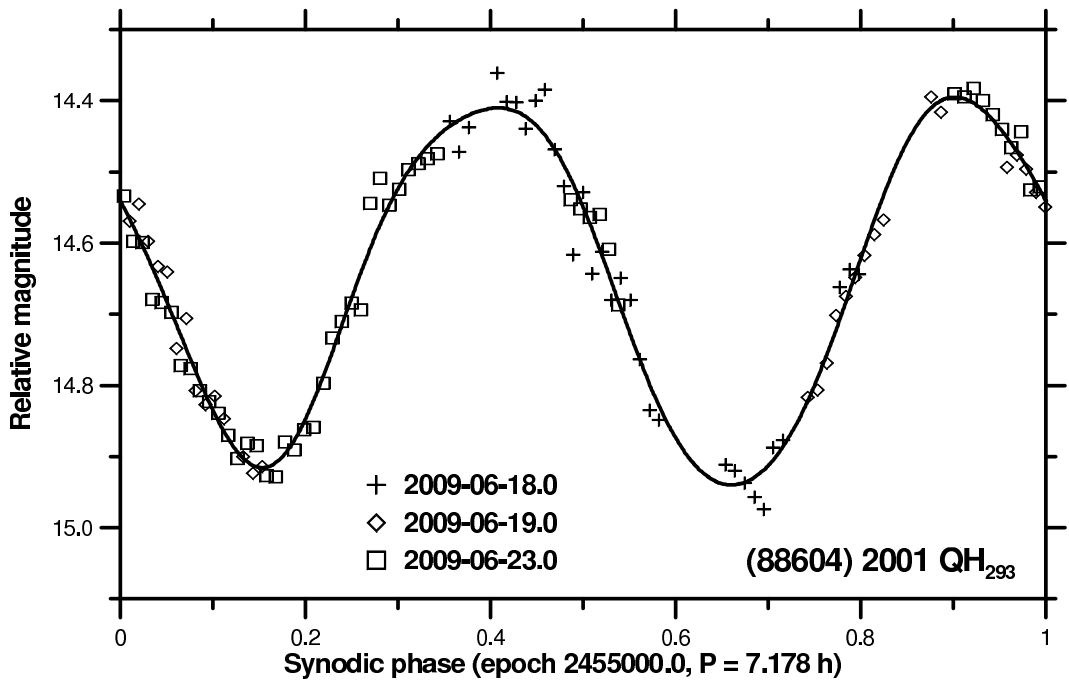

Supplementary Figure 35: Composite lightcurve of (88604) 2001 QH$_{293}$.

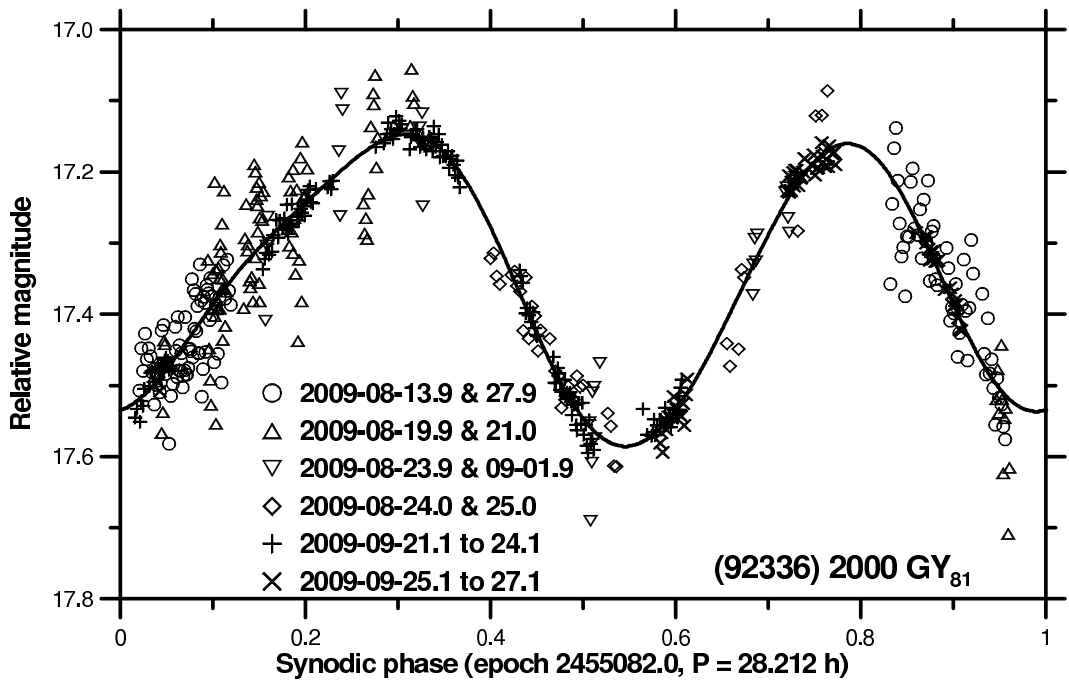

Supplementary Figure 36: Composite lightcurve of (92336) 2000 GY$_{81}$.

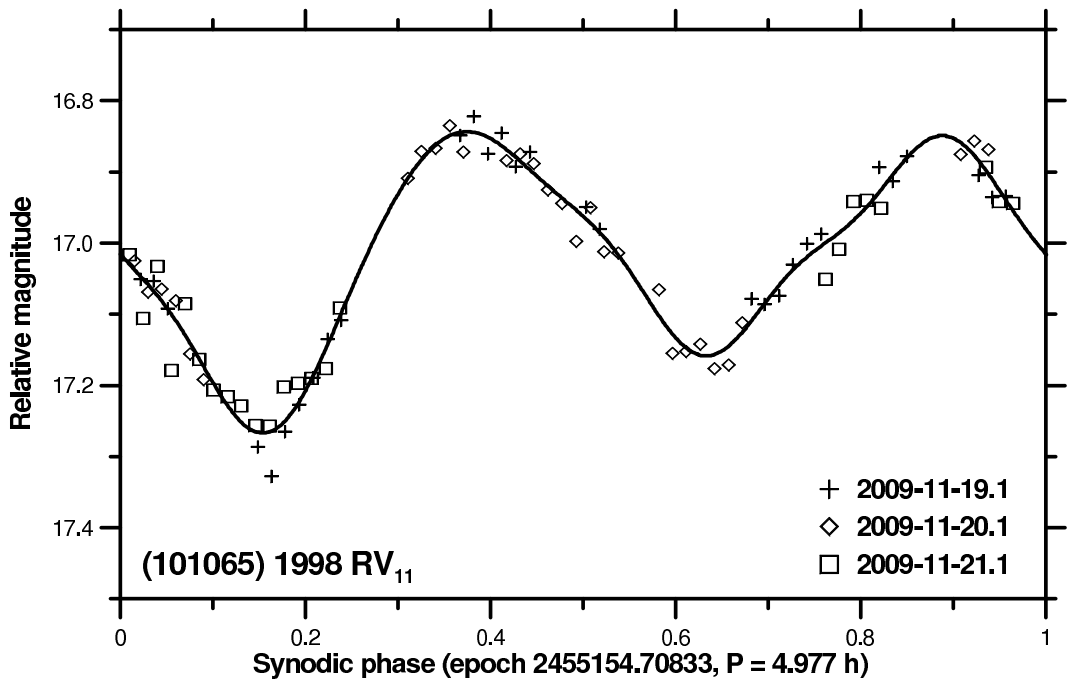

Supplementary Figure 37: Composite lightcurve of (101065) 1998 RV$_{11}$.

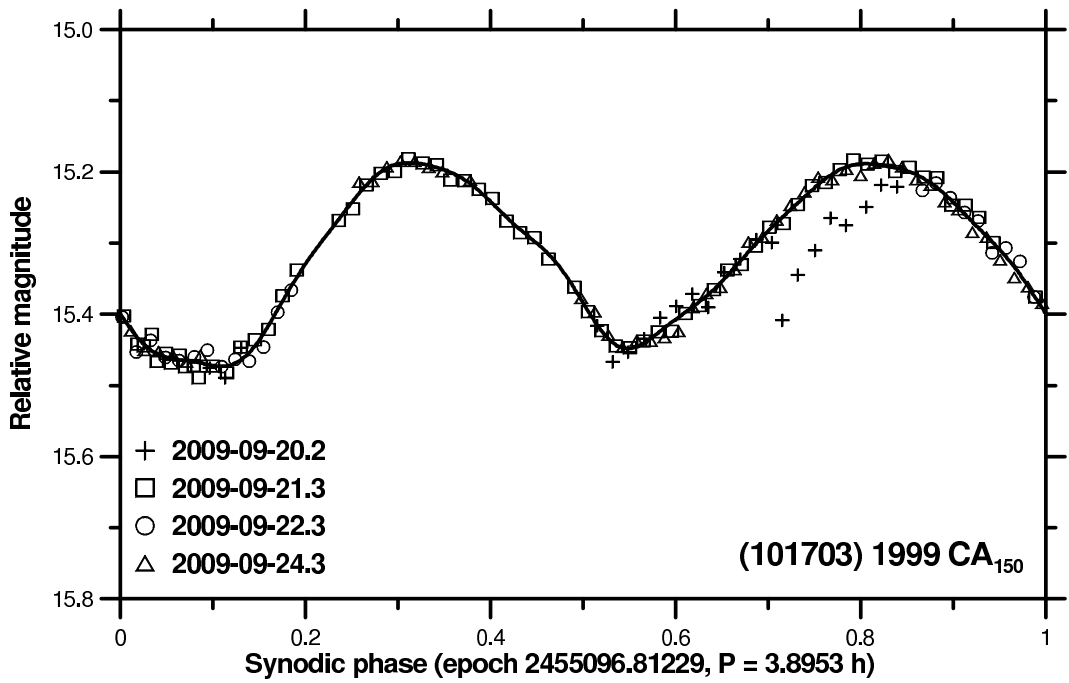

Supplementary Figure 38: Composite lightcurve of (101703) 1999 CA$_{150}$.

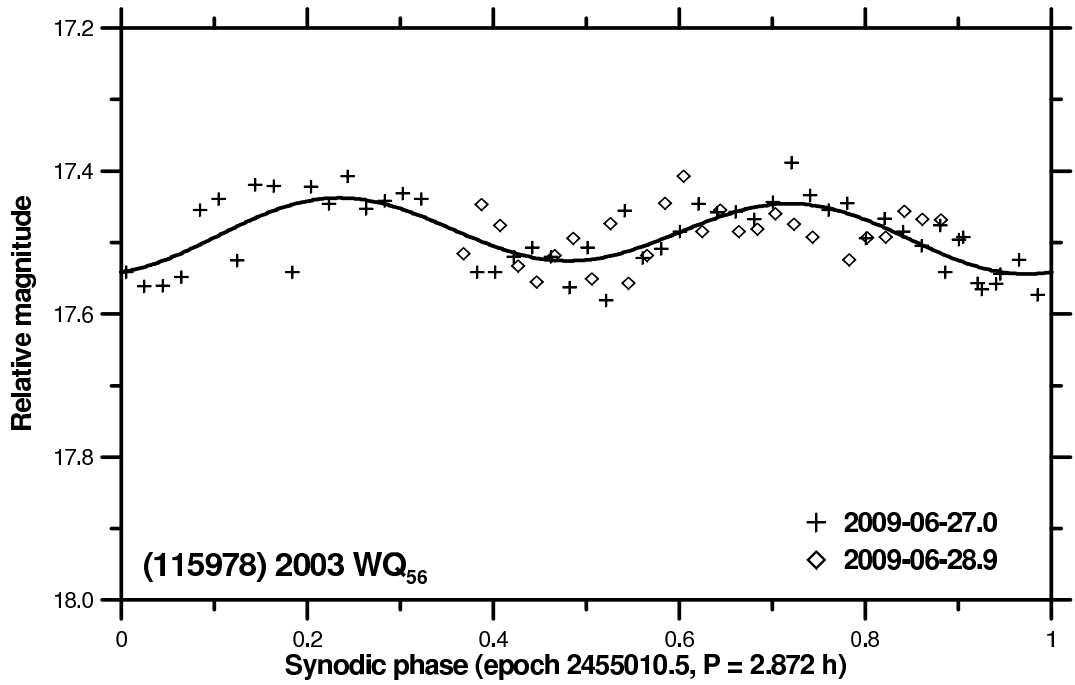

Supplementary Figure 39: Composite lightcurve of (115978) 2003 WQ$_{56}$.

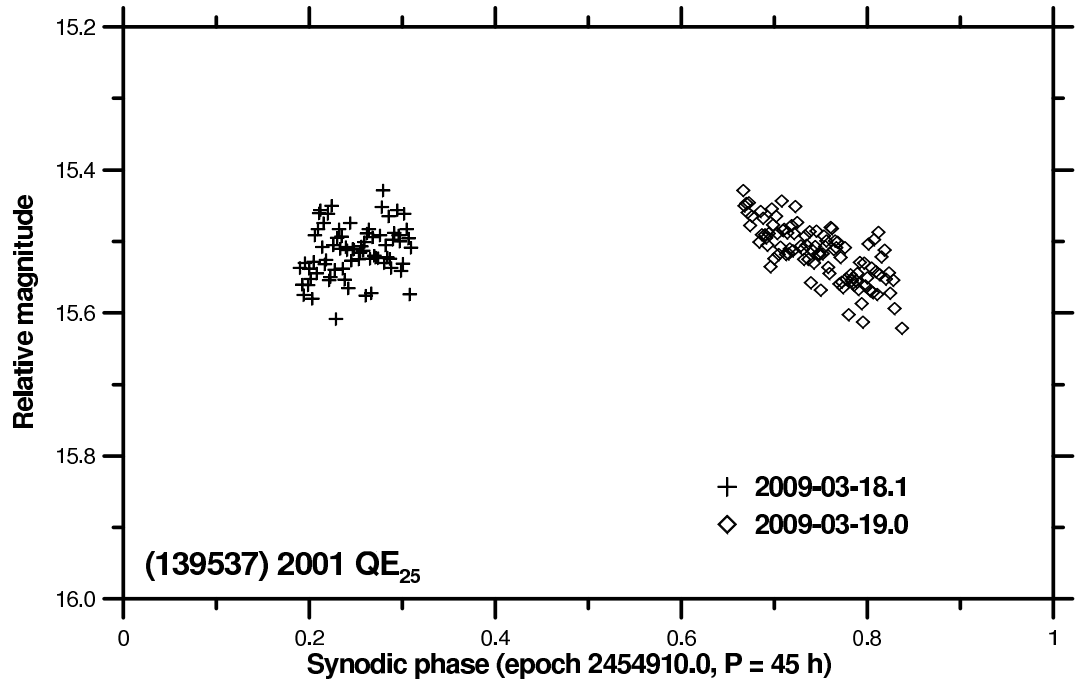

Supplementary Figure 40: Lightcurve data of (139537) 2001 QE$_{25}$. The magnitude scale offset between the two relative nightly sessions is arbitrary. See text for comments on the period estimate.

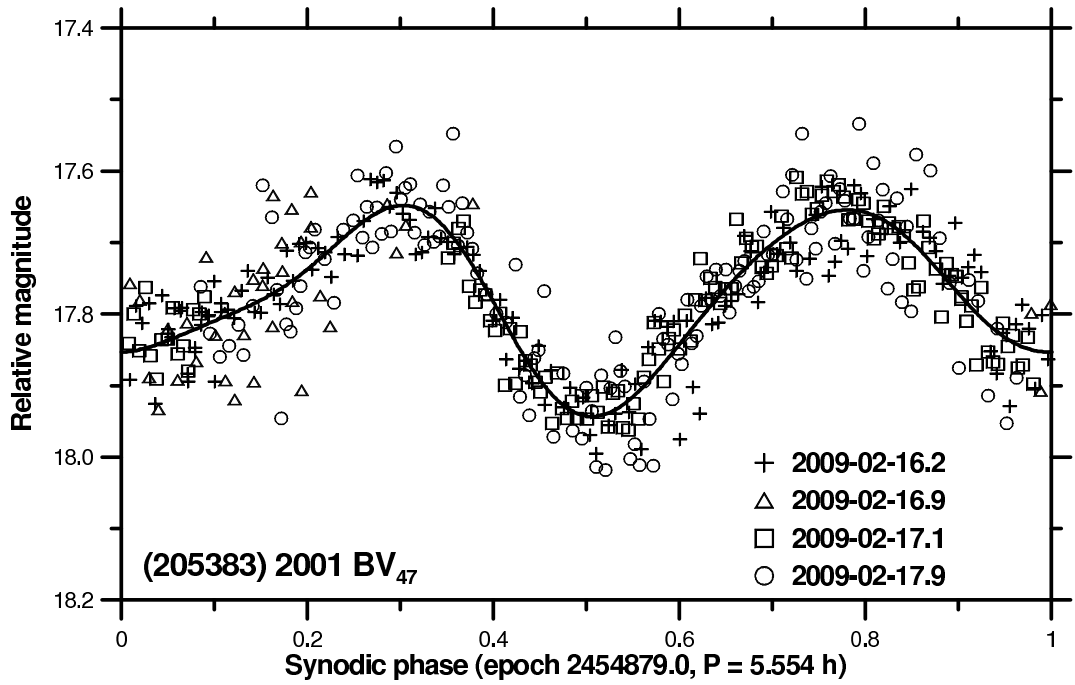

Supplementary Figure 41: Composite lightcurve of (205383) 2001 BV$_{47}$.

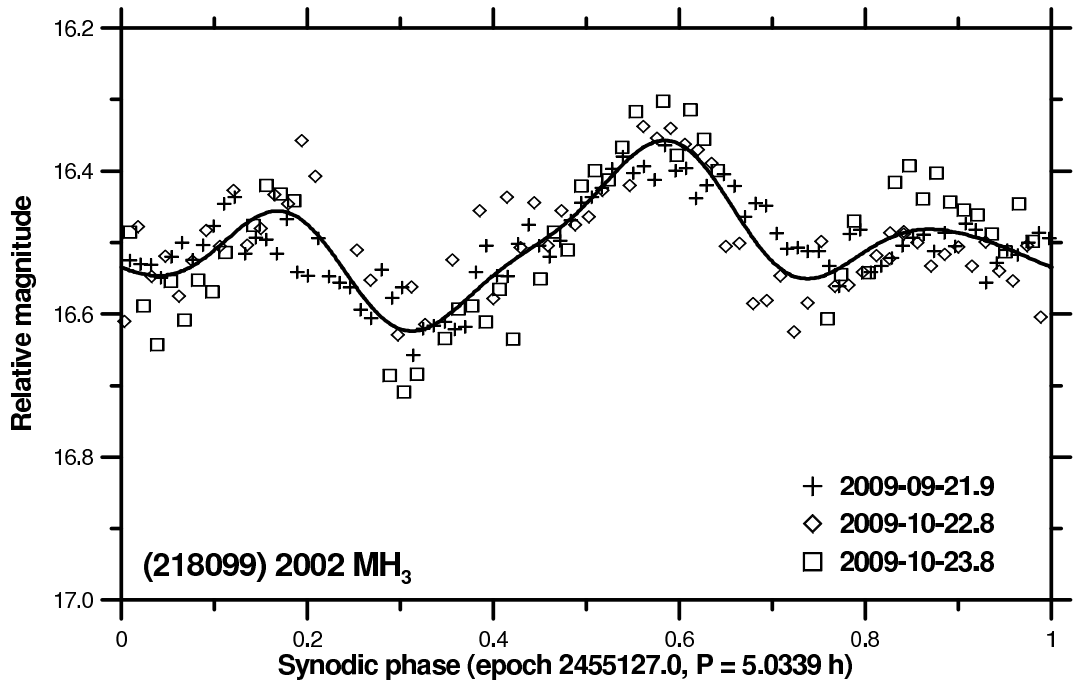

Supplementary Figure 42: Composite lightcurve of (218099) 2002 MH$_3$.

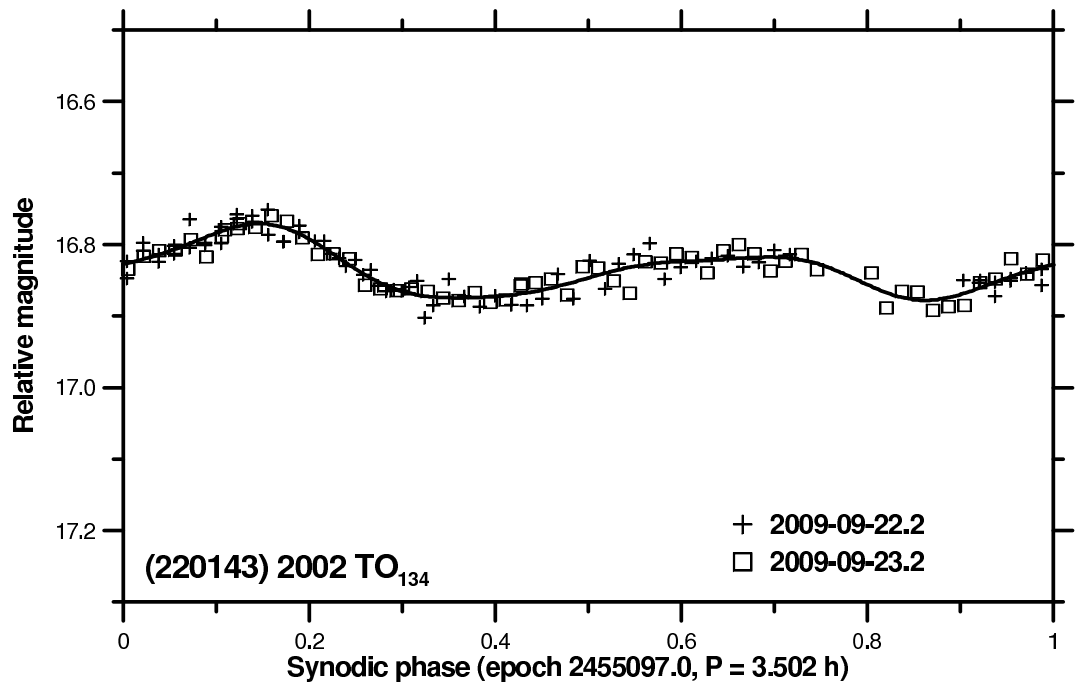

Supplementary Figure 43: Composite lightcurve of (220143) 2002 TO$_{134}$.

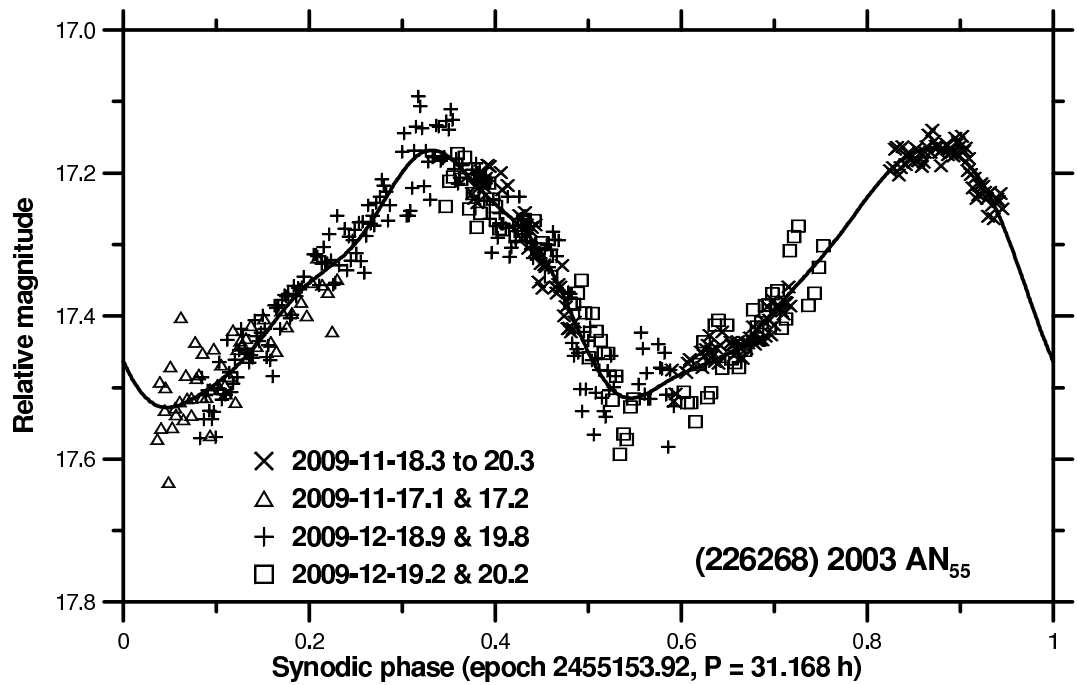

Supplementary Figure 44: Composite lightcurve of (226268) 2003 AN$_{55}$.

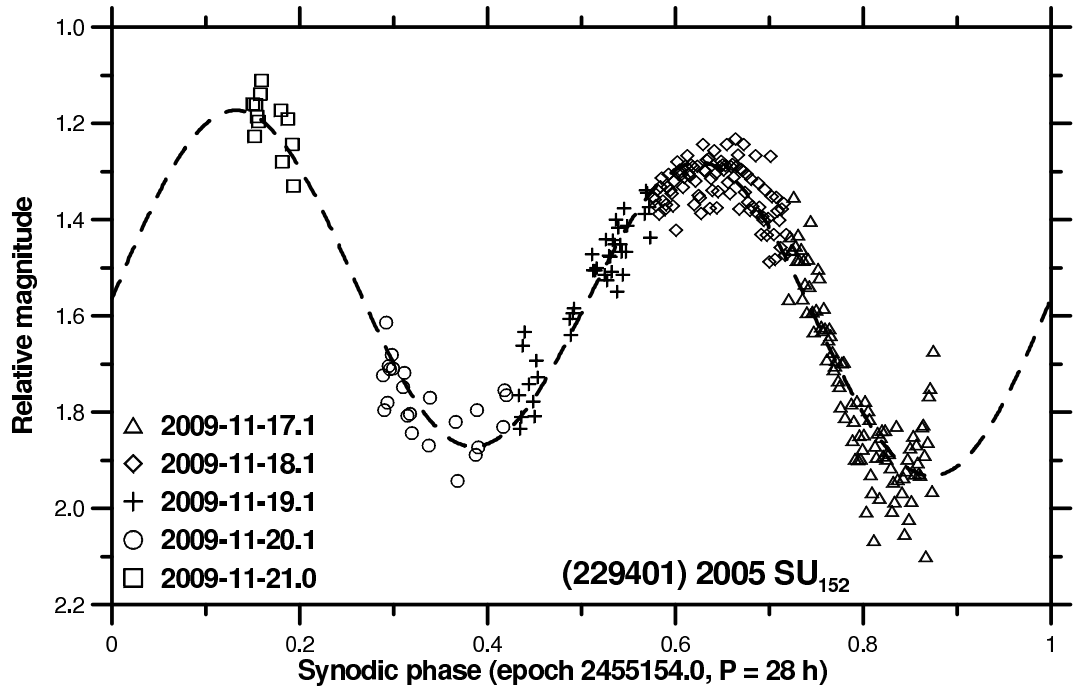

Supplementary Figure 45: Composite lightcurve of (229401) 2005 SU$_{152}$.